\documentclass[aps]{revtex4}
\def\pochh #1#2{{(#1)\raise-4pt\hbox{$\scriptstyle#2$}}}
\def\binom#1#2{\left(\begin{array}{c}#1\\#2\end{array}\right)}
\font\mas=msbm10 \font\mass=msbm7 \textfont8=\mas \scriptfont8=\mass

\usepackage{graphicx} 
\usepackage{dcolumn} 
\usepackage{color}
\usepackage{amsmath}

\begin {document}

\title {Time dependent propagator for an-harmonic oscillator with quartic term in potential. }
\author{J. Boh\' a\v cik}
\email{bohacik@savba.sk} \affiliation{Institute of Physics, Slovak
Academy of Sciences, D\' ubravsk\' a cesta 9, 845 11 Bratislava, Slovakia.}
\author{P. Pre\v snajder}
\email{presnajder@fmph.uniba.sk}
\affiliation{Department of Theoretical
Physics and Physics Education, Faculty of Mathematics, Physics and Informatics, Comenius University, Mlynsk\' a dolina F2, 842 48 Bratislava, Slovakia.}
\author{P. August\' in}
\email{peto1506@gmail.com}
\affiliation{Department of
Theoretical Physics and Physics Education, Faculty of Mathematics, Physics and Informatics, Comenius University, Mlynsk\' a dolina F2, 842 48 Bratislava, Slovakia.}

\begin{abstract}
In this work, we present the analytical approach to the evaluation of the conditional measure Wiener path integral. We consider the time-dependent model parameters. We find the differential equation for the variable, determining the behavior of the harmonic as well the an-harmonic parts of the oscillator. We present the an-harmonic part of the result in the form of the operator function.

 (PACS: 03.65.-w, 03.65Db, 03.65.d, 05.40.Jc)
 (MCS: 8140)
\end{abstract}
\maketitle

\section{Introduction}

Some time ago, we studied the problem of a description of the confinement-deconfinement phase transition in the infrared variant of the pure gluodynamics $SU(2)$ symmetric model at finite temperature. The temperature was inversely proportional to the interval of the periodicity of the gluon's functions in the imaginary time variable.
We followed the conjecture \cite{apel-car} that the infrared segment of the theory possesses the same confinement characteristics as the full theory.  We achieved the infrared sector by integrating out all degrees of freedom with nonzero coefficients at quadratic powers of the gluon fields. The result was the partition function for the infrared segment of the theory.   We proceeded by the Gaussian method of integration, and we abandoned the contributions of the higher powers of gluon's fields as the price for this evaluation tool.

We found that the effective model possesses two distinct parts \cite{plet, hs}.
We found the high-temperature part characterized by zero mass gluon modes. The effective potential possesses the global minimum in the zeroth value of the chromo-electric gluon condensate. The effective potential as the function of the chromo-electric gluon condensate was proportional to the sum of the square of this variable, and a periodic function. The interval of the periodicity of this periodic function was the function of the mean value of the Polyakov loop.

The low-temperature part possesses a more rich structure. The parabola from the effective potential disappeared, and potential was the periodic function of the chromo-electric gluon condensate. The positions of the minima of the periodic function were dependent on the temperature. The interval of the periodicity of this function is a function of the mean value of the Polyakov loop. In this low-temperature part, some gluon degrees of freedom from the high-temperature part disappeared, except the two, which became massive, if we consider in this model the coefficients at second-order power of the gluon's fields as the square of the mass of the gluon.
We find that the square of the mass term on the single interval of periodicity as a function of the mean value of the Polyakov loop is proportional to (minus) second Bernoulli polynomial $B_2$. On the edges of the periodicity interval, the mass square becomes negative. But for the temperatures, where the minima of the effective potential appear for those values
of the Polyakov's loop, the high-temperature scenario is more appropriate than the low-temperature model. A beautiful conjecture of this behavior is the first-order phase transition from $SU(2)$ symmetric high-temperature regime to $Z(2)$ symmetric low-temperature ones. To study this phenomenon, we must be able to take into account the contributions of the higher powers of the gluon fields, abandoned in Gaussian evaluations of the functional integral. The study of this possible phase transitions is our motivation to study the problem of the functional integral beyond the Gaussian approach analytically. Because the study of the simple non-Abelian $SU(2)$ model looks like a long-distance run, let us present the path-integral research of the simpler scalar an-harmonics oscillator.

In physics, one can meet the functional integration as the mathematical tool for study of the Brownian motion and in quantum physics as the tool for evaluations of the amplitudes of transition probabilities.
The problem of the Brownian motion as the classical analog of the quantum mechanical propagator studies allows evaluating the probability of finding a particle in a given time at the given place.
Mehler's formula gives the transition probability for Brownian motion in the external harmonic oscillator force (see \cite{mehler}, \cite{hille}, \cite{dob}):
\begin {equation}
W_{CL}(x_i,t_i;x_f,t_f) =
\left(\frac{k}{2 \pi \sinh{(\nu)}}\right)^{1/2}\ \exp{\left\{-\frac{k(x_i^2+x_f^2)}{2\tanh{(\nu)}}+\frac{k x_i x_f}{\sinh{(\nu)}}\right\}}
\end {equation}

Feynman \cite{fey, feynm} expressed the quantum mechanical amplitude of propagation of the particle for a general potential
$V(x)$ as a path integral of the following form:
\begin {equation}
W_{QM}(q_i, t_i; q_f, t_f) =  \int_{all\ paths}\ \prod_{\tau=t_i}^{t_f}\frac{dx(t)}{\sqrt{\frac{2i\pi\hbar}{m}dt}}
\exp{\left\{\frac{i}{\hbar}\int_{t_i}^{t_f}\ dt\ \left(\frac{1}{2}m\dot{x}^2 - V(x)\right)\right\}}
\label{fkr}
\end {equation}

Later Kac \cite{kac, jaffe} rigorously justified the imaginary time analog of Feynman path integral for a broad class of potentials
$V(x)$. The propagator in the imaginary time formalism is the conditional measure Wiener path integral defined by the continuum limit of the time-sliced finite-dimensional integral.
For the imaginary time and the harmonic potential, the Feynman propagator reproduces the Mehler's formula.

We encounter the same evaluation mechanism, the conditional measure Wiener path integral, both in classical as well as in quantum theory.
The principal difference between the classical description of the Brownian process as a random motion and the quantum particle motion
via path integral inhere in the interpretation of the results. In the classical description of the Brownian motion, we interpret
the path integral results as the probability of the displacement of the particle from the initial position "i" to the final
position "f." In quantum mechanics, we understand the results as the amplitude of the propagation of the system from the initial state "i"
to the final state "f," and this should not be confused with the statistical probability of the classical Brownian motion.

However, there are very few path integrals that allow explicit evaluations. Such are, e.g., systems of the harmonics oscillators
and the free fields. Gaussian path integration methods can evaluate the corresponding transition amplitudes and represent
the multidimensional generalizations of the Mehler's formula.
In the quantum mechanics and quantum field theory, the path integral approach provides a way to effectively derive
the standard perturbative expansions and even indicates steps beyond perturbative methods. (See, e.g., Roepstoff \cite{gert}, Das \cite{das}, Chaichian-Demichev \cite{dem}).

We aim to study the propagator in the symmetric an-harmonic potential with the quartic term, defined by the conditional measure Wiener path integral. We do not decompose the integration variables to the "classical" and the "fluctuation" parts as in the standard approaches to functional integrations. Still, we integrate for each slice point the whole variable, not a fluctuation only.
We study the model with time-dependent mass, frequency, and an-harmonic parameters.
In our model, the frequency of the harmonics force
can be negative, and then the potential has a double-well shape.

The article is organized as follows. In the next section, we explain our evaluation without
going to the details. We send the genuinely interested readers to the Appendices, where are the essential features of the evaluation arts clarified. In the third section, we discuss some results
following from the expansion of the umbral form of the an-harmonic part of the propagator.

\section{Evaluation of the path integral}

We define the propagator in the Euclidean field theory or statistical mechanics by the path integral with fixed initial and final paths points.
We are going to evaluate the Euclidean functional integral in phase space formally written as \cite{das}, \cite{dem}
\begin {equation}
\mathcal{W} = \int[\mathcal{D}\pi(\tau)] [\mathcal{D}\varphi(\tau)]\exp (-\mathcal{S}[\pi, \varphi])\ ,\label{phspace}
\end {equation}
\noindent
where $\pi(\tau)$ and $\varphi(\tau)$ are the phase space coordinates and  $\mathcal{S}[\pi, \varphi]$ denotes the corresponding Euclidean action.
The  quantity $\mathcal{S}[\pi, \varphi])$ is quadratic in the variable $\pi(\tau).$
Formally integrating by Gauss method over variable $\pi(\tau)$ we obtain the continuum \textit{conditional Wiener measure} path integral defined as
\begin {equation}
\mathcal{W} = \int\ \left[D\varphi(\tau)\right]\ \exp{(-E[\varphi])},
\label{pi1}
\end {equation}
where
\begin {equation}
E[\varphi] = \int\limits_0^{\beta}\ d\tau\left[c(\tau)/2\left(\frac{\partial
\varphi(\tau)}{\partial \tau}\right)^2
+b(\tau)\varphi(\tau)^2+a(\tau)\varphi(\tau)^4\right],
\end {equation}
\noindent
and $c(\tau), b(\tau), a(\tau)$ are functions of the time.

\noindent
We follow definition of the path integral (\ref{phspace}), (\ref{pi1})  as the limit of the finite dimensional integral \cite{das}. Integration over conjugate momentum gives a normalization constants for the integrations over the variables $\varphi_i$. In the \textit{conditional Wiener measure} path integral, the values $\varphi(0)=x_i$ and $\varphi(\beta)=x_f$ are fixed by definition. By time-slicing discretization  we define $N-1$ dimensional integral

\begin {equation}
\mathcal{W}_{N}=
\frac{1}{\sqrt{\frac{2\pi\triangle}{c_0}}}\,
 \int\limits
_{-\infty}^{+\infty} \prod \limits _{i=1}^{N-1} \frac{d\varphi_i}{\sqrt{\frac{2\pi\triangle}{c_i}}}
\exp(-E_{N}) , \label{afindim}
\end {equation}
where
\begin {equation}
E_{N}= \sum\limits _{i=1}^N \triangle\left[c_i/2
\left(\frac{\varphi_i-\varphi_{i-1}}{\triangle}\right)^2
+b_i\varphi_i^2+a_i\varphi_i^4\right], \label{afindim1}
\end {equation}
represent the standard time-slice discretization of $E[\varphi].$ The $\triangle= \beta / N$ is the interval between $i th$ and $i+1 th$ time points.
\noindent
The factors $\left(\frac{2\pi\triangle}{c_i}\right)^{-1/2}$ are the result of the gaussian integration over conjugate momentum variable $\pi(\tau_i)$ in the $i-th$ slice point. Theirs number is the same in the case of $N$ point time slicing for the \textit{conditional Wiener measure} as well as \textit{unconditional Wiener measure} path integrals \cite{das}. The \textit{conditional Wiener measure} path integral is defined by the continuum limit of (\ref{afindim}):

$$\mathcal{W} = \lim_{N\rightarrow\infty}\ \mathcal{W}_{N}.$$
This definition of the path integral evade the problem of the indefinite integral measure, because the measure of the finite dimensional integrals is defined correctly.

To evaluate the finite dimensional integrals in (\ref{afindim}) we must solve the problem of evaluation of the one dimensional integral
\begin {equation}
I_1(a,b,c)=\int\limits _{-\infty}^{+\infty}\;dx\;\exp\{-(a x^4+b x^2+c
x)\}
 \label {int1d}
\end {equation}
where $Re\: a>0$.

There are no explicit standard methods to evaluate one-dimensional integrals with a fourth-order polynomial in exponent non-perturbatively.
 However, $I_1(a,b,c)$ is an entire function for any complex $b$ and $c$, since there exist all integrals:
 \[ \partial_c^n\partial_b^m I_1(a,b,c)=(-1)^{n+m}
 \int\limits _{-\infty}^{+\infty}
 \;dx\;x^{2m+n}\exp\{-(a x^4+bx^2+cx)\}
 \]
Consequently, the power expansions of $I_1(a,b,c)$ in $c$
and/or $b$ has an infinite radius of convergence (and in
particular they are uniformly convergent on any compact set of
values of $c$ and/or $b$).
Performing the power expansion
in $c,$ which we shall frequently use, we find:
\begin {equation}
I_1(a,b,c)=\sum\limits _{n=0} ^{\infty} \frac{(-c)^n} {n!}\int\limits
_{-\infty}^{+\infty}\;dx\;x^n\exp\{-(a x^4+b x^2)\}
 \label{sim1}
\end {equation}
This integral can be expressed in terms of the
parabolic cylinder function $D_{\nu}(z)$, $\nu=-m-1/2$, (see, for
instance, \cite {prud}). For $n$ odd, due to symmetry of the
integrand the integral in (\ref{sim1}) is zero, for $n$ even, $n=2m$ we have:
\begin {equation}
  D_{-m-1/2}(z)\ =\
\frac{e^{-z^2/4}}{\Gamma(m+1/2)}\int_0^\infty\;dx\;
 x^{m-1/2}\exp\{-\frac{1}{2}x^2-zx\}
\label{s1}
\end {equation}

\noindent
Explicitly, for Eq.(\ref{sim1}) we have:
 \begin {equation}
 I_1(a,b,c)=\frac{\Gamma(1/2)}{(2a)^{1/4}}\ e^{z^2/4}\ \sum\limits _{m=0} ^{\infty}
 \frac{(\xi)^m}{m!}\ D_{-m-1/2}(z)
 \label{s2}\ ,\ \xi=\frac{c^2}{4\sqrt{2a}}\ ,\ z=\frac{b}{\sqrt{2a}}
\end {equation}
This sum is convergent for any values of $c$, $b$ and $Re\: a$
positive.
 The convergence of the infinite series (\ref{s2}) can be shown as follows.
 For $|z|$ finite, $|z|<\sqrt{|\nu|}$ and
$\mid arg(-\nu)\mid \leq \pi/2$, if $\mid \nu \mid \rightarrow
\infty$, the following asymptotic relation is valid \cite{bateman}:
\begin {equation}
D_{\nu}(z)=\frac{1}{\sqrt{2}}\; \exp\left[\frac{\nu}{2}
(\ln{(-\nu)}-1) -\sqrt{-\nu}\;
z\right]\left[1+O\left(\frac{1}{\sqrt{\mid \nu\mid
}}\right)\right] \label {ass1}
\end {equation}
The $m-th$ term  of the sum (\ref {s2}) possesses the asymptotic
\begin {equation}
\frac{1}{m!}\exp\left[-\frac{m}{2}(\ln{m}-1)
-\sqrt{m}\;z+z^2/4+m\ln\xi \right]
\end {equation}

Following Bolzano-Cauchy's criteria, the sum (\ref {s2}) is not only
absolutely, but uniformly convergent also for the finite values of
the parameters of the integral (\ref {int1d}).

The another attractive possibility to evaluate the integral (\ref {int1d}) is the method
based on the generating function for Hermite polynomials. If we replace the linear and quadratic terms in Eq. (\ref {int1d}) by the generating function for Hermite polynomials $H_n(x)$ (see Bateman \cite{bateman}):
$$\sum_{n=0}^{\infty}\frac{z^n}{n!}\ H_n(x) = \exp{(2zx-z^2)}, \ \
H_n(x) = n!\sum_{j=0}^{\lfloor\frac{n}{2}\rfloor}\ \frac{(-1)^j(2x)^{n-2j}}{(n-2j)!j!},$$
we find:
$$\exp\{-(c x+b x^2)\} = \sum_{n=0}^{\infty}\ \frac{(\sqrt{b}x)^n}{n!}\ H_n\left(\frac{-c}{2\sqrt{b}}\right).$$
Inserting this into the  integral (\ref {int1d}), taking into account that nonzero contributions we find for even summation indexes $n=2m$ only, and by reordering of the summations:
$$\sum_{m=0}^{\infty}\ \sum_{j=0}^{m}\ \rightarrow \
\sum_{\mu=0}^{\infty} \sum_{j=0}^{\infty}(m-j=\mu)\ \ , $$
we find:
$$I_1 = \frac{1}{(2a)^{1/4}}\sum_{\mu=0}^{\infty}
\frac{\left(\frac{c^2}{\sqrt{2a}}\right)^{\mu}}{(2\mu)!}\Gamma(\mu+1/2)\
\sum_{j=0}^{\infty}\ \frac{\left(-\frac{b}{\sqrt{2a}}\right)^j}{j!}\ \pochh{\mu+1/2}{j}\ D_{-j-\mu-1/2}(0)$$
The sum over index $j$ is the Taylor's expansion of the function (see Eq. (\ref{pointcp}))
$$\exp{\left(\frac{b^2}{4(2a)}\right)}D_{-\mu-1/2}\left(\frac{b}{\sqrt{2a}}\right)$$
around argument $0$. We find for $I_1$ by this method of calculation the same result (\ref{s2})
as for method based on the Taylor's expansion of the linear term in the exponent.

To evaluate $N-1$ dimensional integral (\ref{afindim}), we expand the parts linear in the integration variables $\varphi_i$   in the function $\exp{(-E_N)}$ in Eq. (\ref{afindim1}) by Taylor's expansion.
Evaluating the individual integrals in Eq. (\ref{afindim}) by above described method, we find:

\begin {eqnarray}
& &\mathcal{W}_{N} =\frac{1}{\sqrt{\frac{2\pi\triangle}{c_0}}}\ \sum\limits_{\rho=0}^{1}
\sum\limits_{n_0=0}^{\infty}\frac{\left(\frac{c_1}{\triangle} \varphi_0 \sqrt{\sigma_1}\right)^{2n_0+\rho}}{(2 n_0+\rho)!}\label{afindim3} \\
&\times&\prod_{i=1}^{N-2}\left\{\frac{1}{\sqrt{\frac{c_i+c_{i+1}}{c_i}+2 \frac{b_i}{c_i}\triangle^2}}\
\sum\limits_{n_i=0}^{\infty} \frac{\Sigma_i^{n_i+\rho/2}}{n_i!}\ \pochh{n_i+\rho+1/2}{n_{i-1}}\mathcal{D}_{-n_i-\rho-1/2-n_{i-1}}(z_i)\right\}\nonumber\\
&\times&
\left\{\frac{1}{\sqrt{\frac{c_{N-1}+c_{N}}
{c_{N-1}}+2\frac{b_{N-1}}{c_{N-1}}\triangle^2}} \sum\limits_{n_{N-1}=0}^{\infty}
\frac{\left((\frac{c_N}{2\triangle}\varphi_N)^2\sigma_{N-1}\right)^{n_{N-1}+\rho/2}}{n_{N-1}!}
\pochh{n_{N-1}+\rho+1/2}{n_{N-2}}\mathcal{D}_{-n_{N-1}-\rho-1/2-n_{N-2}}(z_{N-1})\right\}\nonumber \\
&\times&\exp{\left\{-a_N\triangle\varphi_N^4 - (\frac{c_N}{2\triangle}+b_N\triangle)\varphi_N^2-\frac{c_1}{2\triangle}\varphi_0^2\right\}}. \nonumber
\end {eqnarray}
The summation index $\rho$ was introduced since an integration variable
 $\varphi_i$ appears in the Taylor's expansions in two consecutive time slice points. Therefore $\varphi_i$ appears in integrand with power equal to the sum of two consecutive summation indexes $n_i+n_{i+1},$ which must be even. It is possible if all $n_i$ are even or odd, and we find two groups of contributions to $N-1$ dimensional integral. Value $\rho = 0$ correspond to the case when all $n_i$ are even and $\rho = 1$ correspond to all $n_i$ are odd.

 The parameters of the model in i-th time slice point are $a_i, b_i, c_i$. To simplify the notation of the equations we introduced the symbols, which are the functions of the parameters of the model:
\begin {equation}
z_i = \frac{\frac{c_i+c_{i+1}}{2\triangle}+b_i\triangle}{\sqrt{2 a_i \triangle}},\ \
\sigma_i = \frac{1}{\frac{c_i+c_{i+1}}{2\triangle}+b_i\triangle},\ \
\Sigma_i = \left(\frac{c_{i+1}}{2\triangle}\right)^2\ \sigma_i\ \sigma_{i+1}.
\label{Sigma}
\end {equation}
The $\mathcal{D}$ function is defined as:
\begin {equation}
\mathcal{D}_{-n_i-\rho-1/2-n_{i-1}}(z_i) = z_i^{n_i+\rho+1/2+n_{i-1}}\ \exp{\left(\frac{z_i^2}{4}\right)}\ D_{-n_i-\rho-1/2-n_{i-1}}(z_i).,
\end {equation}
Let us stress that parameters $a_i$ specifying an-harmonic behaviour appears only in the functions $\mathcal{D}_{\nu}(z_i)$.

The result in Eq. (\ref{afindim3}) is exact. We have not used any approximation in the evaluation. Our aim in evaluations of (\ref{afindim3}) is to sum over all indices $n_i$ and $\rho.$
We decide to use for sum over $n_i$ the summations formula (\ref{pointcp}) derived from the Taylor's expansion formula (\ref{tpc}) for parabolic cylinder functions.
To provide this task, we introduced the \textit{only} approximation in our calculation. We must solve the problem of how to evaluate the sum of the product of two parabolic cylinder functions, with the dependence on the same summation index. We decide to approximate one of the parabolic cylinder function by Poincar\'e type expansion\cite{olver},\cite{temme},\cite{temme2}, to prepare the sum for application of the Taylor's expansion formula.
We postpone the description of these evaluations to the Appendixes A - F. Here we present the result:
\begin {eqnarray}
& &\mathcal{W}_{N}^{leading} =\ \frac{1}{\sqrt{\frac{2\pi\triangle}{c_0}\ \prod_{i=1}^{N-1}(\frac{c_i+c_{i+1}}{c_i}+2 \frac{b_i}{c_i}\triangle^2)\Omega_{i-1}}}\
\sum_{\mu=0}\ \frac{4!}{(4\mu)!}\ (-1)^{\mu} \left(\frac{\triangle\ \varphi_N^4}{\mathcal{X}^4\ Q_{N-1}^4}\right)^{\mu}\label{Ndim8} \\
&\times&
\sum_{p_1=1}^{N-\mu}\ a_{p_1}Q_{p_1}^4\ \sum_{p_2=p_1+1}^{N-\mu+1}\ a_{p_2}Q_{p_2}^4\ .....
\sum_{p_{\mu}=p_{\mu-1}+1}^{N}\ a_{p_{\mu}}Q_{p_{\mu}}^4\nonumber \\
&\times&
\left\{\lim_{\xi_{\mu-1}\rightarrow 1}\ \partial_{\xi_{\mu-1}}^4\right\}\  \cdots \left\{\lim_{\xi_{i}\rightarrow 1}\ \partial_{\xi_{i}}^4\ \xi_{i}^{-4(\mu-i-1)}\right\}
\cdots
\left\{\lim_{\xi_{1}\rightarrow 1}\ \partial_{\xi_{1}}^4\ \xi_{1}^{-4(\mu-2)}\right\}\nonumber \\
&\times&\
\left(\sqrt{-D_1-\xi_1^2 D_2-\cdots -\xi_1^2\xi_2^2.\cdots. \xi_{\mu-2}^2\xi_{\mu-1}^2 D_{\mu}\ }\right)^{4\mu}\nonumber \\
&\times& H_{4\mu}
\left(\frac{D_1+\xi_1 D_2+\cdots +
\xi_1\cdots \xi_{\mu-2}\xi_{\mu-1}(D_{\mu}+\mathcal{X}/2)}{\sqrt{-D_1-\xi_1^2 D_2-\cdots -\xi_1^2\xi_2^2.\cdots. \xi_{\mu-2}^2\xi_{\mu-1}^2 D_{\mu}}}\right)
\nonumber \\
&\times& \exp{(\mathcal{Y}\ -\frac{c_1}{2\triangle}\varphi_0^2\ +\ \mathcal{X})}\
\exp{\left\{-a_N\triangle\varphi_N^4 -
b_N\triangle\varphi_N^2-\left(\frac{c_N}{2}\ \frac{Q_{N-1}-Q_{N-2}}{\triangle\, Q_{N-1}}\right)\ \varphi_N^2
\right\}}\ .\nonumber
\end {eqnarray}

$\mathcal{W}_{N}^{leading}$ is the part of $\mathcal{W}_{N}$, which will survive the continuum limit. In the Appendixes, we discuss the difference $\ \ \mathcal{W}_{N}-\mathcal{W}_{N}^{leading},\ $ which tends to zero in the continuum limit. We stress the appearance of the Hermite polynomial $H_{4\mu}$ in the current result, as well the dependence on the auxiliary variables $\xi.$ The Hermite polynomial $H_{4\mu}$ is defined following Bateman \cite{bateman}.
The symbols $Q_i,\, \Omega_i,\, D_i,\, \mathcal{Y},\, \mathcal{X},\, H_n(x)$ appears to simplify the notation.
They were introduced during the recurrence procedure of the evaluations and
defined as:

$$\Omega_i = 1-\frac{\Sigma_i}{\Omega_{i-1}},\ \ \Omega_0 = 1,\ \  \Omega_1 = 1- \Sigma_1.$$
$\Sigma_i$ ia defined in (\ref{Sigma}). The value $Q_i$ is defined by the equation:
$$\Omega_i = \frac{\psi_{i+1} Q_{i+1}}{Q_i},\ \  \psi_i = \sigma_i\frac{c_{i+1}}{2\triangle}\ .$$
$Q_i$ is the key variable for descriptions of the oscillator problem.
In the continuum limit, we derive the differential equation for $Q(\tau)$ governing the behavior of
the oscillator. $Q(\tau)$ determines both the harmonic and an-harmonic part of the propagator.

The variable $\mathcal{X}$ appearing in (\ref{Ndim8}) depends on initial and final values of the model:

$$\mathcal{X} = \sqrt{\frac{Q_0Q_1}{2\triangle^2}}\ \frac{\sqrt{c_1c_2}}{Q_N}\, \varphi_0\,\varphi_N\ .$$

The important variables are $d_i\ ,$ defined in each time slice point:

$$d_i = \frac{\frac{Q_0Q_1}{2\triangle^2}\ \frac{c_1c_2}{2}\ \varphi_0^2}{c_{i+2}Q_{i+1}Q_{i}}\ \triangle,$$

$\mathcal{Y}$ is the sum of $d_i$ and  is the "global" variable:

$$\mathcal{Y} = d_{N-2}+d_{N-3}+\cdots + d_1+d_0.$$

The value $D_i$ is the sum of $d_i$ between two slice points with the numbers $p_i$ and $p_{i+1}$ in the equation (\ref{Ndim8}):
$$D_i = d_{p_i} + d_{p_i+1} + d_{p_i+2} + \cdots + d_{p_{(i+1)}-1}.$$

Finally, we remind that the ordinary Hermite polynomials \cite{bateman} $H_n(x)$ are defined as:

$$H_n(x) = n!\ \sum_{m=0}^{\lceil n/2 \rceil}\ \frac{(-1)^m(2x)^{n-2m}}{m! (n-2m)!}.$$
In the Appendix F one can find how $H_{4\mu}$ appeared in our results.

When we accomplished the summations over $n_i$, and $\rho$, as the next step we can apply on Eq.(\ref{Ndim8}) the continuum limit $N\rightarrow \infty,$ with additional prescriptions that all values with slice index $i$ will be converted to continuum argument $\tau$ in the interval $<0,\,\beta>$ by:
$$\triangle = \beta / N,\, \tau = \frac{i\,\beta}{N}.$$
The detailed description of this limiting process is done in Appendix G, with the result:

\begin {equation}
\mathcal{W}_{\beta}^{leading} =\ \frac{\mathcal{W}_{\beta}^{harm}}{\sqrt{f(\beta)}}\
\sum_{\mu=0}^{\infty}\ \frac{4!}{(4\mu)!}\ (-1)^{\mu} \left(\frac{1}{4}\right)^{\mu}\ \mathcal{W}(\mu),\label{Ndimtext}
\end {equation}
where the harmonic oscillator contribution to propagator is:
\begin {equation}
\mathcal{W}_{\beta}^{harm} =
\exp{\left(\lim_{\epsilon\rightarrow 0}\left\{\int_{\epsilon}^{\beta}d\tau\frac{1}{c(\tau)Q^2(\tau)}-
\frac{\epsilon}{c(\epsilon)Q^2(\epsilon)}\right\}\frac{c^2(0)\varphi(0)^2}{2}\ \ +\ \frac{c(0)\varphi(0)\varphi(\beta)}{Q(\beta)}
-\frac{\dot{Q}(\beta)}{Q(\beta)}\frac{c(\beta)\varphi(\beta)^2}{2}
\right)}
\label{harm}
\end {equation}
whereas the an-harmonic correction to propagator in continuum limit is:
\begin {eqnarray}
\mathcal{W}(\mu)&=&\nonumber \\
&=&
\int_0^{\beta}d\tau_1\ a(\tau_1)Q(\tau_1)^4\ \int_{\tau_1}^{\beta}d\tau_2\ a(\tau_2)Q(\tau_2)^4\ \cdots\
\int_{\tau_{\mu-1}}^{\beta}d\tau_{\mu}\ a(\tau_{\mu})Q(\tau_{\mu})^4\label{anharm}  \\
&\times&
\left\{\lim_{\xi_{\mu-1}\rightarrow 1}\ \partial_{\xi_{\mu-1}}^4\right\}\  \cdots \left\{\lim_{\xi_{i}\rightarrow 1}\ \partial_{\xi_{i}}^4\ \xi_{i}^{-4(\mu-i-1)}\right\}
\cdots
\left\{\lim_{\xi_{1}\rightarrow 1}\ \partial_{\xi_{1}}^4\ \xi_{1}^{-4(\mu-2)}\right\}\nonumber \\
&\times&\
\left(\sqrt{-\mathcal{I}(\tau_1)-(\xi_1^2-1)\mathcal{I}(\tau_2)-\cdots -\xi_1^2\xi_2^2 \cdot \cdots \cdot \xi_{\mu-2}^2(\xi_{\mu-1}^2-1)\mathcal{I}(\tau_{\mu})\
}\right)^{4\mu}\nonumber \\ &\times& H_{4\mu}
\left(\frac{\frac{c(0)\varphi(0)}{\sqrt{2}}[\mathcal{I}(\tau_1)+(\xi_1-1)\mathcal{I}(\tau_2)+ \cdots +\xi_1\xi_2 \cdot \cdots \cdot
\xi_{\mu-2}(\xi_{\mu-1}-1)\mathcal{I}(\tau_{\mu})]+\xi_1 \cdot \cdots \cdot \xi_{\mu-1}\frac{\varphi(\beta)}{\sqrt{2}Q(\beta)}}
{\sqrt{-\mathcal{I}(\tau_1)-(\xi_1^2-1)\mathcal{I}(\tau_2)-\cdots -\xi_1^2\xi_2^2 \cdot \cdots \cdot \xi_{\mu-2}^2(\xi_{\mu-1}^2-1)\mathcal{I}(\tau_{\mu})\ }}\right).
\nonumber
\end {eqnarray}
The function $\mathcal{I}(\tau)$ are derived from $D_i$ functions in Appendix G:

$$D_i = \mathcal{I}(\tau_i) - \mathcal{I}(\tau_{(i+1)}),\ \ \ \mathcal{I}(\tau_i) = \int_{\tau_i}^{\beta}\  \frac{d\tau}{c(\tau)Q(\tau)^2}\ .$$

Finally the function $f(\beta)$ in (\ref{Ndimtext}) is the solution of differential equation derived by the similar method as was proposed by Gel'fand and Yaglom \cite{gelY} for harmonic oscillator:
\begin {equation}
\ddot{f}(\tau) - \partial_{\tau}\ln{c(\tau)}\ \dot{f}(\tau) - \left(2\frac{b(\tau)}{c(\tau)} + \partial_{\tau}^2\ln{c(\tau)}\right)f(\tau)\ =\ 0,
\label{gyhlav}
\end {equation}
with initial conditions:
\begin {equation}
f(0) = 0,\, \dot{f}(0) = \frac{2\pi}{c(0)}.
\end {equation}
The function $Q(\tau)$ is the solution of the differential equation:
\begin {equation}
\ddot{Q}(\tau) + \partial_{\tau}\ln{c(\tau)}\ \dot{Q}(\tau) - 2\frac{b(\tau)}{c(\tau)}\ Q(\tau) = 0,
\label{deqhlav}
\end {equation}
with the initial conditions:
$$Q(0) = 0, \, \dot{Q}(0) = 1.$$
The derivation of both equations is done in Appendix H.

The function $Q(\tau)$ plays an essential role in evaluations of the harmonics and an-harmonics parts of the propagator. The differential equation (\ref{deqhlav}) depends on the parameters $b(\tau)$ and $c(\tau)$ of the model. It has the same form as the equation of the motion for the coordinate variable for the harmonic oscillator with time-dependent mass and frequency \cite{pedrosa}.
What is the concern of the harmonic part of the propagator
$$\frac{\mathcal{W}_{\beta}^{harm}}{\sqrt{f(\beta)}},$$
there is a formal difference comparing to the results of other authors \cite{grothaus} evaluating the propagator by the method of integration over fluctuations from the classical path. Nevertheless,  for the real models, the results for the propagator of the harmonic oscillator are the same. For the evaluations based on the discretization's algorithms \cite{wipf}, our results for the harmonic oscillator are in coincidence.

To evaluate the an-harmonic part of the propagator in Eq. (\ref{anharm}) we use the idea of the generating functions for Hermite polynomials, (see \cite{bateman}):

$$\sum_{n=0}^{\infty}\ \frac{z^n}{n!}\ H_n(x) = \exp{(2zx-z^2)}.$$
The arguments of the Hermite polynomial and square-root functions possesses the combinations of the integration variables dependent terms.
Our aim is to introduce the factorization to the integrals in (\ref{anharm}).
For the purposes of the evaluations, we introduce the auxiliary function $I_{4\nu}(\mu, \beta)$ defined as:

\begin {eqnarray}
& &I_{4\nu}(\mu, \beta)= \int_0^{\beta}d\tau_1\ a(\tau_1)Q(\tau_1)^4\ \int_{\tau_1}^{\beta}d\tau_2\ a(\tau_2)Q(\tau_2)^4\ \cdots\
\int_{\tau_{\mu-1}}^{\beta}d\tau_{\mu}\ a(\tau_{\mu})Q(\tau_{\mu})^4 \label{her1hlav}\\
&\times&
\left\{\lim_{\xi_{\mu-1}\rightarrow 1}\ \partial_{\xi_{\mu-1}}^4\right\}\  \cdots \left\{\lim_{\xi_{i}\rightarrow 1}\ \partial_{\xi_{i}}^4\ \xi_{i}^{-4(\mu-i-1)}\right\}
\cdots
\left\{\lim_{\xi_{1}\rightarrow 1}\ \partial_{\xi_{1}}^4\ \xi_{1}^{-4(\mu-2)}\right\}\nonumber \\
&\times&\
\left(\sqrt{-\mathcal{I}(\tau_1)-(\xi_1^2-1)\mathcal{I}(\tau_2)-\cdots -\xi_1^2\xi_2^2 \cdot \cdots \cdot \xi_{\mu-2}^2(\xi_{\mu-1}^2-1)\mathcal{I}(\tau_{\mu})\
}\right)^{4\nu}\nonumber \\ &\times& H_{4\nu}
\left(\frac{\frac{c(0)\varphi(0)}{\sqrt{2}}[\mathcal{I}(\tau_1)+(\xi_1-1)\mathcal{I}(\tau_2)+ \cdots +\xi_1\xi_2 \cdot \cdots \cdot
\xi_{\mu-2}(\xi_{\mu-1}-1)\mathcal{I}(\tau_{\mu})]+\xi_1 \cdot \cdots \cdot \xi_{\mu-1}\frac{\varphi(\beta)}{\sqrt{2}Q(\beta)}}
{\sqrt{-\mathcal{I}(\tau_1)-(\xi_1^2-1)\mathcal{I}(\tau_2)-\cdots -\xi_1^2\xi_2^2 \cdot \cdots \cdot \xi_{\mu-2}^2(\xi_{\mu-1}^2-1)\mathcal{I}(\tau_{\mu})\ }}\right)
\nonumber
\end {eqnarray}

The function $I_{4\nu}(\mu, \beta)$ has the same numbers of the integrals and the derivatives as $\mathcal{W}(\mu),$ the difference is in Hermite polynomial index and power of square root function.
Let us stress, that  $I_{4\mu}(\mu, \beta)=\mathcal{W}(\mu).$

The Hermite polynomial in (\ref{her1hlav}) we are going to exploit for evaluation of
a generating function for $I_{4\nu}(\mu, \beta)$.
By multiplying both sides of equation (\ref{her1hlav})
by $t^{4\nu}/(4\nu)!\ ,$ then summing up the equation over index $\nu$ and using the identity:

$$t^n + (-t)^n + (it)^n + (-it)^n = 4t^n\ \delta_{n, 4\nu}$$
we obtain:

\begin {eqnarray}
& &\sum_{\nu=0}^{\infty}\ \frac{t^{4\nu}}{(4\nu)!}\ I_{4\nu}(\mu, \beta) =\label{genfu} \\
&=&\left\{
\sum_{n=0}^{\infty}\ \frac{t^n}{n!}\ I_{n}(\mu, \beta) +
\sum_{n=0}^{\infty}\ \frac{(-t)^n}{n!}\ I_{n}(\mu, \beta) +
\sum_{n=0}^{\infty}\ \frac{(it)^n}{n!}\ I_{n}(\mu, \beta) +
\sum_{n=0}^{\infty}\ \frac{(-it)^n}{n!}\ I_{n}(\mu, \beta)\right\}/4 \nonumber
\end {eqnarray}
To evaluate $\mathcal{W}(\mu)$, we must to find $4\mu-th$ coefficient of the Taylor's expansion of the generating function (\ref{genfu}) in variable $t$ and in the limit $t\rightarrow 0.$

The procedure of the evaluation of (\ref{genfu}) by this art is described in Appendix I. The result for each of the sum is given as:

\begin {eqnarray}
& &\sum_{n=0}^{\infty}\ \frac{z^n}{n!}\ I_{n}(\mu, \beta) = \label{her3hlav}\\
&\times&\int_0^{\beta}d\tau_1\ a(\tau_1)Q(\tau_1)^4\ \exp{\left\{2z\left(\mathcal{I}(\tau_1)\frac{c(0)\varphi(0)}{\sqrt{2}}+\frac{\varphi(\beta)}{\sqrt{2}Q(\beta)}\right)
 + (2z)^2\frac{\mathcal{I}(\tau_1)}{4}\right\}} \nonumber \\
&\times&
\left\{\lim_{\xi_{\mu-1}\rightarrow 1}\ \partial_{\xi_{\mu-1}}^4\right\}\  \cdots \left\{\lim_{\xi_{i}\rightarrow 1}\ \partial_{\xi_{i}}^4\ \xi_{i}^{-4(\mu-i-1)}\right\}
\cdots
\left\{\lim_{\xi_{1}\rightarrow 1}\ \partial_{\xi_{1}}^4\ \xi_{1}^{-4(\mu-2)}\right\}\nonumber \\
&\times&\int_{\tau_1}^{\beta}d\tau_2\ a(\tau_2)Q(\tau_2)^4\ \exp{\left\{2z(\xi_1-1)\left(\mathcal{I}(\tau_2)\frac{c(0)\varphi(0)}{\sqrt{2}}+\frac{\varphi(\beta)}{\sqrt{2}Q(\beta)}\right)
 + (2z)^2(\xi_1^2-1)\frac{\mathcal{I}(\tau_2)}{4}\right\}} \nonumber \\
& &\vdots \nonumber \\
&\times&\int_{\tau_{\mu-1}}^{\beta}d\tau_{\mu}\ a(\tau_{\mu})Q(\tau_{\mu})^4\
\exp{\left\{2z\xi_1\cdot \cdots\cdot \xi_{\mu-2}(\xi_{\mu-1}-1)\left(\mathcal{I}(\tau_{\mu})\frac{c(0)\varphi(0)}{\sqrt{2}}+\frac{\varphi(\beta)}{\sqrt{2}Q(\beta)}\right)
+\right. }\nonumber \\
& &\left. +(2z)^2\xi_1^2\xi_2^2 \cdot \cdots \cdot\xi_{\mu-2}^2(\xi_{\mu-1}^2-1)\frac{\mathcal{I}(\tau_{\mu})}{4}\right\} \nonumber
\end {eqnarray}
where $z = \{t,\,-t,\,it,\,-it\}$.
The generating function is factorized for the integration variables $\tau_i\ .$
The another advantage of the factorization is the structure of the $\xi_i$ dependence, allowing us to handle by derivative operation only one integral.

In the next step, we evaluate the derivatives:

$$\left\{\lim_{\xi_{\mu-1}\rightarrow 1}\ \partial_{\xi_{\mu-1}}^4\right\}\  \cdots \left\{\lim_{\xi_{i}\rightarrow 1}\ \partial_{\xi_{i}}^4\ \xi_{i}^{-4(\mu-i-1)}\right\}
\cdots
\left\{\lim_{\xi_{1}\rightarrow 1}\ \partial_{\xi_{1}}^4\ \xi_{1}^{-4(\mu-2)}\right\}$$
in the Eq. (\ref{her3hlav}). These derivatives thanks the factorization can be easy to survey.
We have shown in Appendixes J-K that this evaluation result in the representation of an-harmonics
part of the propagator by the incomplete Hermite polynomials introduced by Dattoli \cite{dattoli}.

Inserting finally the results from corresponding Appendices to Eq.(\ref{Ndimtext}) we have for the propagator of the an-harmonic oscillator with the quartic term the result:
\begin {equation}
\mathcal{W}_{\beta}^{leading} =\ \frac{\mathcal{W}_{\beta}^{harm}}{\sqrt{f(\beta)}}\
\sum_{\mu=0}\, (-4)^{\mu}(4!)^{\mu}
\left(\sum^4_{\kappa_1,\cdots ,\kappa_{\mu}=0}\  I_{\kappa_1,\cdots ,\kappa_{\mu}}(\beta)\
\mathcal{D}(\kappa_1,\cdots ,\kappa_{\mu})\right)
\label{Ndimtext2}
\end {equation}
where the integral $I_{\kappa_1,\cdots ,\kappa_{\mu}}(\beta)$ is defined as:
\begin {equation}
I_{\kappa_1,\cdots ,\kappa_{\mu}}(\beta) =
\int_0^{\beta}a(\tau_1)Q^4(\tau_1)\mathcal{I}^{\kappa_1}(\tau_{1})
\int_{\tau_1}^{\beta}a(\tau_2)Q^4(\tau_2)\mathcal{I}^{\kappa_2}(\tau_{2})\cdots
\int_{\tau_{\mu-1}}^{\beta}a(\tau_{\mu})Q^4(\tau_{\mu})\mathcal{I}^{\kappa_{\mu}}(\tau_{\mu}).
\label{multiint}
\end {equation}
The essential information on the function $\mathcal{D}(\kappa_1,\cdots ,\kappa_{\mu})$ is given in the Appendix K.

The function $\mathcal{D}$ can be expressed in terms of the functions $\mathcal{H}_{4-\kappa_{\mu},\ \kappa_{\mu}} (\phi_{\beta}, \phi_0 | \gamma)$,  "modified" Dattoli's incomplete Hermite polynomials (see Appendix K), connected to Dattoli's incomplete Hermite polynomials $H_{4-\kappa_{\mu},\ \kappa_{\mu}} (\phi_{\beta}, \phi_0 | \gamma)$ by:

$$\mathcal{H}_{n-\kappa, \kappa}(\phi_{\beta},\phi_0 | \gamma) =
\sum_{k=0}^{\min{(n-\kappa, \kappa)}}\ \frac{\phi^{n-\kappa-k}_{\beta}\phi^{\kappa-k}_0\gamma^k}{(n-\kappa-k)!k!(\kappa-k)!}\,=\,
(n-\kappa)!\kappa!\, H_{n-\kappa, \kappa}(\phi_{\beta},\phi_0 | \gamma).$$
The values $\phi_0$ and $\phi_{\beta}$ are introduced in Appendix J.

To abbreviate the notation we introduce symbol $h_{\kappa}$:

$$h_{\kappa} \equiv -4.4!\ \mathcal{H}_{4-\kappa, \kappa}(\phi_{\beta},\phi_0 | \gamma).$$
Then the function $\mathcal{D}$ defined in the Appendix K can be written in the form:

\begin {eqnarray}
& &(-4)^{\mu}(4!)^{\mu}\mathcal{D}(\kappa_1,\cdots ,\kappa_{\mu}) = \label{recurresult2} \\
&=&\sideset{}{'}\sum^{4}_{\{n_i\}=0}
\left(\partial^{n_1}_{\phi_0}h_{\kappa_1}\right)\partial^{n_1}_{\phi_{\beta}}\left[
\left(\partial^{n_2}_{\phi_0}h_{\kappa_2}\right)\partial^{n_2}_{\phi_{\beta}}\left[
\cdots
\left(\partial^{n_{\mu-2}}_{\phi_0}h_{\kappa_{\mu-2}}\right)\partial^{n_{\mu-2}}_{\phi_{\beta}}
\left[\left(\partial^{n_{\mu-1}}_{\phi_0}h_{\kappa_{\mu-1}}\right)
\left(\partial^{n_{\mu-1}}_{\phi_{\beta}}h_{\kappa_{\mu}}\right)\right]\cdots \right]\right]\nonumber
\end {eqnarray}
where we used the notation:
$$\sideset{}{'}\sum^{4}_{\{n_i\}=0} \equiv \sum^{4}_{n_1=0}\frac{1}{2^{n_1} n_{1}!}
 \sum^{4}_{n_2=0}\frac{1}{2^{n_2} n_{2}!}\ \cdots
 \sum^{4}_{n_{\mu-1}=0}\ \frac{1}{2^{n_{\mu-1}}\ n_{\mu-1}!}$$

 The most l.h.s. term in (\ref{recurresult2}) is the product of derivatives $\left(\partial^{n_{\mu-1}}_{\phi_0}h_{\kappa_{\mu-1}}\right)
\left(\partial^{n_{\mu-1}}_{\phi_{\beta}}h_{\kappa_{\mu}}\right).$
The next block is formed by the derivative of this first term by $\partial^{n_{\mu-2}}_{\phi_{\beta}}$, multiplied by $\left(\partial^{n_{\mu-2}}_{\phi_0}h_{\kappa_{\mu-2}}\right)$.
Following such a recurrence procedure, we can find (\ref{recurresult2}).

We can define the operator affecting the functions $h_{\kappa}$:
 $$\hat{\mathcal{O}}_{\kappa} \equiv \sideset{}{'}\sum^{4}_{n=0}\left(\partial^n_{\phi_0}h_{\kappa}\right)\partial^n_{\phi_{\beta}}\ ,\ \ \ \hat{\mathcal{O}}_{\kappa_i}\ \cdot h_{\kappa_j} =
 \sideset{}{'}\sum^{4}_{n=0}\left(\partial^n_{\phi_0}h_{\kappa_i}\right)
 \left(\partial^n_{\phi_{\beta}}h_{\kappa_j}\right)\ ,\ \ \
 \hat{\mathcal{O}}_{\kappa_i}\cdot 1 = h_{\kappa_i},$$
 then
 $$\mathcal{D}(\kappa_1,\cdots ,\kappa_{\mu}) = \left(\hat{\mathcal{O}}_{\kappa}\right)^{\mu}\cdot 1 = \left(\hat{\mathcal{O}}_{\kappa}\right)^{\mu-1}\cdot h_{\kappa_{\mu}}.$$

By definition of the another operator, $\hat{\mathcal{I}}_{\kappa}$ affecting on the integral (\ref{multiint}) only, by the prescription:
$$\hat{\mathcal{I}}_{\kappa_0}\ \cdot I_{\kappa_1,\cdots ,\kappa_{\mu}} =
I_{\kappa_0, \kappa_1,\cdots ,\kappa_{\mu}},\ \ \ \hat{\mathcal{I}}_{\kappa_0}\ \cdot 1\ =\ I_{\kappa_0}$$
we can formally summed up the an-harmonic part with the result:
\begin {eqnarray}
& &\sum_{\mu=0}\, (-4)^{\mu}(4!)^{\mu}
\left(\sum^4_{\kappa_1,\cdots ,\kappa_{\mu}=0}\  I_{\kappa_1,\cdots ,\kappa_{\mu}}(\beta)\
\mathcal{D}(\kappa_1,\cdots ,\kappa_{\mu})\right) =\nonumber \\
&=&\sum_{\mu=0}\left(\sum_{\kappa=0}^4\hat{\mathcal{O}}_{\kappa}\hat{\mathcal{I}}_{\kappa}\right)^{\mu}=
\frac{1}{1-\sum_{\kappa=0}^4\hat{\mathcal{O}}_{\kappa}\hat{\mathcal{I}}_{\kappa}}\
\label{anharma}
\end {eqnarray}
We represent the function for an-harmonics correction of the propagator by the rational operator function.
This method is familiar in mathematics as  "umbral calculus," \cite{thesis}, when the non-complicated operator functions represents higher transcendental functions.
The expansions of the operator function held in a formal sense only, the convergence must be checked for the final structures, obtained via the action of the operators $\hat{\mathcal{I}}_{\kappa}$ and $\hat{\mathcal{O}}_{\kappa}$ on the functions $I_{\kappa}$ and $h_{\kappa}.$
The significant feature of the umbral calculus is the principle of the formal properties \cite{thesis}. If an umbral correspondence between two different functions will be established, such correspondence can be extended to other operations on these functions (including integration). Then we can evaluate physical quantities connected to the propagator without explicit evaluation of the propagator.
The theory and also the applications of the umbral calculus to the various fields of mathematics are elaborated (see, for instance, \cite{dibucch}).

The $\mu - th$ term of Eq. (\ref{anharma}) with help of the operators can be rewritten in the form:

$$(-4)^{\mu}(4!)^{\mu}I_{\kappa_1,\cdots ,\kappa_{\mu}}(\beta)\ \mathcal{D}(\kappa_1,\cdots ,\kappa_{\mu}) =
\left(\sum_{\kappa=0}^4h_{\kappa}\hat{\mathcal{I}}_{\kappa} + \sum_{\kappa=0}^4\ \hat{A}_{\kappa}\hat{\mathcal{I}}_{\kappa}\right)^{\mu-1}\cdot(\sum_{\kappa=0}^4\ h_{\kappa_{\mu}}I_{\kappa_{\mu}}) $$
where operator $\hat{A}_{\kappa}$ is defined by:
$$\hat{A}_{\kappa} = \sum^{4}_{n=1}{}^{'}\left(\partial^n_{\phi_0}h_{\kappa}\right)\partial^n_{\phi_{\beta}} =
\hat{\mathcal{O}}_{\kappa} - h_{\kappa}.$$
With help of the operator $\hat{A}_{\kappa}$ we have for an-harmonics part of the propagator the result:
\begin {equation}
\sum_{\mu=0}\, (-4)^{\mu}(4!)^{\mu}
\left(\sum^4_{\kappa_1,\cdots ,\kappa_{\mu}=0}\  I_{\kappa_1,\cdots ,\kappa_{\mu}}(\beta)\
\mathcal{D}(\kappa_1,\cdots ,\kappa_{\mu})\right) =
\frac{1}{1-\frac{\hat{A}_{\kappa}\hat{\mathcal{I}}_{\kappa}}{1-\sum_{\kappa=0}^4\ h_{\kappa}\hat{\mathcal{I}}_{\kappa}}}\ \cdot
\left(\exp{(h_{\kappa}I_{\kappa})}\right)
\label{anharmb}
\end {equation}
The equations (\ref{anharma}) and (\ref{anharmb}) can be convert to one to other using definitions of the operators
$\hat{A}_{\kappa}$ and $\hat{\mathcal{O}}_{\kappa}$ and by identity:
$$\left(1-\sum_{\kappa=0}^4\ h_{\kappa}\hat{\mathcal{I}}_{\kappa}\right)\ \cdot \exp{(h_{\kappa}I_{\kappa})} = 1.$$

\section{Discussion}

Let us discus the an-harmonics part of propagator in term of the final functions. The $\mu -th$ term of the an-harmonics part of propagator expanded to the series can be written as:
\begin {equation}
\left(\hat{\mathcal{O}}_{\kappa}\hat{\mathcal{I}}_{\kappa}\right)^{\mu}\ \cdot 1 \equiv \left(h_{\kappa}\hat{\mathcal{I}}_{\kappa} +  \hat{A}_{\kappa}\hat{\mathcal{I}}_{\kappa}\right)^{\mu-1}\cdot( h_{\kappa_{\mu}}I_{\kappa_{\mu}})\ .
\label{D1}
\end {equation}
In (\ref{D1}) we omitted $\sum_{\kappa=0}^4$ and we suppose the summation convention for the repeated indexes. We can rewrite the above equation in the form:
\begin {equation}
\sum_{i=0}^{\mu-1}\ \mathcal{Z}_{\mu,i}\ I_{\kappa_1,\cdots ,\kappa_{\mu}}(\beta)
\label{D11}
\end {equation}
The function $\mathcal{Z}_{\mu,i}$ is the sum of the all terms of the expansion of Eq. (\ref{D1}), with $i$ operators $\hat{A}_{\kappa}$ and $\mu - i$ functions $h_{\kappa}$.
For instance,
$$\mathcal{Z}_{\mu, 0}\ =\ h_{\kappa_1}.h_{\kappa_2}.\cdots . h_{\kappa_{\mu}}\ ,$$
$$\mathcal{Z}_{\mu, \mu-1}\ =\ \hat{A}_{\kappa_1}.\hat{A}_{\kappa_2}.\cdots .
\hat{A}_{\kappa_{\mu-1}}.h_{\kappa_{\mu}}\ ,$$
the $\mathcal{Z}_{\mu, \mu-2}$ is the sum of $\mu - 1$ terms with two functions $h_{\kappa}$ in each of them:
$$\mathcal{Z}_{\mu, \mu-2} =
h_{\kappa_1}.\hat{A}_{\kappa_2}.\cdots .\hat{A}_{\kappa_{\mu-1}}.h_{\kappa_{\mu}}\ +
\hat{A}_{\kappa_1}.h_{\kappa_2}.\cdots .\hat{A}_{\kappa_{\mu-1}}.h_{\kappa_{\mu}}\ +\ \cdots\ +\ \hat{A}_{\kappa_1}.\hat{A}_{\kappa_2}.\cdots .h_{\kappa_{\mu-1}}.h_{\kappa_{\mu}}\ ,
$$
etc. Of cors, in (\ref{D11}) we suppose the summation convention for the repeated indexes $\kappa_i.$

We can write the sum of the all an-harmonic $\mu-th$ contributions in the form:
$$
\sum_{\mu=0}^{\infty}\left(\hat{\mathcal{O}}_{\kappa}\hat{\mathcal{I}}_{\kappa}\right)^{\mu}
= 1 + \sum_{i=1}^{\infty}\ \sum_{\mu=i}^{\infty}\ \mathcal{Z}_{\mu,\mu-i}\ I_{\kappa_1,\cdots ,\kappa_{\mu}}(\beta)\ ,
$$
where $\mathcal{Z}_{\mu,\mu-i}$ can be written in terms of the derivatives of the modified Dattoli's incomplete Hermite polynomials $\mathcal{H}_{n-\kappa, \kappa}(\phi_{\beta},\phi_0 | \gamma)$.

The first nontrivial term for the constant index $i=1$,
$$
\mathcal{P}_1=\sum_{\mu=1}^{\infty}\ \mathcal{Z}_{\mu,\mu-1}\ I_{\kappa_1,\cdots ,\kappa_{\mu}}(\beta)\ ,
$$
is the sum of the series with alternating sign. This sum is in principe finite, iff
$$
\lim_{\mu\rightarrow \infty}\ |\mathcal{Z}_{\mu,\mu-1}\ I_{\kappa_1,\cdots ,\kappa_{\mu}}(\beta)|\ \rightarrow\ 0\ .
$$

The second nontrivial term is
$$
\mathcal{P}_2=\sum_{\mu=2}^{\infty}\ \mathcal{Z}_{\mu,\mu-2}\ I_{\kappa_1,\cdots ,\kappa_{\mu}}(\beta)\ .
$$
If we will use for the derivatives of the product of two functions the prescription:
$$\partial^n_x\left(f.g\right)=f\ \left(\partial^n_x\ g\right)+g\ \left(\partial^n_x\ f\right)+\sum_{i=1}^{n-1}\ \binom{n}{i}\left(\partial_x^{n-i}f\right)\left(\partial_x^{i}g\right).$$
The first two terms of this equation contribute to the sub-group of $\mathcal{Z}_{\mu,\mu-1}$ in the group of the terms $\mathcal{Z}_{\mu,\mu-2},$ and we find
the result for $\mathcal{P}_2$ in the form:

$$
\mathcal{P}_2 = \frac{\mathcal{P}_1^2}{2!} + \frac{1}{1-\hat{A}_{\kappa}\hat{\mathcal{I}}_{\kappa}}\ .
\mathcal{B}\left(h_{\kappa}.\hat{\mathcal{I}}_{\kappa}; \frac{\mathcal{P}_1}{2!}; \mathcal{P}_1\right)
$$
where
\begin {eqnarray}
& &\mathcal{B}\left(h_{\kappa}.\hat{\mathcal{I}}_{\kappa};\ \sum_{\nu=1}^{\infty}\mathcal{Z}_{\nu,\nu-1}\ I_{\kappa_1,\cdots ,\kappa_{\nu}}(\beta);\ \sum_{\rho=1}^{\infty}\mathcal{Z}_{\rho,\rho-1}\ I_{\kappa_{\alpha},\cdots ,\kappa_{\rho}}(\beta)\right) = \\
&=& \left(\partial_{\phi_{0}}^n\ h_{\kappa}\right)\ \sum_{\nu,\rho=1}^{\infty}\left\{ \sum_{i=1}^{n-1}\binom{n}{i}\left(\partial_x^{n-i}\mathcal{Z}_{\nu,\nu-1}\right)
\left(\partial_{\phi_{\beta}}^{i}\mathcal{Z}_{\rho,\rho-1}\right)\ \hat{\mathcal{I}}_{\kappa}\cdot\left(I_{\kappa_1,\cdots ,\kappa_{\nu}}(\beta)\ I_{\kappa_{\alpha},\cdots ,\kappa_{\rho}}(\beta)\right)\right\}\ .\nonumber
\end {eqnarray}

We claim that by the similar evaluation method we find for $\mathcal{P}_i$ the leading term
$$\frac{\mathcal{P}_1^i}{i!}$$
plus, other terms, an arrangement of them is in progress. Summing the terms $\frac{\mathcal{P}_1^i}{i!}$
gives the result in the non-perturbation form:
$$ \exp{(\mathcal{P}_1)}$$
The study of the different contributions to $\mathcal{P}_i$ is still in progress, from the point of view, if this exponential term is the dominant contribution to the an-harmonics part of the propagator.

\appendix

\section{The evaluation of the leading contribution of N-1 dimensional integral}

In this section we will show the strategy of the summation over the Taylor's expansion summation indexes. The each of the summation indexes appears in product of two parabolic cylinder functions.  We are going to sum up the product of two parabolic cylinder functions so that the result was the parabolic cylinder function also. By such method in the recurrence steps we will sum up the product of two parabolic cylinder functions.

As we have shown in  \cite{my},  the multiple sums in Eq (\ref{afindim3}) are uniformly convergent.
 The individual summation over given $n_i$ in Eq.(\ref{afindim3}) is the sum of the products of two functions $\mathcal{D}_{\nu_i}(z_i)$ depending on the summation index $n_i$:
\begin {equation}
\sum\limits_{n_i=0}^{\infty}\ \frac{\Sigma_i^{n_i+\rho/2}}{n_i!}\ \pochh{n_i+\rho+1/2}{n_{i-1}}\ \mathcal{D}_{-n_i-\rho-1/2-n_{i-1}}(z_i)\
\pochh{n_{i+1}+\rho+1/2}{n_i}\ \mathcal{D}_{-n_{i+1}-\rho-1/2-n_i}(z_{i+1})
\label{single}
 that\end {equation}
We did't find in the literature a suitable summation prescription for this case.
To evaluate these summations,  we introduce the first and the only one approximation in our calculation, when one of the parabolic cylinder functions is replaced by the asymptotic Poincar\' e - type expansion  of the parabolic cylinder functions valid for the finite index $n_i$ and large argument $z$:
\begin {equation}
\mathcal{D}_{-\nu-1/2}(z)\; \equiv \;
z^{\nu+1/2}\;e^{z^2/4}\;D_{-\nu-1/2}(z)\;=
\;\sum\limits_{j=0}^{\mathcal{J}}\; (-1)^j
\;\frac{\pochh{\nu+1/2}{2j}}{j!\;(2z^2)^j}
+\varepsilon_{\mathcal{J}}(\nu, z),
\label{poincare}
\end {equation}
where $\varepsilon_{\mathcal{J}}(\nu, z)$ is the remainder of the Poincar\' e - type expansion. Olver \cite{olver} and Temme  \cite{temme2} studied this remainder and evaluated its upper limit.
Poincar\' e - type expansion is valid, iff $z^2\gg |\nu|.$
 The asymptotic expansion is particularly useful in the continuum limit $\triangle \rightarrow 0$, when $z\approx \triangle^{-3/2}$ and functions $\mathcal{D}_{-\nu-1/2}(z) \rightarrow 1.$ The first term in (\ref{poincare}) contributes to the leading part of (\ref{afindim3}), whereas the second part generates the remainder.

Inserting the leading part of (\ref{poincare}) to (\ref{single}),
we perform the summation by the identity, derived by Taylor's expansion formula for parabolic cylinder functions \cite{bateman}:
\begin {equation}
e^{x^2/4}\sum\limits_{k=0}^{\infty}\;
\frac{\pochh{\nu}{k}}{k!}\;t^k\;D_{-\nu-k}(x)\;=\;e^{(x-t)^2/4}\;D_{-\nu}\;(x-t).
\label{pointcp}
\end {equation}
Following Taylor's expansion for parabolic cylinder functions we can prove for the functions $\mathcal{D}_{-\nu}(z)$ the summation formula:
\begin {equation}
\sum_{k=0}^{\infty}\,\frac{\pochh{\nu}{k}}{k!}\,t^k\,\mathcal{D}_{-\nu-k}(z) =
\left(\frac{1}{1-t}\right)^{\nu}\,\mathcal{D}_{-\nu}(z(1-t)).
\label{tpc}
\end {equation}

To evaluate the finite-dimensional integral Eq. (\ref{afindim3}) let us define the object $\mathcal{M}_{\Lambda}$
\begin {eqnarray}
& &\mathcal{M}_{\Lambda} = \nonumber \\
&=&\prod_{i=1}^{\Lambda}\left\{\frac{1}{\sqrt{\frac{c_i+c_{i+1}}{c_i}+2 \frac{b_i}{c_i}\triangle^2}}\
\sum\limits_{n_i=0}^{N_i} \frac{\Sigma_i^{n_i+\rho/2}}{n_i!}\ \pochh{n_i+\rho+1/2}{n_{i-1}}\mathcal{D}_{-n_i-\rho-1/2-n_{i-1}}(z_i)\right\}\nonumber \\
&\times&\pochh{n_{\Lambda+1}+\rho+1/2}{n_{\Lambda}}\mathcal{D}_{-n_{\Lambda+1}-\rho-1/2-n_{\Lambda}}(z_{\Lambda+1})
\label{rec1}
\end {eqnarray}
$\mathcal{M}_{\Lambda}$ differs from (\ref{afindim3}) by finite summations over $n_i.$ Due to uniform convergence of (\ref{afindim3}) we can done (\ref{rec1}) as close to (\ref{afindim3}) as we need, by convenient choice of $N_i$.

We will sum up (\ref{rec1}) by recurrence procedure.
As the first recurrence step we perform the $n_{1}$ summation, replacing $\mathcal{D}_{-n_{1}-\rho-1/2-n_{0}}(z_{1})$ by the Poincar\'e type expansion (\ref{poincare}). We find:
\begin {eqnarray}
& &\sum\limits_{n_{1}=0}^{N_1} \frac{\Sigma_{1}^{n_{1}+\rho/2}}{n_{1}!}\ \pochh{n_{1}+\rho+1/2}{n_{0}}\mathcal{D}_{-n_{1}-\rho-1/2-n_{0}}(z_{1})
\pochh{n_{2}+\rho+1/2}{n_{1}}\mathcal{D}_{-n_{2}-\rho-1/2-n_{1}}(z_{2})\ =\label{rekur}\\
&=&\sum\limits_{j=0}^{\mathcal{J}}
\,\frac{(-1)^j}{j!\;(2z_{1}^2)^j}
\sum\limits_{n_{1}=0}^{N_{1}}\frac{\Sigma_{1}^{n_{1}+\rho/2}}{n_{1}!}
\pochh{n_{2}+\rho+1/2}{n_{1}}\mathcal{D}_{-n_{2}-\rho-1/2-n_{1}}(z_{2})
\,\pochh{n_{1}+\rho+1/2}{n_{0}+2j}\nonumber\ + \\
&+&\varepsilon^{max}_{\mathcal{J}}(-n_{1}-\rho-1/2-n_{0}\, ,\, z_{1})
\sum\limits_{n_{1}=0}^{N_{1}}\frac{\Sigma_{1}^{n_{1}+\rho/2}}{n_{1}!}
\pochh{n_{2}+\rho+1/2}{n_{1}}\mathcal{D}_{-n_{2}-\rho-1/2-n_{1}}(z_{2})
\,\pochh{n_{1}+\rho+1/2}{n_{0}} \nonumber
\end {eqnarray}
where the choice of $N_1$  permits to replace $\mathcal{D}_{-n_{1}-\rho-1/2-n_{0}}(z_{1})$
by the Poincar\'e type expansion of  for all $n_{1}$ .
The symbol $\varepsilon^{max}_{\mathcal{J}}(-n_{1}-\rho-1/2-n_{0}\, ,\, z_{1})$ is the maximal value of the remainders $\varepsilon_{\mathcal{J}}(-n_{1}-\rho-1/2-n_{0}\, ,\, z_{1})$ for all necessary values $n_{1}, \rho, n_{0}$ and $z_1.$
To apply the summation formula Eq. (\ref{tpc}), we must also modify the Pochhammer's $\pochh{n_{1}+\rho+1/2}{n_{0}+2j}$ and $\pochh{n_{1}+\rho+1/2}{n_{0}}.$ We follow the method explained in \cite{my}. We proved the identity:
$$\pochh{n_{1}+\rho+1/2}{n_{0}+2j} =
\sum_{i=0}^{\min(n_{0}+2j\,,\,n_1)}\,
a^{n_{0}+2j}_i(\rho)\frac{n_1!}{(n_1-i)!}$$
where
\begin {equation}
a^{n_{0}+2j}_i(\rho) = \binom{n_{0}+2j}{i}\frac{\pochh{\rho+1/2}{n_{0}+2j}}{\pochh{\rho+1/2}{i}}\,.
\label{poch1}
\end {equation}
Inserting the first identity to Eq. (\ref{rekur})  we find:
\begin {eqnarray}
& &\sum\limits_{n_{1}=0}^{N_1} \frac{\Sigma_{1}^{n_{1}+\rho/2}}{n_{1}!}\ \pochh{n_{1}+\rho+1/2}{n_{0}}\mathcal{D}_{-n_{1}-\rho-1/2-n_{0}}(z_{1})
\pochh{n_{2}+\rho+1/2}{n_{1}}\mathcal{D}_{-n_{2}-\rho-1/2-n_{1}}(z_{2})=\label{rekur1}\\
&=&\sum\limits_{j=0}^{\mathcal{J}}
\,\frac{(-1)^j}{j!\;(2z_{1}^2)^j}
\sum_{i=0}^{n_{0}+2j}\,a^{n_{0}+2j}_i(\rho)
\sum\limits_{n_{1}=i}^{N_{1}}\frac{\Sigma_{1}^{n_{1}+\rho/2}}{(n_{1}-i)!}
\pochh{n_{2}+\rho+1/2}{n_{1}}\mathcal{D}_{-n_{2}-\rho-1/2-n_{1}}(z_{2})\,+\nonumber\\
&+&\varepsilon^{max}_{\mathcal{J}}(-n_{1}-\rho-1/2-n_{0}\, ,\, z_{1})
\sum_{i=0}^{n_{0}}\,a^{n_{0}}_i(\rho)
\sum\limits_{n_{1}=i}^{N_{1}}\frac{\Sigma_{1}^{n_{1}+\rho/2}}{(n_{1}-i)!}
\pochh{n_{2}+\rho+1/2}{n_{1}}\mathcal{D}_{-n_{2}-\rho-1/2-n_{1}}(z_{2})\nonumber
\end {eqnarray}
Extending summation over $n_1$ up to infinity, by summation variable transformation $n_1\rightarrow l=n_1-i,$ we have:
\begin {eqnarray}
& &\sum\limits_{n_{1}=0}^{N_1} \frac{\Sigma_{1}^{n_{1}+\rho/2}}{n_{1}!}\ \pochh{n_{1}+\rho+1/2}{n_{0}}\mathcal{D}_{-n_{1}-\rho-1/2-n_{0}}(z_{1})
\pochh{n_{2}+\rho+1/2}{n_{1}}\mathcal{D}_{-n_{2}-\rho-1/2-n_{1}}(z_{2})=\label{rekur2}\\
&=&\sum\limits_{j=0}^{\mathcal{J}}
\,\frac{(-1)^j}{j!\;(2z_{1}^2)^j}
\sum_{i=0}^{n_{0}+2j}\,a^{n_{0}+2j}_i(\rho)\,\Sigma_{1}^{i+\rho/2}\pochh{n_{2}+\rho+1/2}{i}
\sum\limits_{l=0}^{\infty}\frac{\Sigma_{1}^{l}}{l!}
\pochh{n_{2}+\rho+1/2+i}{l}\,\mathcal{D}_{-n_{2}-\rho-1/2-i-l}(z_{2}) +\nonumber\\
&+&\varepsilon^{max}_{\mathcal{J}}(-n_{1}-\rho-1/2-n_{0}\, ,\, z_{1})
\sum_{i=0}^{n_{0}}\,a^{n_{0}}_i(\rho)\,\Sigma_{1}^{i+\rho/2}\pochh{n_{2}+\rho+1/2}{i}\times \nonumber\\
&\times&
\sum\limits_{l=0}^{\infty}\frac{\Sigma_{1}^{l}}{l!}
\pochh{n_{2}+\rho+1/2+i}{l}\mathcal{D}_{-n_{2}-\rho-1/2-i-l}(z_{2})\nonumber
\end {eqnarray}
In the above equation we omit the remainders of the uniformly convergent series, which sufficiently strongly converge to zero in the continuum limit.

By the summation's formula (\ref{tpc}) we have:
\begin {eqnarray}
& &\sum\limits_{n_{1}=0}^{N_{1}} \frac{\Sigma_{1}^{n_{1}+\rho/2}}{n_{1}!}\ \pochh{n_{1}+\rho+1/2}{n_{0}}\mathcal{D}_{-n_{1}-\rho-1/2-n_{0}}(z_{1})
\pochh{n_{2}+\rho+1/2}{n_{1}}\mathcal{D}_{-n_{2}-\rho-1/2-n_{1}}(z_{2})=\label{rekur3}\\
&=&\frac{1}{\sqrt{1-\Sigma_1}}\,\sum\limits_{j=0}^{\mathcal{J}} (-1)^j
\,\frac{(-1)^j}{j!\;(2z_{1}^2)^j}
\sum_{i=0}^{n_{0}+2j}\,a^{n_{0}+2j}_i(\rho)\,
\left(\frac{\Sigma_{1}}{1-\Sigma_{1}}\right)^{i+\rho/2} \times \nonumber\\
&\times& \left(\frac{1}{1-\Sigma_{1}}\right)^{n_2+\rho/2}\pochh{n_{2}+\rho+1/2}{i}
\mathcal{D}_{-n_{2}-\rho-1/2-i}(z_{2}(1-\Sigma_{1}))\ +\nonumber\\
&+&\frac{1}{\sqrt{1-\Sigma_1}}\,\varepsilon^{max}_{\mathcal{J}}(-n_{1}-\rho-1/2-n_{0}\, ,\, z_{1})\,
\sum_{i=0}^{n_{0}}\,a^{n_{0}}_i(\rho)\,\left(\frac{\Sigma_{1}}{1-\Sigma_{1}}\right)^{i+\rho/2}\ \times \nonumber\\
&\times&\left(\frac{1}{1-\Sigma_{1}}\right)^{n_2+\rho/2}\pochh{n_{2}+\rho+1/2}{i}
\mathcal{D}_{-n_{2}-\rho-1/2-i}(z_{2}(1-\Sigma_{1}))\nonumber
\end {eqnarray}
In this equation, we can see only one parabolic cylinder function in both terms, as was our aim.
The first term on the r.h.s. of the above equation will contribute to the leading term, the second one will contribute to the remainder. From  the point of view of the recurrence summation procedure, in the second steep we will have the sum of the product of the two parabolic cylinder functions in the leading, as well as in remainder terms. The same art of evaluation as in the first step will give us the leading part contribution, plus two terms proportional to
 $\varepsilon_{\mathcal{J}}(-n_{1}-\rho-1/2-n_{0})$ and $\varepsilon_{\mathcal{J}}(-n_{2}-\rho-1/2-n_{1}),$ plus term proportional to product of both remainders.
Providing the full recurrence  summation procedure, where every step is done by described method, we find that for $\mathcal{M}_{\Lambda}$ the next inequality is valid:
$$\mathcal{M}_{\Lambda}^{leading} < \mathcal{M}_{\Lambda} <
\mathcal{M}_{\Lambda}^{leading}(1+\varepsilon_{max})^{\Lambda}\ ,$$
where  $\varepsilon_{max}$ is the maximal value of the set
$$\frac{\mathcal{M}_{i}^{leading}}{\mathcal{M}_{\Lambda}^{leading}}\varepsilon_{\mathcal{J}}(-n_{i}-\rho-1/2-n_{i-1}\, ,\, z_{i}),\,i=1,2,\cdots ,\ \Lambda.$$
The bounds on the remainder:
$$\varepsilon_{\mathcal{J}}(\nu,z) = \mathcal{D}_{-\nu-1/2}(z) -
\;\sum\limits_{j=0}^{\mathcal{J}}\; (-1)^j
\;\frac{\pochh{\nu+1/2}{2j}}{j!\;(2z^2)^j}$$
was estimated by Olver \cite{olver} and improved by Vidumas and Temme \cite{temme2}. In our case we have:
\begin {eqnarray}
\varepsilon_{\mathcal{J}}(\nu,z)&\leq& \frac{2z^2}{z^2+2\nu}\,
\frac{\pochh{\nu+1/2}{2\mathcal{J}}}{\mathcal{J}!(2z^2)^{\mathcal{J}}}
\,\,_2F_1\left(\mathcal{J}/2,\,1/2;\,\mathcal{J}/2+1;\,1-\frac{4\mathcal{J}^2}{z^4}\right)\\
&\times& \exp{\left[\frac{2z^2}{z^2+2\nu}\,\frac{2}{z^2}\,\,
_2F_1\left(1/2,\,1/2;\,1/2+1;\,1-\frac{4\mathcal{J}^2}{z^4}\right)\right]} \nonumber
\end {eqnarray}
where $\nu=n_{i-1} + n_i \leq N_{i-1} + N_i.$ In the continuum limit, where $z^2$ tends to infinity as $d^3$
and $N_{i-1},\, N_i,\, \mathcal{J}$ are finite, the value $\varepsilon_{max} \leq d^{-1-\epsilon}$ i.e. $\varepsilon_{max}$ tends to zero stronger as $d^{-1}.$ Then, in continuum limit
$$\lim_{d\rightarrow\infty}\,(1+\varepsilon_{max})^d\,\rightarrow\, 1$$
and the upper bound in continuum limit on $\mathcal{M}_{\Lambda}$ will be the leading term. Because the natural lower bound on $\mathcal{M}_{\Lambda}$ is leading term also, we can expects that in the continuum limit the leading term determine the value of the conditional Wiener integral $\mathcal{W}.$

Evaluating the next recurrence steps, we meet the combinations of the summations introduced by Poincar\' e - type expansion (\ref{poincare}). We deal with them by the following art:
$$
\sum_{i_1=0}^{\mathcal{J}_1}\frac{1}{i_1!}\ \sum_{i_2=0}^{\mathcal{J}_2}\frac{1}{i_2!}\ \rightarrow\
\sum_{\mu=0}^{\mathcal{J}_1+\mathcal{J}_2}\frac{1}{\mu!}\
\sum_{\max{(0, \mu-\mathcal{J}_2)}}^{\min{(\mu, \mathcal{J}_1)}}\binom{\mu}{i_1},
$$
where $\mu=i_1+i_2$.
Proceeding by the same method described for the first step, we find for $\mathcal{M}_{\Lambda}$ of Eq. (\ref{rec1})
the result:
\begin {eqnarray}
& &\mathcal{M}_{\Lambda} = \frac{1}{\sqrt{\prod_{i=1}^{\Lambda}(\frac{c_i+c_{i+1}}{c_i}+2 \frac{b_i}{c_i}\triangle^2)\Omega_i}}\
\sum_{\mu=0}^{\mathcal{J}_1+\cdots+\mathcal{J}_{\Lambda}}\ \frac{(-1)^{\mu}}{\mu!\ (2 z_{\Lambda}^2\Omega_{\Lambda-1}^2)^{\mu}}\
\sum_{i_{\Lambda}=0}^{n_0+2\mu}\ \left(\frac{\Sigma_{\Lambda}}{\Omega_{\Lambda}\Omega_{\Lambda-1}}\right)^{i_{\Lambda}+\rho/2}\nonumber \\
&\times&\left(\Lambda\right)_{i_{\Lambda}}^{n_0+2\mu}\ \left(\frac{1}{\Omega_{\Lambda}}\right)^{n_{\Lambda+1}+\rho/2}
\pochh{n_{\Lambda+1}+\rho+1/2}{i_{\Lambda}}
\mathcal{D}_{-n_{\Lambda+1}-\rho-1/2-i_{\Lambda}}(z_{\Lambda+1}\Omega_{\Lambda})
\label{ld1}
\end {eqnarray}
where $\mu$ goes to the sum of all values $\mathcal{J}_i$ appeared in the individual recurrence steps.
For the recurrently defined value $\left(\Lambda\right)_{i_{\Lambda}}^{n_0+2\mu}$ we find:
\begin {eqnarray}
& &\left(\Lambda\right)_{i_{\Lambda}}^{n_0+2\mu} =
\sum_{\lambda=\max(0,\,\mu-\mathcal{J}_{\Lambda})}^{\min(\mu,\,\mathcal{J}_1+\cdots+\mathcal{J}_{\Lambda-1})}\ \binom{\mu}{\lambda}\ \left(\frac{z_{\Lambda}\Omega_{\Lambda-1}}{z_{\Lambda-1}\Omega_{\Lambda-2}}\right)^{2\lambda}
\label{ld2} \\
&\times&\sum_{i_{(\Lambda-1)}=max(0;i_{\Lambda}-2\mu+2\lambda)}^{n_0+2\lambda}\ a_{i_{\Lambda}}^{2\mu-2\lambda+i_{(\Lambda-1)}}\ (\rho)\
\left(\frac{\Sigma_{\Lambda-1}}{\Omega_{\Lambda-1}\Omega_{\Lambda-2}}\right)^{i_{\Lambda-1}+\rho/2}
\left(\Lambda-1\right)_{i_{(\Lambda-1)}}^{n_0+2\lambda}\nonumber
\end {eqnarray}
The first step of the recurrence relation is:
\begin {equation}
\left(1\right)_{i_{1}}^{n_0+2\lambda}=a_{i_{1}}^{n_0+2\lambda}\ (\rho)
\end {equation}
In the evaluation we used the identity (\ref{poch1}):
\begin {equation}
a_{m}^{n}\ (\rho) = \binom{n}{m}\
\frac{\pochh{1/2+\rho}{n}}{\pochh{1/2+\rho}{m}}
\end {equation}
To simplify the notations, we defined the variable $\Omega_i$:
\begin {equation}
\Omega_i = 1-\frac{\Sigma_i}{\Omega_{i-1}},\ \ \Omega_0 = 1,\ \ \Omega_1 = 1- \Sigma_1\ ,
\label{omegai}
\end {equation}
where $\Sigma_i$ depends on the models parameters  (\ref{Sigma}):
$$
\Sigma_i = \left(\frac{c_{i+1}}{2\triangle}\right)^2\ \sigma_i\ \sigma_{i+1},
$$
and
$$\sigma_i = \frac{1}{\frac{c_i+c_{i+1}}{2\triangle}+b_i\triangle}.$$

  To evaluate the leading part of $\mathcal{W}_{N},$ we can use the method of evaluation defined for $\mathcal{M}_{\Lambda}$ up to $\Lambda = N-2.$ The last sum  can be read:
\begin {equation}
\mathcal{M}_{N-1}^{leading}=\frac{1}{\sqrt{\frac{c_{N-1}+c_{N}}
{c_{N-1}}+2\frac{b_{N-1}}{c_{N-1}}\triangle^2}}\ 
 \sum\limits_{n_{N-1}=0}^{\infty}
\frac{\left((\frac{c_N}{2\triangle}\varphi_N)^2\sigma_{N-1}\right)^{n_{N-1}+\rho/2}}{n_{N-1}!}\ 
\mathcal{M}_{N-2}
\end {equation}
We use Poicar\'e decomposition for $\mathcal{D}$ in $\mathcal{M}_{N-2}$ as well and we find:
\begin {eqnarray}
& &\mathcal{M}_{N-1}^{leading} = \frac{1}{\sqrt{\prod_{i=1}^{N-1}(\frac{c_i+c_{i+1}}{c_i}+2 \frac{b_i}{c_i}\triangle^2)\Omega_{i-1}}}
\\&\times&
\sum_{\mu=0}^{\mathcal{J}_1+\cdots+\mathcal{J}_{N-1}}\ \frac{(-1)^{\mu}}{\mu!\ (2 z_{N-1}^2\Omega_{N-2}^2)^{\mu}}
\sum_{i_{N-1}}^{n_0+2\mu}\ \left(\frac{(\frac{c_N}{2\triangle}\varphi_N)^2\
\sigma_{N-1}}{\Omega_{N-2}}\right)^{i_{N-1}+\rho/2}
\left(N-1\right)_{i_{N-1}}^{n_0+2\mu}\ \exp{\left(\frac{(\frac{c_N}{2\triangle}\varphi_N)^2\
\sigma_{N-1}}{\Omega_{N-2}}\right)}\nonumber
\end {eqnarray}
Inserting this result for the leading term of multiple summations to Eq.(\ref{afindim3}) we find the leading term for $N-1$ dimensional integral. In the continuum limit this leading term approaches to result $\mathcal{W},$ because the remainder of uniformly convergent series approaches to zero. We find:
\begin {eqnarray}
& &\mathcal{W}_{N}^{leading} =\ \frac{1}{\sqrt{\frac{2\pi\triangle}{c_0}\ \prod_{i=1}^{N-1}(\frac{c_i+c_{i+1}}{c_i}+2 \frac{b_i}{c_i}\triangle^2)\Omega_{i-1}}}\
\sum\limits_{\rho=0}^{1}
\sum\limits_{n_0=0}^{\infty}\frac{\left(\frac{c_1}{\triangle} \varphi_0 \sqrt{\sigma_1}\right)^{2n_0+\rho}}{(2 n_0+\rho)!}\nonumber \\
&\times&\sum_{\mu=0}^{\mathcal{J}_1+\cdots+\mathcal{J}_{N-1}}\ \frac{(-1)^{\mu}}{\mu!\ (2 z_{N-1}^2\Omega_{N-2}^2)^{\mu}}
\sum_{i_{N-1}}^{n_0+2\mu}\ \left(\frac{(\frac{c_N}{2\triangle}\varphi_N)^2\
\sigma_{N-1}}{\Omega_{N-2}}\right)^{i_{N-1}+\rho/2}
\left(N-1\right)_{i_{N-1}}^{n_0+2\mu}
\nonumber \\
&\times&\exp{\left\{-a_N\triangle\varphi_N^4 - (\frac{c_N}{2\triangle}+b_N\triangle)\varphi_N^2+\left(\frac{(\frac{c_N}{2\triangle}\varphi_N)^2\
\sigma_{N-1}}{\Omega_{N-2}}\right)
-\frac{c_1}{2\triangle}\varphi_0^2\right\}}.
\label{Ndim}
\end {eqnarray}
The next task is the analytical evaluation of the recurrently defined function $\left(N-1\right)_{i_{N-1}}^{n_0+2\mu}.$

\section{Evaluation of the $\left(N-1\right)_{i_{N-1}}^{n_0+2\mu}$}

Our task is to establish a reasonable form of the recurrence equation for $\left(N-1\right)_{i_{N-1}}^{n_0+2\mu}$ .
We start from the recurrence equation (\ref{ld2}) for $\left(\Lambda\right)_{i_{\Lambda}}^{n_0+2\mu}$:
\begin {eqnarray}
& &\left(\Lambda\right)_{i_{\Lambda}}^{n_0+2\mu} =
\sum_{\lambda=\max(0,\,\mu-\mathcal{J}_{\Lambda})}^{\min(\mu,\,\mathcal{J}_1+\cdots+\mathcal{J}_{\Lambda-1})}\ \binom{\mu}{\lambda}\ \left(\frac{z_{\Lambda}\Omega_{\Lambda-1}}{z_{\Lambda-1}\Omega_{\Lambda-2}}\right)^{2\lambda}\times
\label{ld3} \\
&\times&\sum_{i_{(\Lambda-1)}=max(0,\,i_{\Lambda}-2\mu+2\lambda)}^{n_0+2\lambda}\ a_{i_{\Lambda}}^{2\mu-2\lambda+i_{(\Lambda-1)}}\ (\rho)\
\left(\frac{\Sigma_{\Lambda-1}}{\Omega_{\Lambda-1}\Omega_{\Lambda-2}}\right)^{i_{\Lambda-1}+\rho/2}
\left(\Lambda-1\right)_{i_{(\Lambda-1)}}^{n_0+2\lambda}\nonumber
\end {eqnarray}

The first recurrence step is:

\begin {equation}
\left(1\right)_{i_{(\Lambda-1)}}^{n_0+2\lambda}=a_{i_{(\Lambda-1)}}^{n_0+2\lambda}\ (\rho)\ .
\end {equation}

We can see that if $i_{\Lambda} < 0,$ then $\left(\Lambda\right)_{i_{\Lambda}}^{n_0+2\mu} = 0$ as the result of definition (\ref{poch1}) of the object $a_{i_{\Lambda}}^{2\mu-2\lambda+i_{(\Lambda-1)}}\ (\rho).$
Therefore $\left(\Lambda-1\right)_{i_{(\Lambda-1)}}^{n_0+2\lambda} = 0$ for $i_{(\Lambda-1)}<0\ $ and the sum over index $i_{(\Lambda-1)}$ can start from the value $i_{\Lambda}-2\mu+2\lambda,$ where $i_{\Lambda-1}$ can be negative for some of the values $i_{\Lambda},\ \mu,\ \lambda.$

To simplify and clarify the recurrence evaluations, we define the new useful  variable $Q_i$ by the equation:
\begin {equation}
\Omega_i = \frac{\psi_{i+1} Q_{i+1}}{Q_i},\ \psi_i = \sigma_i\frac{c_{i+1}}{2\triangle}.
\label{qq}
\end {equation}
Together with previously defined $\Sigma_i$:
\begin {equation}
\Sigma_i = \left(\frac{c_{i+1}}{2\triangle}\right)^2\ \sigma_i\ \sigma_{i+1}\ =\ \frac{c_{i+1}}{c_{i+2}}\psi_i\psi_{i+1}\ ,
\label{newsigma}
\end {equation}
\noindent
we can rewrite the term:

$$
\frac{\Sigma_{\Lambda-1}}{\Omega_{\Lambda-1}\Omega_{\Lambda-2}} =
\frac{c_{\Lambda}Q_{\Lambda-1}Q_{\Lambda-2}}{c_{\Lambda+1}Q_{\Lambda}Q_{\Lambda-1}}\ ,
$$

\noindent
and we find for Eq.(\ref{ld3}) :
\begin {eqnarray}
& &\left(c_{\Lambda+1}Q_{\Lambda}Q_{\Lambda-1}\right)^{i_{\Lambda}+\rho/2}\ \left(\Lambda\right)_{i_{\Lambda}}^{n_0+2\mu} =
\sum_{\lambda=\max(0,\,\mu-\mathcal{J}_{\Lambda})}
^{\min(\mu,\,\mathcal{J}_1+\cdots+\mathcal{J}_{\Lambda-1})} \binom{\mu}{\lambda}\ \left(\frac{z_{\Lambda}\Omega_{\Lambda-1}}{z_{\Lambda-1}\Omega_{\Lambda-2}}\right)^{2\lambda}
\label{rcn}\\
&\times&\sum_{i_{(\Lambda-1)}=i_{\Lambda}-2\mu+2\lambda}^{n_0+2\lambda}\
\frac{a_{i_{\Lambda}}^{2\mu-2\lambda+i_{(\Lambda-1)}}\ (\rho)}{(c_{\Lambda+1}Q_{\Lambda}Q_{\Lambda-1})^{i_{(\Lambda-1)}-i_{\Lambda}}}\
\left(c_{\Lambda}Q_{\Lambda-1}Q_{\Lambda-2}\right)^{i_{(\Lambda-1)}+\rho/2}
\left(\Lambda-1\right)_{i_{(\Lambda-1)}}^{n_0+2\lambda}\nonumber
\end {eqnarray}

\noindent
By summation variables transformations, $i_{(\Lambda-1)} \rightarrow k$ defined by the identity: $$i_{(\Lambda-1)} = i_{\Lambda}-2\mu + 2\lambda + k,$$
and $i_{\Lambda} \rightarrow p$ defined by the identity:
$$i_{\Lambda} = n_0+2\mu - p,$$
and finally by variable transformation $k' \rightarrow p-k$ we find from Eq.(\ref{rcn}) the recurrence relation:
\begin {equation}
C(\Lambda,\mu,p)= \sum_{\lambda=\max(0,\,\mu-\mathcal{J}_{\Lambda})}
^{\min(\mu,\,\mathcal{J}_1+\cdots+\mathcal{J}_{\Lambda-1})}\ \sum_{k'=0}^{p}\ A(\Lambda-1,\mu,p,k')\ C(\Lambda-1,\lambda,k')\ B(\Lambda-1,\mu,\lambda)
\label{rec}
\end {equation}
where we used the new variables defined as:
\begin {equation}
C(\Lambda,\mu,p)=\left(c_{\Lambda+1}Q_{\Lambda}Q_{\Lambda-1}\right)^{n_0+2\mu-p+\rho/2}\ \left(\Lambda\right)_{n_0+2\mu-p}^{n_0+2\mu}
\label{defC}
\end {equation}
\begin {equation}
A(\Lambda-1,\mu,p,k)=\frac{a_{n_0+2\mu-p}^{n_0+2\mu-k}\ (\rho)}{(c_{\Lambda+1}Q_{\Lambda}Q_{\Lambda-1})^{p-k}}
\end {equation}
\begin {equation}
B(\Lambda-1,\mu,\lambda)=\binom{\mu}{\lambda}\ \left(c_{\Lambda+1}Q_{\Lambda}Q_{\Lambda-1}\right)^{2\mu-2\lambda}\
\left(\frac{z_{\Lambda}\Omega_{\Lambda-1}}{z_{\Lambda-1}\Omega_{\Lambda-2}}\right)^{2\lambda}\ .
\end {equation}
Following the recurrence relation defined by Eq.(\ref{rec}) we find the result for $C(N-1,\mu,p):$
\begin {eqnarray}
& &C(N-1,\mu,p)= \label{recur5}\\
&=&\sum_{k_{N-2}}^p\ \sum_{k_{N-3}}^{k_{N-2}}\ . ... .\ \sum_{k_2}^{k_3}\ \sum_{k_1}^{k_2}\ \sum_{\lambda_{N-2}=\max(0,\,\mu-\mathcal{J}_{N-2})}^{\min(\mu,\,\mathcal{J}_{1}+\cdots+\mathcal{J}_{N-3})}\ \sum_{\lambda_{N-3}=\max(0,\,\lambda_{N-2}-\mathcal{J}_{N-3})}^{\min(\lambda_{N-2},\,\mathcal{J}_{1}+\cdots+\mathcal{J}_{N-4})}\ . ... .\
\sum_{\lambda_1=\max(0,\,\lambda_2-\mathcal{J}_2)}^{\min(\lambda_2,\,\mathcal{J}_1)}\nonumber \\
&\times&A(N-2,\mu,p,k_{N-2})\ A(N-3,\lambda_{N-2},k_{N-2},k_{N-3})\ . ... .\ A(2,\lambda_3,k_3,k_2)\ A(1,\lambda_2,k_2,k_1)\nonumber \\
&\times&C(1,k_1,\lambda_1)\nonumber \\
&\times&B(1,\lambda_1,\lambda_2) B(2,\lambda_2,\lambda_3)\ . ... .\ B(N-3,\lambda_{N-3},\lambda_{N-2})\ B(N-2,\lambda_{N-2},\mu)\nonumber
\end {eqnarray}
In the summation procedure over indices $k_i$ we proceed by sum over $k_1$ up to $k_{N-2}.$ The first sum can be read:
\begin {eqnarray}
& &\sum_{k_1=0}^{k_2}\ A(1,\lambda_2,k_2,k_1)C(1,k_1,\lambda_1)= 4^{-k_2}\ (c_2Q_1Q_0)^{n_0+2\lambda_1+\rho/2}\frac{(2n_0+4\lambda_1+\rho)!}{(2n_0+4\lambda_2-2k_2+\rho)!k_2!}\
\\
&\times&\sum_{k_1=0}^{k_2}\binom{k_2}{k_1}\ \frac{(2n_0+4\lambda_2-2k_1+\rho)!}{(2n_0+4\lambda_1-2k_1+\rho)!}
\frac{1}{(c_3Q_2Q_1)^{k_2-k_1}}\frac{1}{(c_2Q_1Q_0)^{k_1}} \nonumber
\end {eqnarray}
We can sum over $k_1$ by the introduction of the derivative operator  by the identity:
$$\frac{(2n_0+4\lambda_2-2k_1+\rho)!}{(2n_0+4\lambda_1-2k_1+\rho)!} =
\lim_{y_1\rightarrow 1}\ \partial_{y_1}^{4\lambda_2-4\lambda_1}\ y_1^{2n_0+4\lambda_2-2k_1+\rho}$$
The derivative operation loose the index $k_1$ from the factorials, left it as the power of the derivative variable, and we can sum over $k_1$. We have:
\begin {eqnarray}
& &\sum_{k_1=0}^{k_2}\ A(1,\lambda_2,k_2,k_1)C(1,k_1,\lambda_1)=\nonumber \\
&=&4^{-k_2}\ (c_2Q_1Q_0)^{n_0+2\lambda_1+\rho/2}\frac{(2n_0+4\lambda_1+\rho)!}{k_2!}\
\\
&\times&\sum_{n_1}^{4\lambda_2-4\lambda_1}\binom{4\lambda_2-4\lambda_1}{n_1}\ \frac{1}{(2n_0+4\lambda_1+n_1-2k_2+\rho)!}
\lim_{y_1\rightarrow 1}\ \partial_{y_1}^{n_1}
\left(\frac{y_1^2}{c_3Q_2Q_1}+\frac{1}{c_2Q_1Q_0}\right)^{k_2} \nonumber
\end {eqnarray}

By this method of evaluation of the summations over indices $k_i$ recurrently, we find:
\begin {eqnarray}
& &\sum_{k_{N-2}=0}^p\ \sum_{k_{N-3}=0}^{k_{N-2}}\ . ... .\ \sum_{k_2=0}^{k_3}\ \sum_{k_1=0}^{k_2}A(N-2,\mu,p,k_{N-2})\ . ... .\  A(1,\lambda_2,k_2,k_1)\ C(1,k_1,\lambda_1)=
\label{prodA} \\
&=&\sum_{n_1=0}^{4\lambda_2-4\lambda_1}\ \binom{4\lambda_2-4\lambda_1}{n_1}\
 . ... .\
\sum_{n_{N-2}=0}^{4\mu-4\lambda_1-n_1-n_2- ...-n_{N-3}}\ \binom{4\mu-4\lambda_1-n_1-n_2- ...-n_{N-3}}{n_{N-2}}\nonumber \\
&\times&4^{-p}\ (c_2Q_1Q_0)^{n_0+2\lambda_1+\rho/2}\ \frac{(2n_0+4\lambda_1+\rho)!}{p!}\ \frac{1}{(2n_0+4\lambda_1+n_1+...+n_{N-2}-2p+\rho)!}\nonumber \\ &\times&\left\{\prod_{j=1}^{N-2}\ \lim_{y_{j}\rightarrow 1}\ \partial_{y_{j}}^{n_j}\right\}
\left(\frac{y_{N-2}^2}{c_{N}Q_{N-1}Q_{N-2}}+\frac{y_{N-3}^2}{c_{N-1}Q_{N-2}Q_{N-3}}+ ... +\frac{y_{1}^2}{c_{3}Q_{2}Q_{1}}+\frac{1}{c_2Q_1Q_0}\right)^p\ . \nonumber
\end {eqnarray}

For the summations of the products of $B$ functions in Eq.(\ref{recur5}) we apply the art:

$$B(\Lambda-1, \mu, \lambda)\ B(\Lambda, \lambda, \nu) \approx
$$
$$
\approx \left(c_{\Lambda+1}Q_{\Lambda}Q_{\Lambda-1}\right)^{2\mu-2\lambda}\
\left(\frac{z_{\Lambda}\Omega_{\Lambda-1}}{z_{\Lambda-1}\Omega_{\Lambda-2}}\right)^{2\lambda}
\left(c_{\Lambda}Q_{\Lambda-1}Q_{\Lambda-2}\right)^{2\lambda-2\nu}\
\left(\frac{z_{\Lambda-1}\Omega_{\Lambda-2}}{z_{\Lambda-2}\Omega_{\Lambda-3}}\right)^{2\nu}= $$
$$
= (z_{\Lambda}\Omega_{\Lambda-1})^{2\mu}\
\left(\frac{c_{\Lambda+1}Q_{\Lambda}Q_{\Lambda-1}}{z_{\Lambda}\Omega_{\Lambda-1}}\right)^{2\mu-2\lambda}
\left(\frac{c_{\Lambda}Q_{\Lambda-1}Q_{\Lambda-2}}{z_{\Lambda-1}\Omega_{\Lambda-2}}\right)^{2\lambda-2\nu}\ \left(\frac{1}{z_{\Lambda-2}\Omega_{\Lambda-3}}\right)^{2\nu}\ .
$$

\noindent
For the product of $B$ terms we find:
\begin {eqnarray}
& &\sum_{\lambda_{N-2}=\max(0,\,\mu-\mathcal{J}_{N-2})}^
{\min(\mu,\,\mathcal{J}_1+\cdots+\mathcal{J}_{N-3})}
\ . ... .\ \sum_{\lambda_1=\max(0,\,\lambda_2-\mathcal{J}_2)}^{\min(\lambda_2,\,\mathcal{J}_1)} B(1,\lambda_1,\lambda_2)\ . ... .\ B(N-3,\lambda_{N-3},\lambda_{N-2})\ B(N-2,\lambda_{N-2},\mu)=\label{prodB}\\
&=&\sum_{\lambda_{N-2}=\max(0,\,\mu-\mathcal{J}_{N-2})}^
{\min(\mu,\,\mathcal{J}_1+\cdots+\mathcal{J}_{N-3})}
\ . ... .\ \sum_{\lambda_1=\max(0,\,\lambda_2-\mathcal{J}_2)}^{\min(\lambda_2,\,\mathcal{J}_1)}  \left(\frac{1}{c_2Q_1Q_0}\right)^{2\lambda_1}\
\left(\frac{c_2Q_1Q_0}{z_1\Omega_0}\right)^{2\lambda_1}\
\binom{\lambda_2}{\lambda_1}\left(\frac{c_3Q_2Q_1}{z_2\Omega_1}\right)^{2\lambda_2-2\lambda_1}
\binom{\lambda_3}{\lambda_2}.....\nonumber \\
& &.....
\binom{\lambda_{i+1}}{\lambda_i}\left(\frac{c_{i+2}Q_{i+1}Q_{i}}{z_{i+1}\Omega_{i}}\right)^{2\lambda_{i+1}-2\lambda_i}
\binom{\lambda_{i+2}}{\lambda_{i+1}}.....\nonumber\\
& &.....\binom{\mu}{\lambda_{N-2}}\left(\frac{c_{N}Q_{N-1}Q_{N-2}}{z_{N-1}\Omega_{N-2}}\right)^{2\mu-2\lambda_{N-2}}
\ (z_{N-1}\Omega_{N-2})^{2\mu} \nonumber
\end {eqnarray}

In the above equation, the factor $$\left(\frac{1}{c_2Q_1Q_0}\right)^{2\lambda_1}$$ take off the same term in Eq.(\ref{prodA}) and $$(z_{N-1}\Omega_{N-2})^{2\mu}$$ kill the same factor in $\mathcal{W}_{N}^{leading}$ (\ref{Ndim}).

We can simplify Eq. (\ref{prodB}) taking into account the definitions of the values $z_i$ (\ref{Sigma}), $\sigma_i$ (\ref{Sigma}), $Q_i$, and $\Omega_i$ (\ref{qq}):
\begin {equation}
\frac{c_{i+2}Q_{i+1}Q_{i}}{z_{i+1}\Omega_{i}} =\ \sqrt{8a_i\triangle^3}\ Q_{i}^2\ .
\label{delta}
\end {equation}
The new variable combination on the r.h.s. (\ref{delta}) is more agreeable for following evaluations. Let us stress the very important result emerging from this representations. As we can see in r.h.s. of (\ref{delta}), the factor $\triangle^{3\mu}$ appear in the Eq. (\ref{prodB}). To obtain the finite result in the continuum limit, this factor must be neutralized. As we will see in Eq. (\ref{triangle}), $\triangle^{2\mu}$ will be destroyed by an analytical evaluations and remaining $\triangle^{\mu}$ will serve to convert of the summations over $\lambda_i$ in (\ref{prodB}) to integrals in the continuum limit. This means, that $\mu-th$ term of the decomposition of the $\mathcal{W}_{N}^{leading}$ must contain just $\mu$ sum over $\lambda_i$, which will be
converted to $\mu-$ple  integral.

To evaluate the leading part of the $N-1$ dimensional integral we need the value $\left(N-1\right)_{i_{N-1}}^{n_0+2\mu}.$ Multiplying together Eq. (\ref{prodA}) and Eq. (\ref{prodB}) we find the value $C(\Lambda,\mu,p)$ (\ref{recur5}). By definition (\ref{defC}) we can finally find the value $\left(N-1\right)_{i_{N-1}}^{n_0+2\mu},$ where $i_{N-1}=n_0+2\mu-p,$ as:
\begin {eqnarray}
& &\left(N-1\right)_{n_0+2\mu-p}^{n_0+2\mu} = C(N-1,\mu,p)\ \left(\frac{1}{c_{N}Q_{N-1}Q_{N-2}}\right)^{n_0+2\mu-p+\rho/2}
\end {eqnarray}
Then for the leading part of  $\mathcal{W}_{N}^{leading}$ (\ref{Ndim}) we read:

\begin {eqnarray}
& &\mathcal{W}_{N}^{leading} =\ \frac{1}{\sqrt{\frac{2\pi\triangle}{c_0}\ \prod_{i=1}^{N-1}(\frac{c_i+c_{i+1}}{c_i}+2 \frac{b_i}{c_i}\triangle^2)\Omega_{i-1}}}\
\sum_{\mu=0}^{\lambda_1+\cdots+\lambda_{N-1}}\ \frac{(-8\triangle^3)^{\mu}}{\mu!\ (2 )^{\mu}}\label{Ndim1} \\
&\times&
\sum_{\lambda_{N-2}=\max(0,\,\mu-\mathcal{J}_{N-2})}^
{\min(\mu,\,\mathcal{J}_1+\cdots+\mathcal{J}_{N-3})}
\ . ... .\ \sum_{\lambda_1=\max(0,\,\lambda_2-\mathcal{J}_2)}^{\min(\lambda_2,\,\mathcal{J}_1)}
\binom{\lambda_2}{\lambda_1}\left(a_1Q_1^4\right)^{\lambda_2-\lambda_1}
\binom{\lambda_3}{\lambda_2}.....
\binom{\lambda_{i+1}}{\lambda_i}\left(a_iQ_{i}^4\right)^{\lambda_{i+1}-\lambda_i}\nonumber\\
& &.....\binom{\mu}{\lambda_{N-2}}\left(a_{N-2}Q_{N-2}^4\right)^{\mu-\lambda_{N-2}}\nonumber \\ &\times&
\sum_{n_1=0}^{4\lambda_2-4\lambda_1}\ \binom{4\lambda_2-4\lambda_1}{n_1}\
 . ... .\
\sum_{n_{N-2}=0}^{4\mu-4\lambda_1-n_1-n_2- ...-n_{N-3}}\ \binom{4\mu-4\lambda_1-n_1-n_2- ...-n_{N-3}}{n_{N-2}}\nonumber \\
&\times&
\sum\limits_{\rho=0}^{1}
\sum\limits_{n_0=0}^{\infty}\frac{\left(\frac{c_1}{\triangle} \varphi_0 \sqrt{\sigma_1}\ \sqrt{c_2Q_1Q_0}\right)^{2n_0+\rho}}{(2 n_0+\rho)!}
\sum_{p=0}^{n_0+2\mu}\ \left(\frac{(\frac{c_N}{2\triangle}\varphi_N)^2\
\sigma_{N-1}}{\Omega_{N-2}\ c_{N}Q_{N-1}Q_{N-2}}\right)^{n_0+2\mu-p+\rho/2}\nonumber\\
&\times&4^{-p}\ \frac{(2n_0+4\lambda_1+\rho)!}{p!}\ \frac{1}{(2n_0+4\lambda_1+n_1+...+n_{N-2}-2p+\rho)!}\nonumber \\ &\times&\left\{\prod_{j=1}^{N-2}\ \lim_{y_{j}\rightarrow 1}\ \partial_{y_{j}}^{n_j}\right\}
\left(\frac{y_{N-2}^2}{c_{N}Q_{N-1}Q_{N-2}}+\frac{y_{N-3}^2}{c_{N-1}Q_{N-2}Q_{N-3}}+ ... +\frac{y_{1}^2}{c_{3}Q_{2}Q_{1}}+\frac{1}{c_2Q_1Q_0}\right)^p
\nonumber \\
&\times&
\exp{\left\{-a_N\triangle\varphi_N^4 -
\left(\frac{c_N}{2\triangle}+b_N\triangle\right)\varphi_N^2+\left(\frac{(\frac{c_N}{2\triangle}\varphi_N)^2\
\sigma_{N-1}}{\Omega_{N-2}}\right)
-\frac{c_1}{2\triangle}\varphi_0^2\right\}}\nonumber.
\end {eqnarray}
Following the definitions of the symbols $\sigma_i,\  \Omega_i$ we find the identities:
\begin {equation}
\frac{c_1}{\triangle}  \sqrt{\sigma_1}\ \sqrt{c_2Q_1Q_0}\ \varphi_0 = \sqrt{c_1 c_2\ Q_0Q_1}\ \  \varphi_0,
\end {equation}

\begin {equation}
\frac{(\frac{c_N}{2\triangle}\varphi_N)^2\
\sigma_{N-1}}{\Omega_{N-2}\ c_{N}Q_{N-1}Q_{N-2}} = \frac{\varphi_N^2}{2\triangle\ Q_{N-1}^2}
\label{triangle}
\end {equation}

Then, we find the identity:

$$
4^{-p}\ \left(\frac{c_1}{\triangle} \varphi_0 \sqrt{\sigma_1}\ \sqrt{c_2Q_1Q_0}\right)^{2n_0+\rho}\
\left(\frac{(\frac{c_N}{2\triangle}\varphi_N)^2\
\sigma_{N-1}}{\Omega_{N-2}\ c_{N}Q_{N-1}Q_{N-2}}\right)^{n_0+2\mu-p+\rho/2} =
$$
$$
= \left(\frac{\varphi_N^4}{4\triangle^2\ Q_{N-1}^4}\right)^{\mu}\
\left(\sqrt{\frac{Q_0Q_1}{2\triangle^2}}\ \frac{\sqrt{c_1c_2}}{Q_{N-1}}\ \varphi_0 \varphi_N\right)^{2n_0-2p+\rho}\
\left(\frac{Q_0Q_1}{2\triangle^2}\ \frac{c_1c_2}{2}\ \varphi_0^2\right)^p\ \triangle^p
$$

For the expression in the exponential factor we prepared the finite factor for the continuum limit $\triangle\rightarrow 0:$
\begin {equation}
-\frac{c_N}{2\triangle} + \frac{(\frac{c_N}{2\triangle})^2\
\sigma_{N-1}}{\Omega_{N-2}} = -\frac{c_N}{2}\ \frac{Q_{N-1}-Q_{N-2}}{\triangle Q_{N-1}},
\end {equation}
the factor $\frac{Q_{N-1}-Q_{N-2}}{\triangle}$ is time derivative of the quantity $Q(\tau)$ in the end point $\beta$ of the time interval.

Inserting these identities to the EQ. (\ref{Ndim1}) we can simplify the leading part of the $N-1$ dimensional integral:

\begin {eqnarray}
& &\mathcal{W}_{N}^{leading} =\ \frac{1}{\sqrt{\frac{2\pi\triangle}{c_0}\ \prod_{i=1}^{N-1}(\frac{c_i+c_{i+1}}{c_i}+2 \frac{b_i}{c_i}\triangle^2)\Omega_{i-1}}}\
\sum_{\mu=0}\ \frac{(-1)^{\mu}}{\mu!} \left(\frac{\triangle\ \varphi_N^4}{Q_{N-1}^4}\right)^{\mu}\label{Ndim3} \\
&\times&
\sum_{\lambda_{N-2}=\max(0,\,\mu-\mathcal{J}_{N-2})}^
{\min(\mu,\,\mathcal{J}_1+\cdots+\mathcal{J}_{N-3})}
\ . ... .\ \sum_{\lambda_1=\max(0,\,\lambda_2-\mathcal{J}_2)}^{\min(\lambda_2,\,\mathcal{J}_1)}
\left(a_0Q_0^4\right)^{\lambda_1}\
\binom{\lambda_2}{\lambda_1}\left(a_1Q_1^4\right)^{\lambda_2-\lambda_1}
.....\binom{\mu}{\lambda_{N-2}}\left(a_{N-2}Q_{N-2}^4\right)^{\mu-\lambda_{N-2}}\nonumber \\ &\times&
\sum_{n_1=0}^{4\lambda_2-4\lambda_1}\ \binom{4\lambda_2-4\lambda_1}{n_1}\
 . ... .\
\sum_{n_{N-2}=0}^{4\mu-4\lambda_1-n_1-n_2- ...-n_{N-3}}\ \binom{4\mu-4\lambda_1-n_1-n_2- ...-n_{N-3}}{n_{N-2}}\nonumber \\
&\times&
\sum\limits_{\rho=0}^{1}
\sum\limits_{n_0=0}^{\infty}
\sum_{p=0}^{n_0+2\mu}\ \left(\sqrt{\frac{Q_0Q_1}{2\triangle^2}}\ \frac{\sqrt{c_1c_2}}{Q_{N-1}}\ \varphi_0 \varphi_N\right)^{2n_0-2p+\rho}\
\frac{(2n_0+4\lambda_1+\rho)!}{p!(2 n_0+\rho)!}\ \frac{1}{(2n_0+4\lambda_1+n_1+...+n_{N-2}-2p+\rho)!}\nonumber \\
&\times&
\left(\frac{Q_0Q_1}{2\triangle^2}\ \frac{c_1c_2}{2}\ \varphi_0^2\right)^p
\left\{\prod_{j=1}^{N-2}\ \lim_{y_{j}\rightarrow 1}\ \partial_{y_{j}}^{n_j}\right\}
\left(\frac{y_{N-2}^2\triangle}{c_{N}Q_{N-1}Q_{N-2}}+\frac{y_{N-3}^2\triangle}{c_{N-1}Q_{N-2}Q_{N-3}}+ ...
+\frac{y_{1}^2\triangle}{c_{3}Q_{2}Q_{1}}+\frac{\triangle}{c_2Q_1Q_0}\right)^p
\nonumber \\
&\times&
\exp{\left\{-a_N\triangle\varphi_N^4 -
b_N\triangle\varphi_N^2-\left(\frac{c_N}{2}\ \frac{Q_{N-1}-Q_{N-2}}{\triangle Q_{N-1}}\right)\ \varphi_N^2 -\frac{c_1}{2\triangle}\varphi_0^2\right\}}\nonumber.
\end {eqnarray}

In the next step we evaluate the derivatives $\partial_{y_{j}}^{n_j}$ in Eq. (\ref{Ndim3})

\section{Evaluation of the derivatives I}

In this section, we introduce the terms $\mathcal{A}_j^n$, which are essential for the appearance of the Hermite polynomials in our evaluation.
Our goal is to evaluate the expression:

\begin {equation}
\left\{\prod_{j=1}^{N-2}\ \lim_{y_{j}\rightarrow 1}\ \partial_{y_{j}}^{n_j}\right\}\left(d_{N-2}y_{N-2}^2+d_{N-1}y_{N-1}^2 + \cdots + d_1y_{1}^2 + d_0\right)^p,
\label{Nder}
\end {equation}
where
\begin {equation}
d_i = \frac{\frac{Q_0Q_1}{2\triangle^2}\ \frac{c_1c_2}{2}\ \varphi_0^2}{c_{i+2}Q_{i+1}Q_{i}}\ \triangle.
\label{defdi}
\end {equation}
Let us stress, that the value $-\frac{c_1}{2\triangle}\varphi_0^2\ ,$ appearing in the exponential factor (\ref{Ndim3}) practically from the beginning of our calculations, is equal to $-2d_0$ in the sense of our definition (\ref{defdi}). Thus, as we will see, we have "naturally" the regularization of the infinite terms in the continuum limit.

To evaluate (\ref{Nder}), we must evaluate the derivative:
\begin {equation}
\lim_{y \rightarrow 1}\partial_y^n\ (d\ y^2+a)^p = \sum_{i=0}^p \binom{p}{i}\ a^{p-i}\ d^i\ \frac{(2i)!}{(2i-n)!}
\label{defder}
\end {equation}
for each derivative variable $y_j$.
Symbol $d$ do not depend on the derivative variables, but in the $a,$ following definition (\ref{Nder}), there are hidden the derivative variables of the succeeding derivative steps.
However, to prepare this result for next derivatives  we define:
\begin {equation}
\frac{(2i)!}{(2i-n)!} = \sum_{j=0}^n\ \mathcal{A}_j^n\ \frac{i!}{(i-j)!}
\label{defcalA}
\end {equation}
Inserting (\ref{defcalA}) to (\ref{defder}), changing the order of summations, we read:
\begin {equation}
\lim_{y \rightarrow 1}\partial_y^n\ (d\ y^2+a)^p = \sum_{j=0}^n \mathcal{A}_j^n\ \frac{p!}{(p-j)!}\ d^j\ (d+a)^{p-j},
\label{finder}
\end {equation}
where $(d+a)^{p-j}$ depends on the following derivative variables $y_j,$ hidden in symbol $a.$

The $\mathcal{A}_j^n$  can be deduced from Eq. (\ref{defcalA}). One can find the equation:
\begin {equation}
\mathcal{A}_j^k = (2j - k + 1)\mathcal{A}_j^{k-1} + 2 \mathcal{A}_{j-1}^{k-1},\ \mathcal{A}_0^0 = 1,\ \mathcal{A}_{k+1}^k = 0,
\end {equation}
Solution of this recurrence equation is:
\begin {equation}
\mathcal{A}_j^k = 2^{2j-k}\ \binom{k}{j}\ \frac{j!}{(2j-k)!}
\label{mathcalA}
\end {equation}
With representation (\ref{finder}) for derivatives we can follow by recurrence procedure and  we can find for (\ref{Nder}) the result:
\begin {eqnarray}
& &\left\{\prod_{j=1}^{N-2}\ \lim_{y_{j}\rightarrow 1}\ \partial_{y_{j}}^{n_j}\right\}\left(d_{N-2}y_{N-2}^2+d_{N-1}y_{N-1}^2 + \cdots + d_1y_{1}^2 + d_0\right)^p =
\label{finder1} \\
&=&\prod_{j=1}^{N-2}\ \left\{\sum_{i_j=\lfloor \frac{n_j+1}{2}\rfloor}^{n_j}\ \mathcal{A}_{i_j}^{n_j}\ d_j^{i_j}\right\}\
\frac{p!}{(p-i_1-i_2- \cdots - i_{N-2})!}\ \left(d_{N-2}+d_{N-3}+\cdots + d_1+d_0\right)^{p-i_1-i_2- \cdots - i_{N-2}}\nonumber
\end {eqnarray}
An important feature of this result follow from the factorial in the denominator. The nonzero contributions to $N-1$ dimensional integral must obey the inequality $$p-i_1-i_2- \cdots - i_{N-2} \geq 0,$$ therefore we have the new lower limit for summation over index $p$ in Eq. (\ref{Ndim3}).

\section{Summations over indices $\rho,\ n_0,$ and $p$}

Inserting (\ref{finder1}) into Eq. (\ref{Ndim3}), replacing the order of summations over indices $\rho,\ n_0,\ p$ and $i_j$ for the leading part of the $N-1$ dimensional integral $\mathcal{W}_{N}^{leading}$ we can read:
\begin {eqnarray}
& &\mathcal{W}_{N}^{leading} =\ \frac{1}{\sqrt{\frac{2\pi\triangle}{c_0}\ \prod_{i=1}^{N-1}(\frac{c_i+c_{i+1}}{c_i}+2 \frac{b_i}{c_i}\triangle^2)\Omega_{i-1}}}\
\sum_{\mu=0}\ \frac{(-1)^{\mu}}{\mu!} \left(\frac{\triangle\ \varphi_N^4}{Q_{N-1}^4}\right)^{\mu}\label{Ndim4} \\
&\times&
\sum_{\lambda_{N-2}=\max(0,\,\mu-\mathcal{J}_{N-2})}^
{\min(\mu,\,\mathcal{J}_1+\cdots+\mathcal{J}_{N-3})}
\ . ... .\ \sum_{\lambda_1=\max(0,\,\lambda_2-\mathcal{J}_2)}^{\min(\lambda_2,\,\mathcal{J}_1)}
\left(a_0Q_0^4\right)^{\lambda_1}\
\binom{\lambda_2}{\lambda_1}\left(a_1Q_1^4\right)^{\lambda_2-\lambda_1}
.....\binom{\mu}{\lambda_{N-2}}\left(a_{N-2}Q_{N-2}^4\right)^{\mu-\lambda_{N-2}}\nonumber \\ &\times&
\sum_{n_1=0}^{4\lambda_2-4\lambda_1}\ \binom{4\lambda_2-4\lambda_1}{n_1}\
 . ... .\
\sum_{n_{N-2}=0}^{4\mu-4\lambda_1-n_1-n_2- ...-n_{N-3}}\ \binom{4\mu-4\lambda_1-n_1-n_2- ...-n_{N-3}}{n_{N-2}}\nonumber \\
&\times&
\sum_{i_1=\lfloor \frac{n_1+1}{2}\rfloor}^{n_1}\ \mathcal{A}_{i_1}^{n_1}\ d_1^{i_1}\
\sum_{i_2=\lfloor \frac{n_2+1}{2}\rfloor}^{n_2}\ \mathcal{A}_{i_2}^{n_1}\ d_2^{i_2}\ \cdots
\sum_{i_{N-2}=\lfloor \frac{n_{N-2}+1}{2}\rfloor}^{n_{N-2}}\ \mathcal{A}_{i_{N-2}}^{n_{N-2}}\ d_{N-2}^{i_{N-2}}\
\nonumber \\
&\times&
\sum\limits_{\rho=0}^{1}
\sum\limits_{n_0=0}^{\infty}
\sum_{p=\mathcal{S}_i}^{n_0+2\mu}\ \frac{(2n_0+4\lambda_1+\rho)!}{(p-\mathcal{S}_i)!(2 n_0+\rho)!}\ \frac{\mathcal{X}^{2n_0-2p+\rho}}{(2n_0+4\lambda_1+\mathcal{S}_n-2p+\rho)!}\
\mathcal{Y}^{p-\mathcal{S}_i}
\nonumber \\
&\times&
\exp{\left\{-a_N\triangle\varphi_N^4 -
b_N\triangle\varphi_N^2-\left(\frac{c_N}{2}\ \frac{Q_{N-1}-Q_{N-2}}{\triangle Q_{N-1}}\right)\varphi_N^2 -\frac{c_1}{2\triangle}\varphi_0^2\right\}}\nonumber,
\end {eqnarray}
where we used the for the summations over indexes $i_j$ and $n_j$ the abbreviations: $$\mathcal{S}_i = i_1+i_1+\cdots + i_{N-2},\ \mathcal{S}_n = n_1+n_2+\cdots + n_{N-2},$$
and we introduced the abbreviations:
$$ \mathcal{X}=\sqrt{\frac{Q_0Q_1}{2\triangle^2}}\ \frac{\sqrt{c_1c_2}}{Q_{N-1}}\ \varphi_0
\varphi_N,$$
\begin {equation}
\mathcal{Y} = d_{N-2}+d_{N-3}+\cdots + d_1+d_0.
\label{defY}
\end {equation}

As we fixed the lower limit in summation over index $p$ from the condition imposed to the denominator in the previous Appendix C,  also the upper limit is fixed by the argument of the factorial in the denominator of Eq. (\ref{Ndim4}), which must be positive: $$2n_0+4\lambda_1+\mathcal{S}_n-2p+\rho \geq 0,$$
therefore upper limit for sum over index $p$ is $n_0+2\lambda_1+\lfloor\frac{\mathcal{S}_n+\rho}{2}\rfloor.$
Let us outstrip the flow of the evaluation. It will be shown, that $Q_0=\triangle.$ This means that in the continuum limit the term $\left(a_0Q_0^4\right)^{\lambda_1}$ in Eq. (\ref{Ndim4}) destroys the contribution to $N-1$ dimensional integral if $\lambda_1
\geq 1.$ For $\lambda_1 = 0$ we find the nonzero contribution and we will exploit this fact to simplify our formulas and from this moment we put $\lambda_1 = 0.$

We perform the summation variable transformation $p\rightarrow q$ by replacement
$p = \mathcal{S}_i + q$ and for summations in Eq. (\ref{Ndim4}) we reads:
\begin {eqnarray}
& &\sum\limits_{\rho=0}^{1}
\sum\limits_{n_0=0}^{\infty}
\sum_{p=\mathcal{S}_i}^{n_0+2\mu}\ \cdots = \\
&=&\mathcal{X}^{-\mathcal{S}_n}\ \sum\limits_{\rho=0}^{1}\
\sum\limits_{n_0=0}^{\infty}\
\sum_{q=0}^{n_0+\lfloor\frac{\mathcal{S}_n+\rho}{2}\rfloor-\mathcal{S}_i}\ \frac{\mathcal{Y}^{q}}{q!}\
\frac{\mathcal{X}^{2n_0+\mathcal{S}_n-2q-2\mathcal{S}_i+\rho}}
{(2n_0+\mathcal{S}_n-2q-2\mathcal{S}_i+\rho)!},\nonumber
\end {eqnarray}

In Eq. (\ref{Ndim4}) for individual sum over index $i_j$ we have the inequality:
$$\lfloor \frac{n_j+1}{2} \rfloor \leq \ i_j\ \leq n_j,$$
and with little algebra it can be proved the inequality:
$$\lfloor\frac{\mathcal{S}_n+\rho}{2}\rfloor\ \leq \ \mathcal{S}_i\ \leq \ \mathcal{S}_n.$$
Because $\lfloor\frac{\mathcal{S}_n+\rho}{2}\rfloor-\mathcal{S}_i$ is non-positive, the sum
over index $n_0$ must start from a positive value $\mathcal{S}_i-\lfloor\frac{\mathcal{S}_n+\rho}{2}\rfloor,$ otherwise the sum over index $q$ is nonsense.
By summation variable transformation $n_0 \rightarrow l_0$ defined by
$$2l_0 = 2n_0+\mathcal{S}_n-2\mathcal{S}_i$$
we can read:
\begin {eqnarray}
& &\sum\limits_{\rho=0}^{1}
\sum\limits_{n_0=0}^{\infty}
\sum_{p=\mathcal{S}_i}^{n_0+2\mu}\ \cdots = \\
&=&\mathcal{X}^{-\mathcal{S}_n}\ \sum\limits_{\rho=0}^{1}\
\sum\limits_{l_0=0}^{\infty}\
\sum_{q=0}^{l_0}\ \frac{\mathcal{Y}^{q}}{q!}\ \frac{\mathcal{X}^{2l_0-2q+\rho}}{(2l_0-2q+\rho)!},\nonumber
\end {eqnarray}
and by replacement of the order of summation: $$\sum\limits_{l_0=0}^{\infty}\
\sum_{q=0}^{l_0}\ \rightarrow \ \sum_{q=0}^{\infty}\ \sum\limits_{l_0=q}^{\infty}\ $$
we find that
\begin {eqnarray}
\sum\limits_{\rho=0}^{1}
\sum\limits_{n_0=0}^{\infty}
\sum_{p=\mathcal{S}_i}^{n_0+2\mu}\ \cdots &\rightarrow& \
\mathcal{X}^{-\mathcal{S}_n}\ \sum_{q=0}^{\infty}\ \sum\limits_{\rho=0}^{1}\ \sum\limits_{l_0=q}^{\infty}\
\frac{\mathcal{Y}^{q}}{q!}\ \frac{\mathcal{X}^{2l_0-2q+\rho}}{(2l_0-2q+\rho)!} = \nonumber \\
&=& \mathcal{X}^{-\mathcal{S}_n}\ \exp{(\mathcal{Y}\ +\ \mathcal{X})}.
\end {eqnarray}

Inserting this result to the Eq.(\ref{Ndim4}) for leading part of $N-1$ dimensional integral we read:
\begin {eqnarray}
& &\mathcal{W}_{N}^{leading} =\ \frac{1}{\sqrt{\frac{2\pi\triangle}{c_0}\ \prod_{i=1}^{N-1}(\frac{c_i+c_{i+1}}{c_i}+2 \frac{b_i}{c_i}\triangle^2)\Omega_{i-1}}}\
\sum_{\mu=0}\ \frac{(-1)^{\mu}}{\mu!} \left(\frac{\triangle\ \varphi_N^4}{Q_{N-1}^4}\right)^{\mu}\label{Ndim5} \\
&\times&
\sum_{\lambda_{N-2}=\max(0,\,\mu-\mathcal{J}_{N-2})}^
{\min(\mu,\,\mathcal{J}_1+\cdots+\mathcal{J}_{N-3})}
\ . ... .\ \sum_{\lambda_1=\max(0,\,\lambda_2-\mathcal{J}_2)}^{\min(\lambda_2,\,\mathcal{J}_1)}
\left(a_1Q_1^4\right)^{\lambda_2}\ \binom{\lambda_3}{\lambda_2}\left(a_2Q_2^4\right)^{\lambda_3-\lambda_2}
.....\binom{\mu}{\lambda_{N-2}}\left(a_{N-2}Q_{N-2}^4\right)^{\mu-\lambda_{N-2}}\nonumber \\ &\times&
\sum_{n_1=0}^{4\lambda_2}\ \binom{4\lambda_2}{n_1}\
 . ... .\
\sum_{n_{N-2}=0}^{4\mu-n_1-n_2- ...-n_{N-3}}\ \binom{4\mu-n_1-n_2- ...-n_{N-3}}{n_{N-2}}\nonumber \\
&\times&\mathcal{X}^{-\mathcal{S}_n}\ \exp{(\mathcal{Y}\ +\ \mathcal{X})}\
\exp{\left\{-a_N\triangle\varphi_N^4 -
b_N\triangle\varphi_N^2-\left(\frac{c_N}{2}\ \frac{Q_{N-1}-Q_{N-2}}{\triangle Q_{N-1}}\right)\ \varphi_N^2 -\frac{c_1}{2\triangle}\varphi_0^2\right\}}\nonumber \\ &\times&
\sum_{i_1=\lfloor \frac{n_1+1}{2}\rfloor}^{n_1}\ \mathcal{A}_{i_1}^{n_1}\ d_1^{i_1}\
\sum_{i_2=\lfloor \frac{n_2+1}{2}\rfloor}^{n_2}\ \mathcal{A}_{i_2}^{n_2}\ d_2^{i_2}\ \cdots
\sum_{i_{N-2}=\lfloor \frac{n_{N-2}+1}{2}\rfloor}^{n_{N-2}}\ \mathcal{A}_{i_{N-2}}^{n_{N-2}}\ d_{N-2}^{i_{N-2}}\ .
\nonumber
\end {eqnarray}
Following definition of the values $d_i$ (\ref{defdi}) we can see that
$$2d_0 = \frac{c_1}{2\triangle}\varphi_0^2.$$

We will see, that this is exactly the term which regularize the integral resulting from continuum limit of the value $\mathcal{Y}:$
$$\mathcal{Y}-2d_0 = d_{N-2}+d_{N-3}+\cdots + d_1-d_0 \rightarrow\ \lim_{\epsilon\rightarrow 0}\left\{\int_{\epsilon}^{\beta}\frac{d\tau}{c(\tau)Q^2(\tau)}-
\frac{\epsilon}{c(\epsilon)Q^2(\epsilon)}\right\}\frac{c^2(0)\varphi_0^2}{2}\ \cdot$$

The next step in our evaluations is the summations over indices $i_j.$

\section{Summations over indices $i_j$}

These summations are very important for the evaluation because we introduce the Hermite polynomials into our results.
Following the definition of the symbol $\mathcal{A}$ in Eq. (\ref{mathcalA}), our goal is to evaluate the sum:
\begin {equation}
\sum_{i=\lfloor \frac{n+1}{2}\rfloor}^{n}\ \mathcal{A}_{i}^{n}\ d^{i}\ = \sum_{i=\lfloor \frac{n+1}{2}\rfloor}^{n}\
2^{2i-n}\ \binom{n}{i}\ \frac{i!}{(2i-n)!}\ d^{i}.
\label{e1}
\end {equation}
Let us remember, that
\begin {equation}
d_i = \frac{\frac{Q_0Q_1}{2\triangle^2}\ \frac{c_1c_2}{2}\ \varphi_0^2}{c_{i+2}Q_{i+1}Q_{i}}\ \triangle.
\end {equation}

We find in  Prudnikov`s tables, \cite{prud} Vol I, Chapter 4.1.7, Eq.(13):
$$
\sum_{k=0}^{n}\frac{(-1)^k}{(2k+\delta)!\ (n-k)!}\ x^{2k+\delta}\ =\
\frac{(-1)^n}{(2n+\delta)!}H_{2n+\delta}\left(\frac{x}{2}\right)\ ,\ \ \ \delta=0, 1.
$$

To apply this identity to Eq. (\ref{e1}) we distinguish two events, when $n=2k$ is even and when $n=2k+1$ is odd. For $n$ even, we read:

\begin {eqnarray}
\sum_{j=k}^{2k}\
2^{2j-2k}\ \binom{2k}{j}\ \frac{j!}{(2j-2k)!}\ d^{j} &=& \
(2k)! (\sqrt{d})^{2k}\ \sum_{m=0}^k\ \frac{(-1)^m(2i\sqrt{d})^{2m}}{(k-m)!((2m)!} =\ (-1)^k(\sqrt{d})^{2k}\ H_{2k}(i\sqrt{d})=\nonumber\\
&=& (\sqrt{-d})^{2k}\ H_{2k}(\sqrt{-d})
\end {eqnarray}

Similarly, for odd $n$ we have:

$$
\sum_{i=k+1}^{2k+1}\ 2^{2i-2k-1}\ \binom{2k+1}{i}\ \frac{i!}{(2i-2k-1)!}\ d^{i} = \ (-\sqrt{-d})^{2k+1}\ H_{2k+1}(\sqrt{-d})
$$
When we compare both results, we read for the sum in question:
\begin {equation}
\sum_{i=\lfloor \frac{n+1}{2}\rfloor}^{n}\ \mathcal{A}_{i}^{n}\ d^{i}\ = \ (-\sqrt{-d})^{n}\ H_{n}(\sqrt{-d}),
\end {equation}
where $H_n(x)$ are ordinary Hermite polynomials \cite{bateman} defined as
$$H_n(x) = n!\ \sum_{m=0}^{\lceil n/2 \rceil}\ \frac{(-1)^m(2x)^{n-2m}}{m! (n-2m)!}$$

Inserting this result to Eq. (\ref{Ndim5}) for $\mathcal{W}_{N}^{leading}$ we read:

\begin {eqnarray}
& &\mathcal{W}_{N}^{leading} =\ \frac{1}{\sqrt{\frac{2\pi\triangle}{c_0}\ \prod_{i=1}^{N-1}(\frac{c_i+c_{i+1}}{c_i}+2 \frac{b_i}{c_i}\triangle^2)\Omega_{i-1}}}\
\sum_{\mu=0}\ \frac{(-1)^{\mu}}{\mu!} \left(\frac{\triangle\ \varphi_N^4}{Q_{N-1}^4}\right)^{\mu}\label{Ndim6} \\
&\times&
\sum_{\lambda_{N-2}=\max(0,\,\mu-\mathcal{J}_{N-2})}^
{\min(\mu,\,\mathcal{J}_1+\cdots+\mathcal{J}_{N-3})}
\ . ... .\ \sum_{\lambda_2=\max(0,\,\lambda_3-\mathcal{J}_3)}^{\min(\lambda_3,\,\mathcal{J}_1+\mathcal{J}_2)}
\left(a_1Q_1^4\right)^{\lambda_2}\ \binom{\lambda_3}{\lambda_2}\left(a_2Q_2^4\right)^{\lambda_3-\lambda_2}.....\nonumber\\
&\times&
.....\binom{\lambda_{i+1}}{\lambda_i}\left(a_iQ_{i}^4\right)^{\lambda_{i+1}-\lambda_i}
.....\binom{\mu}{\lambda_{N-2}}\left(a_{N-2}Q_{N-2}^4\right)^{\mu-\lambda_{N-2}}\nonumber \\ &\times&
\sum_{n_1=0}^{4\lambda_2}\ \binom{4\lambda_2}{n_1}\ \left(\frac{-\sqrt{-d_1}}{\mathcal{X}}\right)^{n_1}\ H_{n_1}(\sqrt{-d_1})
\sum_{n_2=0}^{4\lambda_3-n_1}\ \binom{4\lambda_3-n_1}{n_2}\ \left(\frac{-\sqrt{-d_2}}{\mathcal{X}}\right)^{n_2}\ H_{n_2}(\sqrt{-d_2})\nonumber \\
&\times& . ... .\
\sum_{n_{N-2}=0}^{4\mu-n_1-n_2- ...-n_{N-3}}\ \binom{4\mu-n_1-n_2- ...-n_{N-3}}{n_{N-2}}
\left(\frac{-\sqrt{-d_{N-2}}}{\mathcal{X}}\right)^{n_{N-2}}\ H_{n_{N-2}}(\sqrt{-d_{N-2}})
\nonumber \\
&\times& \exp{(\mathcal{Y}\ +\ \mathcal{X})}\
\exp{\left\{-a_N\triangle\varphi_N^4 -
b_N\triangle\varphi_N^2-\left(\frac{c_N}{2}\ \frac{Q_{N-1}-Q_{N-2}}{\triangle Q_{N-1}}\right)\ \varphi_N^2 -\frac{c_1}{2\triangle}\varphi_0^2\right\}}\ .\nonumber
\end {eqnarray}

In the above equation, we have the summations of the products of the Hermite polynomials. We will to simplify this equation by
the summations over indexes $n_j$.

\section{Summations over indices $n_j$}

To approaching the continuum limit, the number of the slice points $N$ will be greater than steeps of the decomposition $\mu$. Because the maximal number of the summations indexes $\lambda_i$ in (\ref{Ndim6}) is $\mu$, some of the summations indexes $\lambda_i, \cdots, \lambda_{N-2}$  must be equal.
First of all, let us discuss the case, when two neighboring indexes $\lambda_i,$ $\lambda_{i+1}$ are equal.

In the first part of the Eq. (\ref{Ndim6}), we find $$\binom{\lambda_{i+1}}{\lambda_i}\left(a_iQ_{i}^4\right)^{\lambda_{i+1}-\lambda_i} = 1,$$

and in the second part of this equation we are going to evaluate the summations: $$\sum_{n_i=0}^{K}\ \binom{K}{n_i}\ \left(\frac{-\sqrt{-d_i}}{\mathcal{X}}\right)^{n_i}\ H_{n_i}(\sqrt{-d_i})\ \sum_{n_{i+1}=0}^{K-n_i}\ \binom{K-n_i}{n_{i+1}}\
\left(\frac{-\sqrt{-d_{i+1}}}{\mathcal{X}}\right)^{n_{i+1}}\ H_{n_{i+1}}(\sqrt{-d_{i+1}}).$$

Following the characteristics of the binomials, we can rewrite the above term to the form: $$\sum_{n_{ii}=0}^{K}\ (-1)^{n_{ii}}\binom{K}{n_{ii}}\ \sum_{n_i=0}^{n_{ii}}\binom{n_{ii}}{n_i}\ \left(\frac{\sqrt{-d_i}}{\mathcal{X}}\right)^{n_i}\
\left(\frac{\sqrt{-d_{i+1}}}{\mathcal{X}}\right)^{n_{ii}-n_i}\ H_{n_i}(\sqrt{-d_i})\ H_{n_{ii}-n_i}(\sqrt{-d_{i+1}}),$$
here the symbol $n_{ii}$ is defined as
$n_{ii} = n_i + n_{i+1}.$

By summation identity \cite{prud}
\begin {equation}
\sum_k^n\ \binom{n}{k}\ t^k\ H_k(x)\ H_{n-k}(y)\ =\ \left(\sqrt{t^2+1}\right)^n\ H_n\left(\frac{tx+y}{\sqrt{t^2+1}}\right)\,,
\label{sumH}
\end {equation}

the double summations is reduced to single sum with the argument of the resulting square-root and the Hermite polynomial equal to the sum of the original arguments:
\begin {equation}
\sum_{n_{ii}=0}^{K}\ \binom{K}{n_{ii}}\ \left(-\frac{\sqrt{-d_i-d_{i+1}}}{\mathcal{X}}\right)^{n_{ii}}\ H_{n_{ii}}(\sqrt{-d_i-d_{i+1}})
\end {equation}

We can conclude that for a group of equal indexes $\lambda_i$, we will have only one sum by the index $n_{ii}$, and the argument of the corresponding Hermite polynomial is the combination of single arguments of the whole group. The number of the sums over indices $n_i$ is, therefore, equal to the number of the groups of equal $\lambda_i$. The same is valid for the first part of the Eq.(\ref{Ndim6}), where we find the non-trivial contribution to the sum only for the index $i,$ characterizing the passage from one equal value $\lambda_i$ group to next equal value $\lambda_j$ group.
To the non-zero leading term of the $N-1$ dimensional integral therefore contributes the all
ordered products of the terms $a_iQ_{i}^4,$ where $i$ is the sequence number of the slice point, where the group of the values $\lambda_i$ is changed. Due to reason to neutralize the factor $\triangle^{\mu}$ we must construct the object giving in the continuum $\mu-$ dimensional integral. It will be done if we choice
just $\mu$ groups of nonzero $\lambda_i$, therefore the difference between neighbouring groups is equal to one. Therefore we can replace the summations
of the product of $\mu$ ordered terms $a_iQ_{i}^4$
over $\lambda_i$ by $\mu$ summations over the points where this change take place:

\begin {eqnarray}
& &\sum_{\lambda_{N-2}=\max(0,\,\mu-\mathcal{J}_{N-2})}^
{\min(\mu,\,\mathcal{J}_1+\cdots+\mathcal{J}_{N-3})}
\ . ... .\ \sum_{\lambda_2=\max(0,\,\lambda_3-\mathcal{J}_3)}^{\min(\lambda_3,\,\mathcal{J}_1+\mathcal{J}_2)}
\left(a_1Q_1^4\right)^{\lambda_2}\ \binom{\lambda_3}{\lambda_2}\left(a_2Q_2^4\right)^{\lambda_3-\lambda_2}.....\nonumber \\
& &.....\binom{\lambda_{i+1}}{\lambda_i}\left(a_iQ_{i}^4\right)^{\lambda_{i+1}-\lambda_i}
.....\binom{\mu}{\lambda_{N-2}}\left(a_{N-2}Q_{N-2}^4\right)^{\mu-\lambda_{N-2}} = \nonumber \\ &=& \mu!\sum_{p_1=1}^{N-\mu}\ a_{p_1}Q_{p_1}^4\ \sum_{p_2=p_1+1}^{N-\mu+1}\ a_{p_2}Q_{p_2}^4\ .....
\sum_{p_{\mu}=p_{\mu-1}+1}^{N}\ a_{p_{\mu}}Q_{p_{\mu}}^4
\end {eqnarray}
Let us stress that $\mu!$ term in the above result cancel  out the same term in the denominator of Eq. (\ref{Ndim6}).

The arguments of the Hermite functions in (\ref{Ndim6}) for equal values of $\lambda_i$ possesses the sum od $d_i$.
Let us denote by $D_i$ the sum of $d_j$ for the group of equal $\lambda_j = i:$
$$D_i\,=\, d_{p_{i}} + d_{p_{i}+1} + \cdots + d_{p_{(i+1)}-1}.$$

Then for summations over indices $n_i$ in the Eq.(\ref{Ndim6}) we reads:
\begin {eqnarray}
& &
\sum_{n_1=0}^{4\lambda_2}\ \binom{4\lambda_2}{n_1}\ \left(\frac{-\sqrt{-d_1}}{\mathcal{X}}\right)^{n_1}\ H_{n_1}(\sqrt{-d_1})
\sum_{n_2=0}^{4\lambda_3-n_1}\ \binom{4\lambda_3-n_1}{n_2}\ \left(\frac{-\sqrt{-d_2}}{\mathcal{X}}\right)^{n_2}\ H_{n_2}(\sqrt{-d_2})\cdots \nonumber \\
&\times& . ... .\
\sum_{n_{N-2}=0}^{4\mu-n_1-n_2- ...-n_{N-3}}\ \binom{4\mu-n_1-n_2- ...-n_{N-3}}{n_{N-2}}
\left(\frac{-\sqrt{-d_{N-2}}}{\mathcal{X}}\right)^{n_{N-2}}\ H_{n_{N-2}}(\sqrt{-d_{N-2}})=\nonumber \\
&=&\sum_{n_1=0}^{4}\ \binom{4}{n_1}\ \left(\frac{-\sqrt{-D_1}}{\mathcal{X}}\right)^{n_1}\ H_{n_1}(\sqrt{-D_1})\ \cdots
\ \sum_{n_i=0}^{4i-n_1\cdots -n_{i-1}}\ \binom{4i-n_1\cdots -n_{i-1}}{n_i}\ \left(\frac{-\sqrt{-D_i}}{\mathcal{X}}\right)^{n_i}\ H_{n_i}(\sqrt{-D_i})\ \nonumber \\
&\times& \cdots\  \sum_{n_{\mu}=0}^{4\mu-n_1\cdots -n_{\mu-1}}\ \binom{4\mu-n_1\cdots -n_{\mu-1}}{n_{\mu}}\ \left(\frac{-\sqrt{-D_{\mu}}}{\mathcal{X}}\right)^{n_{\mu}}\ H_{n_{\mu}}(\sqrt{-D_{\mu}})
\label{E64}
\end {eqnarray}
Let us stress, that through variable $D_i$ we introduced the dependence on the summation indices $p_i$ to this second part, containing the Hermite polynomials.
Another essential feature of this summation are the different modes of the sum in the first step, where we sum one Hermite polynomial only and the following summations, where we sum the product of the Hermite polynomials.

To summing up the equation (\ref{E64}), we start by summation over index $n_{\mu}$ by help of the summation identity for Hermite polynomials \cite{prud}:
\begin {equation}
\sum_k^n\ \binom{n}{k}\ t^k\ H_k(x)\ =\ t^n\ H_n\left(x+\frac{1}{2t}\right)
\label{sumH2}
\end {equation}
By abbreviations $$t_i = \frac{-\sqrt{-D_i}}{\mathcal{X}},\ x_i = \sqrt{-D_{i}}$$ the result is:
\begin {equation}
\sum_{n_{\mu}=0}^{4\mu-n_1\cdots -n_{\mu-1}}\binom{4\mu-n_1\cdots -n_{\mu-1}}{n_{\mu}}\ t_{\mu}^{n_{\mu}}H_{n_{\mu}}\left(x_{\mu}\right) = \left(t_{\mu}\right)^{4\mu-n_1\cdots -n_{\mu-1}}\
H_{4\mu-n_1\cdots -n_{\mu-1}}\left(x_{\mu}+\frac{1}{2t_{\mu}}\right)
\label{singleH}
\end {equation}
The next step is the sum over index $n_{\mu-1}.$ Inserting above result (\ref{singleH}) to this sum one can read:
\begin {eqnarray}
& &\left(t_{\mu}\right)^{4\mu-n_1\cdots -n_{\mu-2}}\ \
\sum_{n_{\mu-1}=0}^{4(\mu-1)-n_1\cdots -n_{\mu-2}}\ \binom{4(\mu-1)-n_1\cdots -n_{\mu-2}}{n_{\mu-1}}\label{f7} \\
&\times&\left(\frac{t_{\mu-1}}{t_{\mu}}\right)^{n_{\mu-1}} H_{n_{\mu-1}}\left(x_{\mu-1}\right)\ H_{4\mu-n_1\cdots -n_{\mu-1}}\left(x_{\mu}+\frac{1}{2t_{\mu}}\right)\nonumber
\end {eqnarray}
We would like to utilize for this sum the identity (\ref{sumH}). To proceed, we must:

a) extend summation over $n_{\mu-1}$ up to $4\mu-n_1\cdots -n_{\mu-2}.$ This means, that we add four terms to this sum. Due to the binomial factor we add by this extension zero-valued terms.

b) to replace the binomial in above equation by the identity: $$\binom{4(\mu-1)-n_1\cdots -n_{\mu-2}}{n_{\mu-1}}\ =\ \frac{(4(\mu-1)-n_1\cdots -n_{\mu-2})!}{(4\mu-n_1\cdots -n_{\mu-2})!}\
\binom{4\mu-n_1\cdots -n_{\mu-2}}{n_{\mu-1}}\ \frac{(4\mu-n_1\cdots -n_{\mu-1})!}{(4(\mu-1)-n_1\cdots -n_{\mu-1})!}$$

c) an unpleasant feature of the above identity are the factorials dependent on summations variable $n_{\mu-1}.$  To replace the division of the factorials with $n_{\mu-1}$ summation variable by a power of an auxiliary variable we utilize the identity:
$$\frac{(4\mu-n_1\cdots -n_{\mu-1})!}{(4(\mu-1)-n_1\cdots -n_{\mu-1})!} = \lim_{\xi_{\mu-1}}\ \partial_{\xi_{\mu-1}}^4\
\left(\xi_{\mu-1}\right)^{4\mu-n_1\cdots -n_{\mu-1}}$$
Finally, for Eq. (\ref{f7}) we have:
\begin {eqnarray}
& &\frac{(4(\mu-1)-n_1\cdots -n_{\mu-2})!}{(4\mu-n_1\cdots -n_{\mu-2})!}\ \lim_{\xi_{\mu-1}}\ \partial_{\xi_{\mu-1}}^4\ \left\{
\left(\xi_{\mu-1}t_{\mu}\right)^{4\mu-n_1\cdots -n_{\mu-2}}\ \
\sum_{n_{\mu-1}=0}^{4\mu-n_1\cdots -n_{\mu-2}}\ \binom{4\mu-n_1\cdots -n_{\mu-2}}{n_{\mu-1}}\right. \nonumber \\
&\times& \left. \left(\frac{t_{\mu-1}}{\xi_{\mu-1}\ t_{\mu}}\right)^{n_{\mu-1}} H_{n_{\mu-1}}\left(x_{\mu-1}\right)\ H_{4\mu-n_1\cdots -n_{\mu-1}}\left(x_{\mu}+\frac{1}{2t_{\mu}}\right)\right\} .
\end {eqnarray}
Now we are prepared to provide the sum over index $n_{\mu-1}.$ Following the identity (\ref{sumH}) the final result for Eq. (\ref{f7}) is:
\begin {eqnarray}
& &\frac{(4(\mu-1)-n_1\cdots -n_{\mu-2})!}{(4\mu-n_1\cdots -n_{\mu-2})!}\ \lim_{\xi_{\mu-1}}\ \partial_{\xi_{\mu-1}}^4\ \left\{
\left(\sqrt{t_{\mu-1}^2+t_{\mu}^2\xi_{\mu-1}^2}\right)^{4\mu-n_1\cdots -n_{\mu-2}}\right. \\
&\times&\left. H_{4\mu-n_1\cdots -n_{\mu-2}}\left(\frac{t_{\mu-1}x_{\mu-1}+(t_{\mu}x_{\mu}+1/2)\xi_{\mu-1}}{\sqrt{t_{\mu-1}^2+t_{\mu}^2\xi_{\mu-1}^2}}\right)\right\}\nonumber
\end {eqnarray}

The next step is the sum over index $n_{\mu-2},$ inserting the above result to this sum we have:
\begin {eqnarray}
& &\lim_{\xi_{\mu-1}}\ \partial_{\xi_{\mu-1}}^4\
\left(\sqrt{t_{\mu-1}^2+t_{\mu}^2\xi_{\mu-1}^2}\right)^{4\mu-n_1\cdots -n_{\mu-3}}\
\sum_{n_{\mu-2}=0}^{4(\mu-2)-n_1\cdots -n_{\mu-3}}\ \binom{4(\mu-2)-n_1\cdots -n_{\mu-3}}{n_{\mu-2}}\nonumber\\
&\times& \frac{(4(\mu-1)-n_1\cdots -n_{\mu-2})!}{(4\mu-n_1\cdots -n_{\mu-2})!}\ \left(\frac{t_{\mu-2}}{\sqrt{t_{\mu-1}^2+t_{\mu}^2\xi_{\mu-1}^2}}\right)^{n_{\mu-2}}\ H_{n_{\mu-2}}(x_{\mu-2})\nonumber \\ &\times&H_{4\mu-n_1\cdots -n_{\mu-2}}\left(\frac{t_{\mu-1}x_{\mu-1}+(t_{\mu}x_{\mu}+1/2)\xi_{\mu-1}}{\sqrt{t_{\mu-1}^2+t_{\mu}^2\xi_{\mu-1}^2}}\right)
\label{f10}
\end {eqnarray}
Following by the same method as above we must:

a) extend summation over $n_{\mu-2}$ up to $4\mu-n_1\cdots -n_{\mu-1};$

b) to replace the binomial in above equation we use the identity: $$\binom{4(\mu-2)-n_1\cdots -n_{\mu-3}}{n_{\mu-2}}\ \frac{(4(\mu-1)-n_1\cdots -n_{\mu-2})!}{(4\mu-n_1\cdots -n_{\mu-2})!}\ = $$ $$=\ \frac{(4(\mu-2)-n_1\cdots -n_{\mu-3})!}{(4\mu-n_1\cdots -n_{\mu-3})!}\
\binom{4\mu-n_1\cdots -n_{\mu-3}}{n_{\mu-2}}\ \frac{(4(\mu-1)-n_1\cdots -n_{\mu-2})!}{(4(\mu-2)-n_1\cdots -n_{\mu-2})!}$$

c) to replace the division of factorials in $n_{\mu-2}$ by a power of dummy variable we utilize the identity: $$\frac{(4(\mu-1)-n_1\cdots -n_{\mu-2})!}{(4(\mu-2)-n_1\cdots -n_{\mu-2})!} = \lim_{\xi_{\mu-2}}\ \partial_{\xi_{\mu-2}}^4\ \left(\xi_{\mu-2}^{-4}\
\left(\xi_{\mu-2}\right)^{4\mu-n_1\cdots -n_{\mu-2}}\right)$$
Now we can use identity (\ref{sumH}) in the expression (\ref{f10}):
\begin {eqnarray}
& &\left\{\lim_{\xi_{\mu-1}}\ \partial_{\xi_{\mu-1}}^4\right\}\ \left\{\lim_{\xi_{\mu-2}}\ \partial_{\xi_{\mu-2}}^4\ \xi_{\mu-2}^{-4}\right\}\
\left(\sqrt{t_{\mu-1}^2\xi_{\mu-2}^2+t_{\mu}^2\xi_{\mu-1}^2\xi_{\mu-2}^2}\ \right)^{4\mu-n_1\cdots -n_{\mu-3}}\
\nonumber\\
&\times&\frac{(4(\mu-2)-n_1\cdots -n_{\mu-3})!}{(4\mu-n_1\cdots -n_{\mu-3})!}\ \sum_{n_{\mu-2}=0}^{4(\mu-2)-n_1\cdots -n_{\mu-3}}\
\left(\frac{t_{\mu-2}}{\sqrt{t_{\mu-1}^2\xi_{\mu-2}^2+t_{\mu}^2\xi_{\mu-1}^2\xi_{\mu-2}^2}\ }\right)^{n_{\mu-2}}\ \nonumber \\
&\times&H_{n_{\mu-2}}(x_{\mu-2})\ H_{4\mu-n_1\cdots -n_{\mu-2}}\left(\frac{t_{\mu-1}x_{\mu-1}+(t_{\mu}x_{\mu}+1/2)\xi_{\mu-1}}{\sqrt{t_{\mu-1}^2+t_{\mu}^2\xi_{\mu-1}^2}}\right)
\end {eqnarray}
with the final result for the second step of the evaluations:
\begin {eqnarray}
& &\left\{\lim_{\xi_{\mu-1}\rightarrow 1}\ \partial_{\xi_{\mu-1}}^4\right\}\ \left\{\lim_{\xi_{\mu-2}\rightarrow 1}\ \partial_{\xi_{\mu-2}}^4\ \xi_{\mu-2}^{-4}\right\}\
\left(\sqrt{t_{\mu-2}^2+t_{\mu-1}^2\xi_{\mu-2}^2+t_{\mu}^2\xi_{\mu-1}^2\xi_{\mu-2}^2}\ \right)^{4\mu-n_1\cdots -n_{\mu-3}}\\
&\times&\frac{(4(\mu-2)-n_1\cdots -n_{\mu-3})!}{(4\mu-n_1\cdots -n_{\mu-3})!}\ H_{4\mu-n_1\cdots -n_{\mu-3}}
\left(\frac{t_{\mu-2}x_{\mu-2}+t_{\mu-1}x_{\mu-1}\xi_{\mu-2}+(t_{\mu}x_{\mu}+1/2)\xi_{\mu-2}\xi_{\mu-1}}{\sqrt{t_{\mu-2}^2+t_{\mu-1}^2\xi_{\mu-2}^2+t_{\mu}^2\xi_{\mu-1}^2\xi_{\mu-2}^2}}\right)\nonumber
\end {eqnarray}

A vital change comparing to the preceding step is the presence of the factor $\xi_{\mu-2}^{-4}$ together with derivative $\partial_{\xi_{\mu-2}}^4.$


By the recurrence procedure we find for summations over indexes $n_i$ in Eq. (\ref{E64}):

\begin {eqnarray}
& &\frac{4!}{(4\mu)!}\ \left\{\lim_{\xi_{\mu-1}\rightarrow 1}\ \partial_{\xi_{\mu-1}}^4\right\}\  \cdots \left\{\lim_{\xi_{i}\rightarrow 1}\ \partial_{\xi_{i}}^4\
\xi_{i}^{-4(\mu-i-1)}\right\} \cdots
\left\{\lim_{\xi_{1}\rightarrow 1}\ \partial_{\xi_{1}}^4\ \xi_{1}^{-4(\mu-2)}\right\}\\
&\times&\
\left(\sqrt{t_1^2+\xi_1^2t_2^2+\cdots +\xi_1^2\xi_2^2\cdots \xi_{\mu-2}^2\xi_{\mu-1}^2t_{\mu}^2}\ \right)^{4\mu}\nonumber \\
&\times& H_{4\mu}
\left(\frac{t_1x_1+\xi_1t_2x_2+\cdots +
\xi_1\cdots \xi_{\mu-2}\xi_{\mu-1}(t_{\mu}x_{\mu}+1/2)}{\sqrt{t_1^2+\xi_1^2t_2^2+\cdots +\xi_1^2\xi_2^2\cdots \xi_{\mu-2}^2\xi_{\mu-1}^2t_{\mu}^2}}\right)\nonumber
\end {eqnarray}

Taking into account that $t_i=\left(\frac{-\sqrt{-D_i}}{\mathcal{X}}\right)$ and $x_i=\sqrt{-D_i},$ we insert the above result to Eq. (\ref{Ndim6}) and we obtain for the leading part of the $N-1$ dimensional integral:
\begin {eqnarray}
& &\mathcal{W}_{N}^{leading} =\ \frac{1}{\sqrt{\frac{2\pi\triangle}{c_0}\ \prod_{i=1}^{N-1}(\frac{c_i+c_{i+1}}{c_i}+2 \frac{b_i}{c_i}\triangle^2)\Omega_{i-1}}}\
\sum_{\mu=0}\ \frac{4!}{(4\mu)!}\ (-1)^{\mu} \left(\frac{\triangle\ \varphi_N^4}{\mathcal{X}^4\ Q_{N-1}^4}\right)^{\mu}\label{Ndim7} \\
&\times&
\sum_{p_1=1}^{N-\mu}\ a_{p_1}Q_{p_1}^4\ \sum_{p_2=p_1+1}^{N-\mu+1}\ a_{p_2}Q_{p_2}^4\ .....
\sum_{p_{\mu}=p_{\mu-1}+1}^{N}\ a_{p_{\mu}}Q_{p_{\mu}}^4\nonumber \\
&\times&
\left\{\lim_{\xi_{\mu-1}\rightarrow 1}\ \partial_{\xi_{\mu-1}}^4\right\}\  \cdots \left\{\lim_{\xi_{i}\rightarrow 1}\ \partial_{\xi_{i}}^4\ \xi_{i}^{-4(\mu-i-1)}\right\}
\cdots
\left\{\lim_{\xi_{1}\rightarrow 1}\ \partial_{\xi_{1}}^4\ \xi_{1}^{-4(\mu-2)}\right\}\nonumber \\
&\times&\
\left(\sqrt{-D_1-\xi_1^2 D_2-\cdots -\xi_1^2\xi_2^2.\cdots. \xi_{\mu-2}^2\xi_{\mu-1}^2 D_{\mu}\ }\right)^{4\mu}\nonumber \\
&\times& H_{4\mu}
\left(\frac{D_1+\xi_1 D_2+\cdots +
\xi_1\cdots \xi_{\mu-2}\xi_{\mu-1}(D_{\mu}+\mathcal{X}/2)}{\sqrt{-D_1-\xi_1^2 D_2-\cdots -\xi_1^2\xi_2^2.\cdots. \xi_{\mu-2}^2\xi_{\mu-1}^2 D_{\mu}}}\right)
\nonumber \\
&\times& \exp{(\mathcal{Y}\ -\frac{c_1}{2\triangle}\varphi_0^2\ +\ \mathcal{X})}\
\exp{\left\{-a_N\triangle\varphi_N^4 -
b_N\triangle\varphi_N^2-\left(\frac{c_N}{2}\ \frac{Q_{N-1}-Q_{N-2}}{\triangle Q_{N-1}}\right)\ \varphi_N^2
\right\}}\ .\nonumber
\end {eqnarray}
By definition the values $D_i$ implicitly depends on two consecutive summation indices $p_{i-1}, p_{i}.\ $ Explicitly,we can see:
$$D_1 = d_{p_1}+\cdots+d_{p_2-1},\ D_2 = d_{p_2}+\cdots+d_{p_3-1},\ \cdots,\ D_{\mu} = d_{p_{\mu}}+\cdots+d_{N-2}$$
To proceed, we can apply on the Eq. (\ref{Ndim7}) the methods of the generating functions of the Hermite polynomials.

\section{Continuum limit of Eq. (\ref{Ndim8})}

We are going to apply on Eqs. (\ref{Ndim8}), (\ref{Ndim7}) the limit $N\rightarrow \infty.$ We define the continuum variable $\tau$ for $i-th$ slicing point by the prescription:
$$\triangle = \beta / N,\, \tau = \frac{i\,\beta}{N}.$$
In the definitions of the variables $d_i,$ $\mathcal{X}$, we can find the term $\frac{Q_0Q_1}{2\triangle^2}$
Following definitions of the variables $\Omega_i,\ \psi_i, Q_i$ and theirs values for indices $i= 0,\ 1$:

$$\Omega_0 = 1, \ \Omega_0 = \frac{Q_1\psi_1}{Q_0},\ \psi_1 = \frac{1}{2} + \mathcal{O}(\triangle),\ Q_1 = Q_0 + \dot{Q_0}\triangle + \mathcal{O}(\triangle^2),$$

we find:

$$\frac{Q_0Q_1}{2\triangle^2} = \frac{Q_0^2 + Q_0\dot{Q_0}\triangle}{2\triangle^2}.$$

The result of this equation is nonzero and finite in continuum limit, if $Q_0 = \triangle$.
Rigorously speaking, the general solution is $\tilde{Q}_0 = c \triangle$, where $c$ is non-zero constant. We can prove by the simple algebra that by the variable transformation $\tilde{Q}_i = c Q_i$ the results based on the solution
$\tilde{Q}_0 = c \triangle$ are the same as we start from $Q_0 = \triangle$.

From equation $\Omega_0 = 1$ follows that $\dot{Q_0} = 1$ and  the continuum limit reads:
$$
\lim_{\triangle\rightarrow 0}\ \frac{Q_0Q_1}{2\triangle^2}\ \rightarrow\ 1\ .
$$
Then, we can find that:

1. The summations over $p_i$ indexes of the ordered product of $\mu$ terms in Eq. (\ref{Ndim7}) will be converted to multiple integrals:

\begin {eqnarray}
& &\lim_{N\rightarrow \infty}\triangle^{\mu}\left(\sum_{p_1=1}^{N-\mu}\ a_{p_1}Q_{p_1}^4\ \sum_{p_2=p_1+1}^{N-\mu+1}\ a_{p_2}Q_{p_2}^4\ .....
\sum_{p_{\mu}=p_{\mu-1}+1}^{N}\ a_{p_{\mu}}Q_{p_{\mu}}^4\right)\ \rightarrow\ \nonumber \\
& &\int_0^{\beta}d\tau_1\ a(\tau_1)Q(\tau_1)^4\ \int_{\tau_1}^{\beta}d\tau_2\ a(\tau_2)Q(\tau_2)^4\ \cdots\
\int_{\tau_{\mu-1}}^{\beta}d\tau_{\mu}\ a(\tau_{\mu})Q(\tau_{\mu})^4\
\end {eqnarray}

2. We can replace by continuum regularized expression the value:

\begin {eqnarray}
& &\mathcal{Y}\ -\frac{c_1}{2\triangle}\varphi_0^2\ =\\
&=& d_{N-2}+d_{N-}+\cdots+d_2+d_1 - d_0
\rightarrow\ \lim_{\epsilon\rightarrow 0}\left\{\int_{\epsilon}^{\beta}\frac{1}{c(\tau)Q^2(\tau)}-
\frac{\epsilon}{c(\epsilon)Q^2(\epsilon)}\right\}\frac{c^2(0)\varphi(0)^2}{2}\ ,
\nonumber
\end {eqnarray}
because by the definition

$$\frac{c_1}{2\triangle}\varphi_0^2\ =\ 2d_0\ .$$

3. Following definitions of the variables $\mathcal{X},\ Q_i$  we have:

$$\mathcal{X}\rightarrow\ \frac{c(0)\varphi(0)\varphi(\beta)}{Q(\beta)}\ ,$$
and in the exponential term we have:
$$\left(\frac{c_N}{2}\ \frac{Q_{N-1}-Q_{N-2}}{\triangle Q_{N-1}}\right)\ \varphi_N^2
\rightarrow\ \frac{\dot{Q}(\beta)}{Q(\beta)}\frac{c(\beta)\varphi(\beta)^2}{2}\ .$$
The square-root argument is treated as a function $f(\beta)$:
$$\frac{2\pi\triangle}{c_0}\ \prod_{i=1}^{N-1}\left(\frac{c_i+c_{i+1}}{c_i}+2 \frac{b_i}{c_i}\triangle^2\right)\Omega_{i-1}\ \rightarrow f(\beta)\ ,$$
where the functions $f(\tau)$ and $Q(\tau)$ are defined in the Appendix H.

4. In the continuum limit, following definition of the values $d_i$ and $D_i,$ we have:

$$D_i = \sum_{j=p_{i}}^{p_{(i+1)}-1}\ d_j\rightarrow\ \frac{c(0)^2\varphi(0)^2}{2}\ \int_{\tau_{i}}^{\tau_{i+1}}\  \frac{d\tau}{c(\tau)Q(\tau)^2} = \frac{c(0)^2\varphi(0)^2}{2}\ \left(\mathcal{I}(\tau_{i}) - \mathcal{I}(\tau_{i+1})\right)\ ,$$

where

$$\mathcal{I}(\tau_i) = \int_{\tau_i}^{\beta}\  \frac{d\tau}{c(\tau)Q(\tau)^2}, \ \ \ \ \tau_{0} = 0,\ \ \tau_{\mu+1}=\beta.$$
Term $\frac{c(0)^2\varphi(0)^2}{2}$ is the same for all $D_i$, we take it out of square-root function in Eq. (\ref{Ndim7}) and we find the identity, which significantly simplify the an-harmonic correction of the propagator:
$$\left(\frac{c(0)\varphi(0)}{\sqrt{2}}\right)^{4\mu}\
\left(\frac{\triangle\ \varphi_N^4}{\mathcal{X}^4\ Q_{N-1}^4}\right)^{\mu}\rightarrow
\left(\frac{1}{4}\right)^{\mu}\ .$$

Finally, in continuum limit the Eqs. (\ref{Ndim8}), (\ref{Ndim7}) can be read:
\begin {equation}
\mathcal{W}_{\beta}^{leading} =\ \frac{\mathcal{W}_{\beta}^{harm}}{\sqrt{f(\beta)}}\
\sum_{\mu=0}\ \frac{4!}{(4\mu)!}\ (-1)^{\mu} \left(\frac{1}{4}\right)^{\mu}\ \mathcal{W}(\mu),\label{Ndimfin}
\end {equation}
where
\begin {equation}
\mathcal{W}_{\beta}^{harm} =
\exp{\left(\lim_{\epsilon\rightarrow 0}\left\{\int_{\epsilon}^{\beta}d\tau\frac{1}{c(\tau)Q^2(\tau)}-
\frac{\epsilon}{c(\epsilon)Q^2(\epsilon)}\right\}\frac{c^2(0)\varphi(0)^2}{2}\ \ +\ \frac{c(0)\varphi(0)\varphi(\beta)}{Q(\beta)}
-\frac{\dot{Q}(\beta)}{Q(\beta)}\frac{c(\beta)\varphi(\beta)^2}{2}
\right)\ ,}
\end {equation}
function $f(\beta)$ is the solution of the differential equation defined in the next Appendix H.

The an-harmonic part of the propagator in the continuum limit can be read:
\begin {eqnarray}
& &\mathcal{W}(\mu)=
\int_0^{\beta}d\tau_1\ a(\tau_1)Q(\tau_1)^4\ \int_{\tau_1}^{\beta}d\tau_2\ a(\tau_2)Q(\tau_2)^4\ \cdots\
\int_{\tau_{\mu-1}}^{\beta}d\tau_{\mu}\ a(\tau_{\mu})Q(\tau_{\mu})^4 \label{contAH} \\
&\times&
\left\{\lim_{\xi_{\mu-1}\rightarrow 1}\ \partial_{\xi_{\mu-1}}^4\right\}\  \cdots \left\{\lim_{\xi_{i}\rightarrow 1}\ \partial_{\xi_{i}}^4\ \xi_{i}^{-4(\mu-i-1)}\right\}
\cdots
\left\{\lim_{\xi_{1}\rightarrow 1}\ \partial_{\xi_{1}}^4\ \xi_{1}^{-4(\mu-2)}\right\}\nonumber \\
&\times&\
\left(\sqrt{-\mathcal{I}(\tau_1)-(\xi_1^2-1)\mathcal{I}(\tau_2)-\cdots -\xi_1^2\xi_2^2 \cdot \cdots \cdot \xi_{\mu-2}^2(\xi_{\mu-1}^2-1)\mathcal{I}(\tau_{\mu})\
}\right)^{4\mu}\nonumber \\
&\times& H_{4\mu}
\left(\frac{\frac{c(0)\varphi(0)}{\sqrt{2}}[\mathcal{I}(\tau_1)+(\xi_1-1)\mathcal{I}(\tau_2)+ \cdots +\xi_1\xi_2 \cdots
\xi_{\mu-2}(\xi_{\mu-1}-1)\mathcal{I}(\tau_{\mu})]+\xi_1 \cdots \xi_{\mu-1}\left(\frac{\varphi(\beta)}{\sqrt{2}Q(\beta)}\right)}
{\sqrt{-\mathcal{I}(\tau_1)-(\xi_1^2-1)\mathcal{I}(\tau_2)-\cdots -\xi_1^2\xi_2^2 \cdot \cdots \xi_{\mu-2}^2(\xi_{\mu-1}^2-1)\mathcal{I}(\tau_{\mu})\ }}\right).
\nonumber
\end {eqnarray}

\section{Evaluation of the differential equations}

Let us evaluate two relevant recurrence equations that characterize the resultant formula for the continuum limit of the $N-1$ dimensional integral.
 The first is connected with evaluations of the denominator in Eqs. (\ref{Ndim}), (\ref{Ndim7}):
 $$\frac{1}{\sqrt{\frac{2\pi\triangle}{c_0}\ \prod_{i=1}^{N-1}(\frac{c_i+c_{i+1}}{c_i}+2 \frac{b_i}{c_i}\triangle^2)\Omega_{i-1}}}$$
For harmonic oscillator the same evaluations were proposed by Gel'fand and Yaglom \cite{gelY}. We define the function:
\begin {equation}
f(n) = \frac{2\pi\triangle}{c_0}\ \prod_{i=1}^{n}\left(\frac{c_i+c_{i+1}}{c_i}+2 \frac{b_i}{c_i}\triangle^2\right)\Omega_{i-1}.
\end {equation}
By recurrence relation (\ref{omegai})

$$\Omega_i = 1-\frac{\Sigma_i}{\Omega_{i-1}},\ \Omega_0 = 1,$$

we can show that
\begin {eqnarray}
& &f(n+1) = \left(\frac{c_{n+1}+c_{n+2}}{c_{n+1}}+2 \frac{b_{n+1}}{c_{n+1}}\triangle^2\right)\Omega_{n}\ f(n)\ =\\ &=&\left(\frac{c_{n+1}+c_{n+2}}{c_{n+1}}+2 \frac{b_{n+1}}{c_{n+1}}\triangle^2\right)\ f(n)\ -\
\Sigma_n\ \left(\frac{c_{n+1}+c_{n+2}}{c_{n+1}}+2 \frac{b_{n+1}}{c_{n+1}}\triangle^2\right)\left(\frac{c_{n}+c_{n+1}}{c_n}+2 \frac{b_{n}}{c_{n}}\triangle^2\right)\
f(n-1)\nonumber
\end {eqnarray}
By definitions of $\Sigma_i,$ Eqs. (\ref{Sigma}), (\ref{newsigma}) we find the equation:

$$f(n+1) = \left(\frac{c_{n+1}+c_{n+2}}{c_{n+1}}+2 \frac{b_{n+1}}{c_{n+1}}\triangle^2\right)\ f(n)\ -\ \frac{c_{n+1}}{c_n}\ f(n-1),$$

which can be expressed as:

$$c_{n}c_{n+1}\left(f(n+1)-f(n)\right) = c_{n}c_{n+2}f(n) - c_{n+1}^2f(n-1) + 2b_{n+1}c_{n}\triangle^2f(n).$$

Now we express $c_{n}$ and $c_{n+2}$ by help of $c_{n+1}$ up to second order in difference $\triangle$ $$c_{n} = c_{n+1} - \dot{c}_{n+1}\triangle + \ddot{c}_{n+1}\frac{\triangle^2}{2!}$$ and $$c_{n+2} = c_{n+1} + \dot{c}_{n+1}\triangle + \ddot{c}_{n+1}\frac{\triangle^2}{2!},$$ where $\dot{c}_{n+1}$ and $\ddot{c}_{n+1}$ are first and second derivatives of the function $c(\tau)$ in the point $\tau = (n+1) \triangle.$

Taking into account terms up to $\triangle^2$ we obtain the difference equation: $$f(n+1)-2f(n)+f(n-1)\ -\ \frac{\dot{c}_{n+1}}{c_{n+1}}\left(f(n+1)-f(n)\right)\triangle =
\left(2\frac{b_{n+1}}{c_{n+1}} + \frac{\ddot{c}_{n+1}}{c_{n+1}} - \left(\frac{\dot{c}_{n+1}}{c_{n+1}}\right)^2\right)\ f(n)\triangle^2$$
with initial conditions: $$f(1) = \frac{4\pi\triangle}{c_0},\ f(2) = \frac{6\pi\triangle}{c_0}.$$ In continuum limit $\triangle \rightarrow 0,$ we will have the differential equation:
\begin {equation}
\ddot{f}(\tau) - \partial_{\tau}\ln{c(\tau)}\ \dot{f}(\tau) - \left(2\frac{b(\tau)}{c(\tau)} + \partial_{\tau}^2\ln{c(\tau)}\right)f(\tau)\ =\ 0,
\label{gy}
\end {equation}
with initial conditions up to first order in $\triangle$:
\begin {equation}
f(0) = 0,\ \dot{f}(0) = \frac{2\pi}{c(0)}.
\end {equation}
Eq. (\ref{gy}) can be converted to normal form by substitution: $$f(\tau) = \sqrt{c(\tau)}\ w(\tau).$$ The equation for function $w(\tau)$ can be read:
\begin {equation}
\ddot{w}(\tau) - \left(2\frac{b(\tau)}{c(\tau)} + \frac{1}{2}\ \partial_{\tau}^2\ln{c(\tau)} +
\frac{1}{4}\ (\partial_{\tau}\ln{c(\tau)})^2\right)w(\tau)\ =\ 0.
\label{nfgy}
\end {equation}

Another recurrence relation behind a difference equation is Eq. (\ref{omegai})
$$\Omega_i = 1-\frac{\Sigma_i}{\Omega_{i-1}},\ \Omega_0 = 1,\ \Omega_1 = 1-\Sigma_1.$$
We defined the new function $Q_i$ connected with $\Omega_i$ by prescription (\ref{qq}):
$$\Omega_i = \frac{\psi_{i+1} Q_{i+1}}{Q_i},\ \psi_i = \frac{c_{i+1}}{c_{i+1}+c_{i} + 2b_i\triangle^2}.$$
Taking into account the definition of $\Sigma_i$ as (\ref{newsigma}): $$\Sigma_i = \frac{c_{i+1}}{c_{i+2}}\psi_{i}\psi_{i+1},$$

we convert the recurrence equation for $\Omega_i$ to difference equation for $Q:$ $$c_{i+2}Q_{i+1} = (c_{i+2}+c_{i+1}+2b_{i+1}\triangle^2)Q_i - c_{i+1}Q_{i-1}$$

We can use the decomposition: $$c_{i+2} = c_{i+1} + \dot{c}_{i+1}\triangle + \ddot{c}_{i+1}\frac{\triangle^2}{2!} $$

to obtain the difference equation in the form:
\begin {equation}
Q_{i+1}-2Q_i+Q_{i-1} + \frac{\dot{c}_{i+1}}{c_{i+1}}(Q_{i+1}-Q_i)\ \triangle - 2\frac{b_{i+1}}{c_{i+1}}Q_i\ \triangle^2 = 0,
\label{difQ}
\end {equation}
with initial conditions $$Q_0 = \triangle, \ \dot{Q}_0 = 1,$$

derived from values
$$\Omega_0 = 1, \ \Omega_0 = \frac{Q_1\psi_1}{Q_0},\ \psi_1 = \frac{1}{2} + \mathcal{O}(\triangle).$$
For the reasonable behaviour of the function $c(\tau)$ the equation

$$
1 = \Omega_0 = \frac{\psi_1(Q_0+\dot{Q_0}\triangle)}{Q_0}
$$

possess the solutio $Q_0=\triangle$ and $\dot{Q_0}=1\ .$

In the continuum limit the difference equation (\ref{difQ}) Can be converted to the differential equation for the function $Q(\tau)$:
\begin {equation}
\ddot{Q}(\tau) + \partial_{\tau}\ln{c(\tau)}\ \dot{Q}(\tau) - 2\frac{b(\tau)}{c(\tau)}\ Q(\tau) = 0,
\label{deq}
\end {equation}
with the initial conditions: $$Q(0) = 0, \ \dot{Q}(0) = 1.$$ By substitution $$Q(\tau) = \frac{w(\tau)}{\sqrt{c(\tau)}}$$ we can convert Eq. (\ref{deq}) to a normal differential equation for the function $w(\tau)$ also:
\begin {equation}
\ddot{w}(\tau) - \left(2\frac{b(\tau)}{c(\tau)} + \frac{1}{2}\ \partial_{\tau}^2\ln{c(\tau)} +
\frac{1}{4}\ (\partial_{\tau}\ln{c(\tau)})^2\right)\ w(\tau) = 0,
\label{deq2}
\end {equation}
with initial conditions: $$w(0) = 0, \ \dot{w}(0) = \sqrt{c(0)}.$$

\section{Factorization of the multiple integrals}

We follow with the evaluation of the an-harmonic part of the propagator in the continuum limit.
The Hermite polynomial and the square-root in the Eqs. (\ref{her1hlav}), (\ref{contAH}) possesses the non-trivial dependence on the integration variables of the multiple integral.
We are going to explain how it is possible to factorize these functions.
To factorize the multiple integrals in Eqs. (\ref{her1hlav}), (\ref{contAH}), we can utilize the method of the generating function for Hermite polynomials. Our aim is to evaluate the expression:
\begin {eqnarray}
& &I_{4\nu}(\mu, \beta)= \int_0^{\beta}d\tau_1\ a(\tau_1)Q(\tau_1)^4\ \int_{\tau_1}^{\beta}d\tau_2\ a(\tau_2)Q(\tau_2)^4\ \cdots\
\int_{\tau_{\mu-1}}^{\beta}d\tau_{\mu}\ a(\tau_{\mu})Q(\tau_{\mu})^4 \label{her1}\\
&\times&
\left\{\lim_{\xi_{\mu-1}\rightarrow 1}\ \partial_{\xi_{\mu-1}}^4\right\}\  \cdots \left\{\lim_{\xi_{i}\rightarrow 1}\ \partial_{\xi_{i}}^4\ \xi_{i}^{-4(\mu-i-1)}\right\}
\cdots
\left\{\lim_{\xi_{1}\rightarrow 1}\ \partial_{\xi_{1}}^4\ \xi_{1}^{-4(\mu-2)}\right\}\nonumber \\
&\times&\
\left(\sqrt{-\mathcal{I}(\tau_1)-(\xi_1^2-1)\mathcal{I}(\tau_2)-\cdots -\xi_1^2\xi_2^2 \cdot \cdots \cdot \xi_{\mu-2}^2(\xi_{\mu-1}^2-1)\mathcal{I}(\tau_{\mu})
}\right)^{4\nu}\nonumber \\
&\times& H_{4\nu}
\left(\frac{\frac{c(0)\varphi(0)}{\sqrt{2}}[\mathcal{I}(\tau_1)+(\xi_1-1)\mathcal{I}(\tau_2)+ \cdots +\xi_1\xi_2 \cdots
\xi_{\mu-2}(\xi_{\mu-1}-1)\mathcal{I}(\tau_{\mu})]+\xi_1\cdots \xi_{\mu-1}\left(\frac{\varphi(\beta)}{\sqrt{2}Q(\beta)} \right)}
{\sqrt{-\mathcal{I}(\tau_1)-(\xi_1^2-1)\mathcal{I}(\tau_2)-\cdots -\xi_1^2\xi_2^2 \cdot \cdots \cdot \xi_{\mu-2}^2(\xi_{\mu-1}^2-1)\mathcal{I}(\tau_{\mu}) }}\right)
\nonumber
\end {eqnarray}

Multiplying both sides of Eq. (\ref{her1}) by $t^{4\nu}/(4\nu)!\ ,$ summing up and by identity: $$t^n + (-t)^n + (it)^n + (-it)^n = 4t^n\ \delta_{n, 4\nu}$$

we read:

$$\sum_{\nu=0}^{\infty}\ \frac{t^{4\nu}}{(4\nu)!}\ I_{4\nu}(\mu, \beta) = \left\{
\sum_{n=0}^{\infty}\ \frac{t^n}{n!}\ I_{n}(\mu, \beta) +
\sum_{n=0}^{\infty}\ \frac{(-t)^n}{n!}\ I_{n}(\mu, \beta) +
\sum_{n=0}^{\infty}\ \frac{(it)^n}{n!}\ I_{n}(\mu, \beta) +
\sum_{n=0}^{\infty}\ \frac{(-it)^n}{n!}\ I_{n}(\mu, \beta)\right\}/4\ .$$
We show that we are able to write the generating function for $I_{4\nu}(\mu, \beta)$ as the sum of the simpler generating functions for the Hermite polynomials.
Let us go to evaluate these generating functions. We define for the arbitrary sum in above equation:
\begin {eqnarray}
& &\sum_{n=0}^{\infty}\ \frac{z^n}{n!}\ I_{n}(\mu, \beta) = \int_0^{\beta}d\tau_1\ a(\tau_1)Q(\tau_1)^4\ \int_{\tau_1}^{\beta}d\tau_2\ a(\tau_2)Q(\tau_2)^4\ \cdots\
\int_{\tau_{\mu-1}}^{\beta}d\tau_{\mu}\ a(\tau_{\mu})Q(\tau_{\mu})^4 \label{her2}\\
&\times&
\left\{\lim_{\xi_{\mu-1}\rightarrow 1}\ \partial_{\xi_{\mu-1}}^4\right\}\  \cdots \left\{\lim_{\xi_{i}\rightarrow 1}\ \partial_{\xi_{i}}^4\ \xi_{i}^{-4(\mu-i-1)}\right\}
\cdots
\left\{\lim_{\xi_{1}\rightarrow 1}\ \partial_{\xi_{1}}^4\ \xi_{1}^{-4(\mu-2)}\right\}\nonumber \\
& &\nonumber\\
&\times&\ \sum_{n=0}^{\infty}\ \frac{1}{n!}
\left(z\
\sqrt{-\mathcal{I}(\tau_1)-(\xi_1^2-1)\mathcal{I}(\tau_2)-\cdots -\xi_1^2\xi_2^2 \cdot \cdots \cdot \xi_{\mu-2}^2(\xi_{\mu-1}^2-1)\mathcal{I}(\tau_{\mu-1})}
\right)^n\nonumber \\
&\times& H_n
\left(\frac{\frac{c(0)\varphi(0)}{\sqrt{2}}[\mathcal{I}(\tau_1)+(\xi_1-1)\mathcal{I}(\tau_2)+ \cdots +\xi_1\xi_2\cdots
\xi_{\mu-2}(\xi_{\mu-1}-1)\mathcal{I}(\tau_{\mu})]+\xi_1 \cdot \cdots \cdot \xi_{\mu-1}\left(\frac{\varphi(\beta)}{\sqrt{2}Q(\beta)}\right)}
{\sqrt{-\mathcal{I}(\tau_1)-(\xi_1^2-1)\mathcal{I}(\tau_2)-\cdots -\xi_1^2\xi_2^2 \cdot \cdots \cdot \xi_{\mu-2}^2(\xi_{\mu-1}^2-1)\mathcal{I}(\tau_{\mu})\
 }}\right)
\nonumber
\end {eqnarray}

Following the definition of the generating function for Hermite polynomials, namely\cite{bateman}

$$\sum_{n=0}^{\infty}\ \frac{u^n}{n!}\ H_n(x) = \exp{(2ux-u^2)},$$

for the summation part of Eq. (\ref{her2}) we read:

\begin {eqnarray}
& &\sum_{n=0}^{\infty}\ \frac{1}{n!}
\left(z\
\sqrt{-\mathcal{I}(\tau_1)-(\xi_1^2-1)\mathcal{I}(\tau_2)-\cdots -\xi_1^2\xi_2^2 \cdot \cdots \cdot \xi_{\mu-2}^2(\xi_{\mu-1}^2-1)\mathcal{I}(\tau_{\mu})
}
\right)^n\nonumber \\
&\times& H_n
\left(\frac{\frac{c(0)\varphi(0)}{\sqrt{2}}[\mathcal{I}(\tau_1)+(\xi_1-1)\mathcal{I}(\tau_2)+ \cdots +\xi_1\xi_2\cdots
\xi_{\mu-2}(\xi_{\mu-1}-1)\mathcal{I}(\tau_{\mu})]+\xi_1\cdots \xi_{\mu-1}\left(\frac{\varphi(\beta)}{\sqrt{2}Q(\beta)}\right)}
{\sqrt{-\mathcal{I}(\tau_1)-(\xi_1^2-1)\mathcal{I}(\tau_2)-\cdots -\xi_1^2\xi_2^2\cdots \xi_{\mu-2}^2(\xi_{\mu-1}^2-1)\mathcal{I}(\tau_{\mu})\
 }}\right) =
\nonumber \\
\nonumber \\
&=&\exp{}\left(2z\ \frac{c(0)\varphi(0)}{\sqrt{2}}\left[\mathcal{I}(\tau_1)+(\xi_1-1)\mathcal{I}(\tau_2)+ \cdots +\xi_1\cdots\xi_{\mu-2}(\xi_{\mu-1}-1)\mathcal{I}(\tau_{\mu})\right]+2z \xi_1\cdots \xi_{\mu-1}\left(\frac{\varphi(\beta)}{\sqrt{2}Q(\beta)}\right)-\right.\nonumber \\
&-&z^2 \left. [-\mathcal{I}(\tau_1)-(\xi_1^2-1)\mathcal{I}(\tau_2)-\cdots -\xi_1^2\xi_2^2 \cdot \cdots \cdot \xi_{\mu-2}^2(\xi_{\mu-1}^2-1)\mathcal{I}(\tau_{\mu})\
]\right). \label{her3}
\end {eqnarray}
This result enable the factorization of the exponent following the integral variables.
Inserting the identity:
$$2z \xi_1 \cdot \cdots \cdot \xi_{\mu-1} =
2z \xi_1 \cdot \cdots \cdot \xi_{\mu-2}(\xi_{\mu-1}-1)
+ 2z \xi_1 \cdot \cdots \cdot \xi_{\mu-3}(\xi_{\mu-2}-1) + \cdots +
2z (\xi_1-1) + 2z$$
to Eq. (\ref{her3}),
we can read:
\begin {eqnarray}
& &\sum_{n=0}^{\infty}\ \frac{z^n}{n!}\ I_{n}(\mu, \beta) = \label{her4}\\
&\times&\int_0^{\beta}d\tau_1\ a(\tau_1)Q(\tau_1)^4\ \exp{\left\{2z\left(\mathcal{I}(\tau_1)\frac{c(0)\varphi(0)}{\sqrt{2}}+\frac{\varphi(\beta)}{\sqrt{2}Q(\beta)}\right)
 + (2z)^2\frac{\mathcal{I}(\tau_1)}{4}\right\}} \nonumber \\
&\times&
\left\{\lim_{\xi_{\mu-1}\rightarrow 1}\ \partial_{\xi_{\mu-1}}^4\right\}\  \cdots \left\{\lim_{\xi_{i}\rightarrow 1}\ \partial_{\xi_{i}}^4\ \xi_{i}^{-4(\mu-i-1)}\right\}
\cdots
\left\{\lim_{\xi_{1}\rightarrow 1}\ \partial_{\xi_{1}}^4\ \xi_{1}^{-4(\mu-2)}\right\}\nonumber \\
&\times&\int_{\tau_1}^{\beta}d\tau_2\ a(\tau_2)Q(\tau_2)^4\ \exp{\left\{2z(\xi_1-1)\left(\mathcal{I}(\tau_2)\frac{c(0)\varphi(0)}{\sqrt{2}}+\frac{\varphi(\beta)}{\sqrt{2}Q(\beta)}\right)
 + (2z)^2(\xi_1^2-1)\frac{\mathcal{I}(\tau_2)}{4}\right\}} \nonumber \\
& &\vdots \nonumber \\
&\times&\int_{\tau_{\mu-1}}^{\beta}d\tau_{\mu}\ a(\tau_{\mu})Q(\tau_{\mu})^4\
\exp{\left\{2z\xi_1\cdot \cdots\cdot \xi_{\mu-2}(\xi_{\mu-1}-1)\left(\mathcal{I}(\tau_{\mu})\frac{c(0)\varphi(0)}{\sqrt{2}}+\frac{\varphi(\beta)}{\sqrt{2}Q(\beta)}\right)
+\right. }\nonumber \\
& &\left. +(2z)^2\xi_1^2\xi_2^2 \cdot \cdots \cdot\xi_{\mu-2}^2(\xi_{\mu-1}^2-1)\frac{\mathcal{I}(\tau_{\mu})}{4}\right\} \nonumber
\end {eqnarray}
We achieved the factorization the functions following the integrations variables. Now, we perform the derivatives following the variables $\xi_i\ ,$ and then we evaluate from the generating function the functions $I_{n}(\mu, \beta)$.

\section{Evaluations of the derivatives II}

In this Appendix we will use another definition of the Hermite polynomials, as we used following Bateman \cite{bateman}. We will use multi-variable, multi-index Hermite polynomials and we will spend a some time to show why and how they appeared in our evaluations.

As the first step, we find the derivative variable $\xi_{\mu-1}$ in one integrand in Eq. (\ref{her4}) only. We start with the derivative
$$\lim_{\xi_{\mu-1}\rightarrow 1}\ \partial_{\xi_{\mu-1}}^4 $$
of the object
$$\exp{\left\{2z\sigma(\mu-2)(\xi_{\mu-1}-1)\left(\mathcal{I}(\tau_{\mu})\frac{c(0)\varphi(0)}{\sqrt{2}}+\frac{\varphi(\beta)}{\sqrt{2}Q(\beta)}\right)
+(2z)^2\sigma^2(\mu-2)(\xi_{\mu-1}^2-1)\frac{\mathcal{I}(\tau_{\mu})}{4}\right\}}$$
where we defined the function:
$$\sigma(i) = \xi_1\cdot\cdots\cdot\xi_i\ ,\,\, \sigma(0) = 1 .$$
We follow by the identity:
\begin {equation}
\partial_{\xi}^n \exp{(a\xi+b\xi^2)} = H_n(a+2b\xi,\ b)\ \exp{(a\xi+b\xi^2)},
\label{idJ1}
\end {equation}
where two variable Hermite polynomial $H_n(x,y)$ defined by the generating function \cite{appell}
\begin {equation}
\sum_{n=0}^{\infty}\ \frac{z^n}{n!}H_n(x,y)\ = \ e^{xz+yz^2}\label{h2v}
\end {equation}

is explicitly  done as:
\begin {equation}
H_n(x,y) = n!\ \sum_{k=0}^{\lfloor n/2 \rfloor}\ \frac{x^{n-2k}y^k}{(n-2k)!k!}\ .
\label{hn2}
\end {equation}

For the derivative in question we can show by direct evaluation:
\begin {eqnarray}
& &\lim_{\xi_{\mu-1}\rightarrow 1}\partial_{\xi_{\mu-1}}^4\left[
\exp{\left\{2z\sigma(\mu-2)(\xi_{\mu-1}-1)\left(\mathcal{I}(\tau_{\mu})\frac{c(0)\varphi(0)}{\sqrt{2}}+\frac{\varphi(\beta)}{\sqrt{2}Q(\beta)}\right)
+(2z)^2\sigma^2(\mu-2)(\xi_{\mu-1}^2-1)\frac{\mathcal{I}(\tau_{\mu})}{4}\right\}}
\right]=\nonumber \\
& & \nonumber \\
&=&(2z\sigma(\mu-2))^4\ H_4\left(\mathcal{I}(\tau_{\mu})\frac{c(0)\varphi(0)}{\sqrt{2}}+\frac{\varphi(\beta)}{\sqrt{2}Q(\beta)}+2(2z\sigma(\mu-2))
\frac{\mathcal{I}(\tau_{\mu})}{4},\ \frac{\mathcal{I}(\tau_{\mu})}{4}\right) .\label{her41}
\end {eqnarray}

Thanks to $\lim_{\xi_{\mu-1}\rightarrow 1}$, the exponential function disappeared, leaving us the Hermite polynomial for the following evaluations. The Hermite polynomial in (\ref{her41}) possesses in the argument the product of the all other derivative variables. Therefore in the next steps we will derivative the product of the Hermite polynomial and the exponential function.
An important feature of the result (\ref{her41}) is the power $\sigma^4(\mu-2),$ which simplifies the next derivatives, killing the negative powers of the derivative variables accompanying the derivative operators. Thus every next derivative will be simply as just described one. The multiplicative factor $z^4$ in final calculation significantly simplifies the evaluation of the corresponding Taylor's coefficients $I_{n}(\mu, \beta)$ of the generating function in $t$ variable.

The next step is the derivative $\lim_{\xi_{\mu-2}\rightarrow 1}\ \partial_{\xi_{\mu-2}}^4 \xi_{\mu-2}^{4(\mu-(\mu-2)-1)}$
on the product of the corresponding Hermite polynomial and the exponential function,
which looks, due to variable $\xi_{\mu-2}^4$ in $\sigma^4(\mu-2),$ as in the precedent derivative step. Let us stress, that in following evaluations we will derive the product of the exponential function and Hermite polynomial.
We find:
\begin {equation}
\partial^n_{\xi} \left(H_n(a+2b \xi,b)e^{(a_1 \xi+b_1 \xi^2)}\right) =  n!n!\ \sum_{i=0}^n\ \frac{(2b)^i}{i!(n-i)!(n-i)!}
H_{n-i}(a+2b \xi,b)\ H_{n-i}(a_1+2b_1 \xi,b_1)e^{(a_1 \xi+b_1 \xi^2)},
\label{idJ2}
\end {equation}
because
\begin {equation}
\partial^i_{\xi}\ H_n(a+2b\xi,b) = (2b)^i\ \frac{n!}{(n-i)!}\ H_{n-i}(a+2b\xi,b).\label{idJ3}
\end {equation}
Applying the $\lim_{\xi \rightarrow 0}$ on the Eq. (\ref{idJ2}), on the r.h.s. we recognize the four variable two index Hermite polynomial $H_{n,n}$ defined  \cite{dattoli} as:
$$H_{m,n}(x,y;z,w|\tau)=m!n!\sum_{s=0}^{min[m,n]}\ \frac{\tau^s\ H_{m-s}(x,y)H_{n-s}(z,w)}{s!(m-s)!(n-s)!}\ .$$

The generating function of this type Hermite polynomials can be read:

$$\sum_{m,n=0}^{\infty}\ \frac{u^m v^n}{m! n!} H_{m,n}(x,y;z,w|\tau)=
\exp{(xu+yu^2 + zv+wz^2 + \tau uv)}.$$

The next step is the evaluation of the derivative of the product an exponential function and
the four variable two index Hermite polynomial:

\begin {eqnarray}
& &\partial_{\xi}^n\left(H_{n,n}(a+2b\xi,b;\ a_1+2b_1\xi,b_1|\tau)e^{(a_2\xi+b_2\xi^2)}\right)=\\
&=&\sum_j^n \binom{n}{j}\ \partial_{\xi}^jH_{n,n}(a+2b\xi,b;\ a_1+2b_1\xi,b_1|\tau)\ H_{n-j}(a_2+2b_2\xi,b_2)\ e^{(a_2\xi+b_2\xi^2)}\ .\nonumber
\end {eqnarray}

The derivative of the function $H_{n,n}$ is the derivative of the product of two functions:

$$\partial_{\xi}^j\left[H_{n-i}(a+2b\xi,b)H_{n-i}(a_1+2b_1\xi,b_1)\right]=$$
$$=\sum_q^j\binom{j}{q}\frac{(n-i)!}{(n-i-j+q)!}\frac{(n-i)!}{(n-i-q)!}(2b)^{j-q}(2b_1)^q
H_{n-i-j+q}(a+2b\xi,b)H_{n-i-q}(a_1+2b_1\xi,b_1).$$
Applying the $\lim_{\xi \rightarrow 0}$
we can show by some algebra, that this result is $H_{n,n,n}$ defined by the generating function for multi-index and multi-variable Hermite polynomials \cite{dattoli}:

$$\sum_{n_1,n_2,n_3}^{\infty} \frac{u_1^{n_1}u_2^{n_2}u_3^{n_3}}{n_1!n_2!n_3!}
H_{n_1,n_2,n_3}(\{x_i\},\{y_i\}|\tau_{1,2}, \tau_{1,3}, \tau_{2,3})=
\exp{\left(\sum_{i=1}^3(x_iu_i+y_iu_i^2)+u_1\tau_{1,2}\,u_2+u_1\tau_{1,3}\,u_3+u_2\tau_{2,3}\,u_3\right)},$$
where $\tau_{1,2}=\tau_{1,3}=2b,$ and $\tau_{2,3}=2b_1.$

Now, proceeding by the recurrence procedure, we can write the result of the derivative operations in Eq.(\ref{her3hlav},\ \ref{her4}):
\begin {eqnarray}
& &\sum_{n=0}^{\infty}\ \frac{z^n}{n!}\ I_{n}(\mu, \beta) = \label{her5}\\
&=&(2z)^{4\mu-4}\int_0^{\beta}d\tau_1\ a(\tau_1)Q(\tau_1)^4\
\int_{\tau_1}^{\beta}d\tau_2\ a(\tau_2)Q(\tau_2)^4\
\cdots
\int_{\tau_{\mu}-1}^{\beta}d\tau_{\mu}\ a(\tau_{\mu})Q(\tau_{\mu})^4\nonumber \\
&\times& H_{4,4,\cdots,4}\left(\left\{\frac{\varphi(\beta)}{\sqrt{2}Q(\beta)} + \mathcal{I}(\tau_{i})\left(\frac{c(0)\varphi(0)}{\sqrt{2}}\right)+(2z)\frac{\mathcal{I}(\tau_{i})}{2} \right\}_{i=2}^{\mu}, \left\{ \frac{\mathcal{I}(\tau_{i})}{4}\right\}_{i=2}^{\mu} | \tau_{i,k}=\frac{\mathcal{I}(\tau_{i})}{2}\right)\nonumber\\
&\times&\exp{\left\{2z\left(\frac{\varphi(\beta)}{\sqrt{2}Q(\beta)}+\frac{c(0)\varphi(0)}{\sqrt{2}}\mathcal{I}(\tau_1)\right) + 4z^2\frac{\mathcal{I}(\tau_1)}{4} \right\}} \nonumber
\end {eqnarray}
On the r.h.s. of the above equation is the generating function of $I_{n}(\mu, \beta).$
The multi-indexes Hermite polynomial $H_{4,4,\cdots,4}$ possesses $\mu-1$ equal indexes. In principe, $H_{4,4,\cdots,4}$ is  a combination of the products of $\mu-1$ two variables Hermite polynomials $H_j$ (\ref{hn2}).

We are going to evaluate $I_{4\mu}(\mu, \beta)$ applying $\lim_{z\rightarrow 0}\partial_z^{4\mu}$ on the above equation. Due to the term $(2z)^{4\mu-4}$ the derivative of the product and the limit $z\rightarrow 0$ of two functions possesses one term only, we find that
\begin {eqnarray}
& & I_{4\mu}(\mu, \beta) = \label{her51}\\
&\times&\binom{4\mu}{4\mu-4}\ (4\mu-4)!\ 2^{4\mu-4}\int_0^{\beta}d\tau_1\ a(\tau_1)Q(\tau_1)^4\
\int_{\tau_1}^{\beta}d\tau_2\ a(\tau_2)Q(\tau_2)^4\
\cdots
\int_{\tau_{\mu}-1}^{\beta}d\tau_{\mu}\ a(\tau_{\mu})Q(\tau_{\mu})^4\nonumber \\
&\times&2^4\ \lim_{z\rightarrow 0}\partial^4_{2z}\ \left[ H_{4,4,\cdots,4}\left(\left\{\frac{\varphi(\beta)}{\sqrt{2}Q(\beta)} + \mathcal{I}(\tau_{i})\left(\frac{c(0)\varphi(0)}{\sqrt{2}}\right)+(2z)\frac{\mathcal{I}(\tau_{i})}{2} \right\}_{i=2}^{\mu}, \left\{ \frac{\mathcal{I}(\tau_{i})}{4}\right\}_{i=2}^{\mu} | \tau_{i,k}=\frac{\mathcal{I}(\tau_{i})}{2}\right)\right.\nonumber\\
&\times& \left. \exp{\left\{2z\left(\frac{\varphi(\beta)}{\sqrt{2}Q(\beta)}+\frac{c(0)\varphi(0)}{\sqrt{2}}\mathcal{I}(\tau_1)\right) + 4z^2\frac{\mathcal{I}(\tau_1)}{4} \right\}}\right] \nonumber
\end {eqnarray}
But the operation $\lim_{z\rightarrow 0}\partial^4_{2z}$ on the product of the exponential function and multi-index Hermite polynomial is the same operation as we studied upper, and the result is the Hermite polynomial with $\mu$ equal indexes, $4\mu$ main variables and $\mu(\mu+1)(\mu+2)/3$ variables $\tau_{i,k}$.

We obtain for an-harmonic part of propagator the result:
\begin {eqnarray}
& &\mathcal{W}(\mu) =  I_{4\mu}(\mu, \beta) = \label{her61}\\
& & \nonumber \\
&=&\frac{(4\mu)!}{4!}\ 2^{4\mu}\int_0^{\beta}d\tau_1\ a(\tau_1)Q(\tau_1)^4\
\int_{\tau_1}^{\beta}d\tau_2\ a(\tau_2)Q(\tau_2)^4\
\cdots
\int_{\tau_{\mu}-1}^{\beta}d\tau_{\mu}\ a(\tau_{\mu})Q(\tau_{\mu})^4\nonumber \\
& & \nonumber \\
&\times& H_{4,4,\cdots,4}\left(\left\{\frac{\varphi(\beta)}{\sqrt{2}Q(\beta)} + \mathcal{I}(\tau_{i})\left(\frac{c(0)\varphi(0)}{\sqrt{2}}\right) \right\}_{i=1}^{\mu}, \left\{ \frac{\mathcal{I}(\tau_{i})}{4}\right\}_{i=1}^{\mu} | \tau_{i,k}=\frac{\mathcal{I}(\tau_{i})}{2}\right)\nonumber
\end {eqnarray}
The Hermite function in this final result $H_{4,4,\cdots,4}$ possesses $\mu$ equal indexes.
This means, that $H_{4,4,\cdots,4}$ is a combination of the products of $\mu$ Hermite two variable functions (\ref{hn2}), with specific arguments:

$$
H_{n}\left(\frac{\varphi(\beta)}{\sqrt{2}Q(\beta)} + \mathcal{I}(\tau_{i})\left(\frac{c(0)\varphi(0)}{\sqrt{2}}\right),\ \frac{\mathcal{I}(\tau_{i})}{4}\right)\ .
$$

By help of the explicit form the Hermite polynomial (\ref{hn2}), expanding the sum in the first argument,
we find:

\begin {equation}
\frac{1}{n!} H_n(\phi_{\beta} + \phi_0\mathcal{I}(\tau) , \gamma \mathcal{I}(\tau)) =
\sum_{\kappa=0}^n\ \mathcal{I}^{\kappa}(\tau)\ \mathcal{H}_{n-\kappa, \kappa}(\phi_{\beta},
\phi_0 | \gamma)\label{J10}
\end {equation}
where
$$\mathcal{H}_{n-\kappa, \kappa}(\phi_{\beta},\phi_0 | \gamma) =
\sum_{k=0}^{\min{(n-\kappa, \kappa)}}\ \frac{\phi^{n-\kappa-k}_{\beta}\phi^{\kappa-k}_0\gamma^k}{(n-\kappa-k)!k!(\kappa-k)!}\,=\,
(n-\kappa)!\kappa!\, H_{n-\kappa, \kappa}(\phi_{\beta},\phi_0 | \gamma).
$$
we defined the new symbols:
\begin {equation}
\phi_{\beta} = \frac{\varphi(\beta)}{\sqrt{2}Q(\beta)},
\end {equation}
\begin {equation}
\phi_0 = \frac{c(0)\varphi(0)}{\sqrt{2}}.
\end {equation}
The functions $H_{n-\kappa, \kappa}(\phi_{\beta},\phi_0 | \gamma)$ were introduced by Dattoli \cite{dattoli} as non-complete Hermite polynomials. For purposes of our evaluations it convenient to use the form $\mathcal{H}_{n-\kappa, \kappa}(\phi_{\beta},\phi_0 | \gamma),$ because derivative of
$\mathcal{H}_{n-\kappa, \kappa}(\phi_{\beta},\phi_0 | \gamma)$ simply lower the corresponding
index of $\mathcal{H}_{n-\kappa, \kappa}$.
The art described above (\ref{J10}) enables us to separate the integration variables dependent functions $\mathcal{I}(\tau_{i})$ out of the arguments of the Hermite polynomials.
This evaluations procedure enables us to present $\mu-th$ term of the an-harmonic part of the propagator in the closed form.
Let us demonstrate this procedure of separations for $\mu=4,$ extension to arbitrary $\mu$ is straightforward.

\section{The check of above evaluations  for $\mu = 4$}

In the preceding Appendix J, we have shown the construction of the $\mu$ multi-index, multi-variable Hermite polynomial by the recurrence derivative of the product of the exponential function and the $\mu-1$ multi-index, multi-variable Hermite polynomial. We shown that the polynomial is a linear combination of the products of $\mu$ one index two-variable polynomials (\ref{hn2}).  At the same time we have shown that in the particular case when the first variable of the two variables Hermite polynomial (\ref{hn2}) is the sum of two terms where one of them possesses the connection to the second variable, we found the conversion of $H_n$ to the combination of the non-complete Hermite polynomials (\ref{J10}). Can we combine both these characteristics together?
The affirmative answer we are going to demonstrate on the contribution to the series defining the an-harmonic correction for $\mu = 4$ term in (\ref{her61}). We will evaluate the derivatives for arbitrary $n$, instead of the real $n=4.$

The key term of evaluations is the Hermite polynomial
$$H_{n,n,n,n}(x_1, M_{11}, x_2, M_{22}, x_3, M_{33}, x_4, M_{44} | M_{12}, M_{12}, M_{12}, M_{23}, M_{23}, M_{34})$$
defined by the derivatives of the generating function:
\begin {eqnarray}
& &\exp(u_1^2 M_{11} + u_1 x_1 +\\
        & &u_2^2 M_{22} + u_2 x_2 + u_1 M_{12} u_2 +\nonumber \\
        & &u_3^2 M_{33} + u_3 x_3 + u_1 M_{12} u_3 + u_2 M_{23} u_3 +\nonumber \\
        & &u_4^2 M_{44} + u_4 x_4 + u_1 M_{12} u_4 + u_2 M_{23} u_4 + u_3 M_{34} u_4)\nonumber
\end {eqnarray}

Applying the derivatives in the order:
$$\partial_{u_1}^n(\partial_{u_2}^n(\partial_{u_3}^n(\partial_{u_4}^n(\exp{(...)}))))$$
following the identities (\ref{idJ1}), (\ref{idJ2}) and by limiting procedure $u_i \rightarrow 0,\ i=1, 2, 3, 4$
we find:
\begin {eqnarray}
& &\sum_{k_3=0}^n \sum_{k_2=0}^n \sum_{k_1=0}^n \binom{n}{k_3} \binom{n}{k_2} \binom{n}{k_1}
 M_{34}^{k_3} M_{23}^{k_2} M_{12}^{k_1} \frac{n!}{(n-k_3)!}\label{exsmu4} \\
&\times&\sum_{m=0}^{k_2} \binom{k_2}{m} \sum_{p=0}^{k_1} \binom{k_1}{p} \sum_{q=0}^{p} \binom{p}{q}\ H_{n-k_1}(x_1,M_{11})
 \frac{(n-k_2)!}{(n-k_1-k_2+p)!}H_{n-k_2-(k_1-p)}(x_2, M_{22})\nonumber \\
 &\times& \frac{(n-k_3)!}{(n-k_2-k_3+m-p+q)!}H_{n-k_3-(k_2-m)-(p-q)}(x_4, M_{44})\
 \frac{(n-k_3)!}{(n-k_3-m-q)!}  H_{n-k_3-m-q}(x_3, M_{33})\nonumber
\end {eqnarray}

Let us stress that by choosing the order of the derivatives, we only choose the "coordinate system" in summation space.
$H_s(x,y)$ are the two variable Hermite polynomials \cite{appell} mentioned in preceding Appendix (\ref{hn2}).

To simplify the notations we use:
$$H_{n}(x_i, M_{ii}) =
H_{n}\left(\frac{\varphi(\beta)}{\sqrt{2}Q(\beta)} + \mathcal{I}(\tau_{i})\left(\frac{c(0)\varphi(0)}{\sqrt{2}}\right),\ \mathcal{I}(\tau_{i})/4\right)\,=\,H_n(\phi_{\beta} + \phi_0\mathcal{I}(\tau_i) , \gamma \mathcal{I}(\tau_i))$$

In the following evaluations we use the identity between Hermite polynomials $H_n(\phi_{\beta} + \phi_0\mathcal{I}(\tau_i) , \gamma \mathcal{I}(\tau_i))$ and Dattoli's non-complete polynomials $H_{n-\kappa, \kappa}(\phi_{\beta},\phi_0 | \gamma)$. We proved such relation by replacement of the first variable in the definition of $H_n(a+b, c|\tau)$ by the sum of variables. Expanding the power of the sum in the first argument of $H_n$, and replacing the order of the summations we find:
$$\frac{1}{n!} H_n(\phi_{\beta} + \phi_0\mathcal{I}(\tau) , \gamma \mathcal{I}(\tau)) =
\sum_{\kappa=0}^n\ \mathcal{I}^{\kappa}(\tau)\ \mathcal{H}_{n-\kappa, \kappa}(\phi_{\beta},
\phi_0 | \gamma)$$
where
$$\mathcal{H}_{n-\kappa, \kappa}(\phi_{\beta},\phi_0 | \gamma) =
\sum_{k=0}^{\min{(n-\kappa, \kappa)}}\ \frac{\phi^{n-\kappa-k}_{\beta}\phi^{\kappa-k}_0\gamma^k}{(n-\kappa-k)!k!(\kappa-k)!}\,=\,
(n-\kappa)!\kappa!\, H_{n-\kappa, \kappa}(\phi_{\beta},\phi_0 | \gamma).$$
We can see, that this result take off the function $\mathcal{I}^{\kappa}(\tau)$ from the argument of the Hermite function. Therefore the Hermite functions will be tied to the multi-integral by the summations index $\kappa$ only.
Let us stress that derivative over variables $\phi_{\beta}$ or $\phi_{0}$ brings down the first, or the second index of $\mathcal{H}_{n-\kappa, \kappa}(\phi_{\beta},\phi_0 | \gamma).$ We will use this in summations when we replace the cumbersome dependence on the summation indexes by the derivatives.
Taking into account the definitions:
$$2M_{12}=\mathcal{I}(\tau_1),\ 2M_{23}=\mathcal{I}(\tau_2),\ 2M_{34}=\mathcal{I}(\tau_3), $$

we can read for Eq. (\ref{exsmu4}):
\begin {eqnarray}
& &(n!)^4\ \sum^n_{\kappa_1, \kappa_2, \kappa_3, \kappa_4=0}\
\sum_{k_3=0}^n\frac{1}{2^{k_3}k_3!} \sum_{k_2=0}^n\frac{1}{2^{k_2}k_2!} \sum_{k_1=0}^n\frac{1}{2^{k_1}k_1!}\\
&\times& \mathcal{I}^{\kappa_1+k_1}(\tau_1)\mathcal{I}^{\kappa_2+k_2}(\tau_2)
\mathcal{I}^{\kappa_3+k_3}(\tau_3)\mathcal{I}^{\kappa_4}(\tau_4)
\sum_{m=0}^{k_2} \binom{k_2}{m} \sum_{p=0}^{k_1} \binom{k_1}{p} \sum_{q=0}^{p} \binom{p}{q}\mathcal{H}_{n-(k_1+\kappa_1), \kappa_1}(\phi_{\beta}, \phi_0 | \gamma)\nonumber\\
&\times&
\mathcal{H}_{n-(k_2+\kappa_2)-(k_1-p), \kappa_2}(\phi_{\beta}, \phi_0 | \gamma)
\mathcal{H}_{n-(k_3+\kappa_3)-m-q, \kappa_3}(\phi_{\beta}, \phi_0 | \gamma)
\mathcal{H}_{n-k_3-(k_2-m)-(p-q)-\kappa_4, \kappa_4}(\phi_{\beta}, \phi_0 | \gamma)\nonumber
\end {eqnarray}
We proceed further by the summation indices transformations namely:
$$\kappa_1+k_1 = \overline{\kappa}_1,$$
$$\kappa_2+k_2 = \overline{\kappa}_2,$$
$$\kappa_3+k_3 = \overline{\kappa}_3,$$
$$\kappa_4 = \overline{\kappa}_4,$$
then the above equation is converted to
\begin {eqnarray}
& &(n!)^4\ \sum^n_{\overline{\kappa}_1, \overline{\kappa}_2, \overline{\kappa}_3, \overline{\kappa}_4=0}\
\sum_{k_3=0}^{\overline{\kappa}_3}\frac{1}{2^{k_3}k_3!} \sum_{k_2=0}^{\overline{\kappa}_2}\frac{1}{2^{k_2}k_2!} \sum_{k_1=0}^{\overline{\kappa}_1}\frac{1}{2^{k_1}k_1!}\\
&\times& \mathcal{I}^{\overline{\kappa}_1}(\tau_1)\mathcal{I}^{\overline{\kappa}_2}(\tau_2)
\mathcal{I}^{\overline{\kappa}_3}(\tau_3)\mathcal{I}^{\overline{\kappa}_4}(\tau_4)
\sum_{m=0}^{k_2} \binom{k_2}{m} \sum_{p=0}^{k_1} \binom{k_1}{p} \sum_{q=0}^{p} \binom{p}{q}\mathcal{H}_{n-\overline{\kappa}_1, \overline{\kappa}_1-k_1}(\phi_{\beta}, \phi_0 | \gamma)\nonumber\\
&\times&
\mathcal{H}_{n-\overline{\kappa}_2-(k_1-p), \overline{\kappa}_2-k_2}(\phi_{\beta}, \phi_0 | \gamma)
\mathcal{H}_{n-\overline{\kappa}_3-m-q, \overline{\kappa}_3-k_3}(\phi_{\beta}, \phi_0 | \gamma)
\mathcal{H}_{n-k_3-(k_2-m)-(p-q)-\overline{\kappa}_4, \overline{\kappa}_4}(\phi_{\beta}, \phi_0 | \gamma)\nonumber
\label{recurresult}
\end {eqnarray}
To fix the limits of the summations we use the characteristics of the polynomials $\mathcal{H},$ which are nonzero for non-negative valued indexes.
To summing up over indices $q, p, m$ we can profit the identities:
\begin {eqnarray}
& &\sum_{q=0}^{p} \binom{p}{q}\ \sum_{m=0}^{k_2} \binom{k_2}{m}
\mathcal{H}_{n-\overline{\kappa}_3-m-q, \overline{\kappa}_3-k_3}(\phi_{\beta}, \phi_0 | \gamma)
\mathcal{H}_{n-k_3-(k_2-m)-(p-q)-\overline{\kappa}_4, \overline{\kappa}_4}(\phi_{\beta}, \phi_0 | \gamma) =\nonumber \\
&=&\sum_{q=0}^{p} \binom{p}{q}\ \sum_{m=0}^{k_2} \binom{k_2}{m}
\left(\partial^{k_3}_{\phi_{0}}\mathcal{H}_{n-\overline{\kappa}_3-m-q, \overline{\kappa}_3}(\phi_{\beta}, \phi_0 | \gamma)\right)
\left(\partial^{k_3}_{\phi_{\beta}}\mathcal{H}_{n-(k_2-m)-(p-q)-\overline{\kappa}_4, \overline{\kappa}_4}(\phi_{\beta}, \phi_0 | \gamma)\right) =\nonumber \\
&=&\partial^p_{\phi_{\beta}}\left[ \partial^{k_2}_{\phi_{\beta}}\left[
\left(\partial^{k_3}_{\phi_{0}}\mathcal{H}_{n-\overline{\kappa}_3, \overline{\kappa}_3}(\phi_{\beta}, \phi_0 | \gamma)\right)
\left(\partial^{k_3}_{\phi_{\beta}}\mathcal{H}_{n-\overline{\kappa}_4, \overline{\kappa}_4}(\phi_{\beta}, \phi_0 | \gamma)\right)\right]\ \right]\nonumber
\end {eqnarray}
and also
\begin {eqnarray}
& & \sum_{p=0}^{k_1} \binom{k_1}{p}
\mathcal{H}_{n-\overline{\kappa}_2-(k_1-p), \overline{\kappa}_2-k_2}(\phi_{\beta}, \phi_0 | \gamma)
\partial^p_{\phi_{\beta}}\left[ \partial^{k_2}_{\phi_{\beta}}\left[
\left(\partial^{k_3}_{\phi_{0}}\mathcal{H}_{n-\overline{\kappa}_3, \overline{\kappa}_3}(\phi_{\beta}, \phi_0 | \gamma)\right)
\left(\partial^{k_3}_{\phi_{\beta}}\mathcal{H}_{n-\overline{\kappa}_4, \overline{\kappa}_4}(\phi_{\beta}, \phi_0 | \gamma)\right)\right]\ \right] =\nonumber \\
&=&\sum_{p=0}^{k_1} \binom{k_1}{p}\left[
\left(\partial^{k_2}_{\phi_{0}}\mathcal{H}_{n-\overline{\kappa}_2-(k_1-p), \overline{\kappa}_2}(\phi_{\beta}, \phi_0 | \gamma)\right)
\partial^{p}_{\phi_{\beta}}\partial^{k_2}_{\phi_{\beta}}\left[
\left(\partial^{k_3}_{\phi_{0}}\mathcal{H}_{n-\overline{\kappa}_3, \overline{\kappa}_3}(\phi_{\beta}, \phi_0 | \gamma)\right)
\left(\partial^{k_3}_{\phi_{\beta}}\mathcal{H}_{n-\overline{\kappa}_4, \overline{\kappa}_4}(\phi_{\beta}, \phi_0 | \gamma)\right)\right]\right] = \nonumber\\
&=&\partial^{k_1}_{\phi_{\beta}}\left[
\left(\partial^{k_2}_{\phi_{0}}\mathcal{H}_{n-\overline{\kappa}_2, \overline{\kappa}_2}(\phi_{\beta}, \phi_0 | \gamma)\right)
\partial^{k_2}_{\phi_{\beta}}\left[
\left(\partial^{k_3}_{\phi_{0}}\mathcal{H}_{n-\overline{\kappa}_3, \overline{\kappa}_3}(\phi_{\beta}, \phi_0 | \gamma)\right)
\left(\partial^{k_3}_{\phi_{\beta}}\mathcal{H}_{n-\overline{\kappa}_4, \overline{\kappa}_4}(\phi_{\beta}, \phi_0 | \gamma)\right)\right]\right]\nonumber
\end {eqnarray}
We arrived to final equation:
\begin {eqnarray}
(n!)^4\ \sum^n_{\overline{\kappa}_1, \overline{\kappa}_2, \overline{\kappa}_3, \overline{\kappa}_4=0}\
\mathcal{I}^{\overline{\kappa}_1}(\tau_1)\mathcal{I}^{\overline{\kappa}_2}(\tau_2)
\mathcal{I}^{\overline{\kappa}_3}(\tau_3)\mathcal{I}^{\overline{\kappa}_4}(\tau_4)
&&\sum_{k_3=0}^{\overline{\kappa}_3}\frac{1}{2^{k_3}k_3!} \sum_{k_2=0}^{\overline{\kappa}_2}\frac{1}{2^{k_2}k_2!} \sum_{k_1=0}^{\overline{\kappa}_1}\frac{1}{2^{k_1}k_1!}\times\\
 \left(\partial^{k_1}_{\phi_{0}}\mathcal{H}_{n-\overline{\kappa}_1, \overline{\kappa}_1}(\phi_{\beta}, \phi_0 | \gamma)\right)
\partial^{k_1}_{\phi_{\beta}}\left[
\left(\partial^{k_2}_{\phi_{0}}\mathcal{H}_{n-\overline{\kappa}_2, \overline{\kappa}_2}(\phi_{\beta}, \phi_0 | \gamma)\right)\partial^{k_2}_{\phi_{\beta}}\right.
&&\left.\left[\left(\partial^{k_3}_{\phi_{0}}\mathcal{H}_{n-\overline{\kappa}_3, \overline{\kappa}_3}(\phi_{\beta}, \phi_0 | \gamma)\right)
\left(\partial^{k_3}_{\phi_{\beta}}\mathcal{H}_{n-\overline{\kappa}_4, \overline{\kappa}_4}(\phi_{\beta}, \phi_0 | \gamma)\right)\right]\right] \nonumber
\end {eqnarray}

For purposes of the next evaluations we introduce the definition:
\begin {eqnarray}
\mathcal{D}(\overline{\kappa}_1,\overline{\kappa}_2,\overline{\kappa}_3,\overline{\kappa}_4) =
&&(n!)^4\ \sum_{k_3=0}^{\overline{\kappa}_3}\frac{1}{2^{k_3}k_3!} \sum_{k_2=0}^{\overline{\kappa}_2}\frac{1}{2^{k_2}k_2!} \sum_{k_1=0}^{\overline{\kappa}_1}\frac{1}{2^{k_1}k_1!}\times\\ \label{dfc}
&& \times \left(\partial^{k_1}_{\phi_{0}}\mathcal{H}_{n-\overline{\kappa}_1, \overline{\kappa}_1}(\phi_{\beta}, \phi_0 | \gamma)\right)
\partial^{k_1}_{\phi_{\beta}}\left[
\left(\partial^{k_2}_{\phi_{0}}\mathcal{H}_{n-\overline{\kappa}_2, \overline{\kappa}_2}(\phi_{\beta}, \phi_0 | \gamma)\right)\partial^{k_2}_{\phi_{\beta}}\right.\nonumber \\
&&\times \left.\left[\left(\partial^{k_3}_{\phi_{0}}\mathcal{H}_{n-\overline{\kappa}_3, \overline{\kappa}_3}(\phi_{\beta}, \phi_0 | \gamma)\right)
\left(\partial^{k_3}_{\phi_{\beta}}\mathcal{H}_{n-\overline{\kappa}_4, \overline{\kappa}_4}(\phi_{\beta}, \phi_0 | \gamma)\right)\right]\right] \nonumber
\end {eqnarray}

By the same method of evaluation, we can construct the function $\mathcal{D}$ for an arbitrary number of arguments.

Following algebraic evaluations described in the preceding section, we can conclude that the an-harmonic part of propagator (\ref{her61}) can be displayed as:
\begin {equation}
\mathcal{W}(\mu) = \frac{(4\mu)!}{4!}\ 2^{4\mu}(n!)^{\mu}
\sum^n_{\kappa_1,\cdots ,\kappa_{\mu}=0}\ I_{\kappa_1,\cdots ,\kappa_{\mu}}(\beta)\
\mathcal{D}(\kappa_1,\cdots ,\kappa_{\mu})
\end {equation}
The integral $I_{\kappa_1,\cdots ,\kappa_{\mu}}(\beta)$ is defined as:
\begin {equation}
I_{\kappa_1,\cdots ,\kappa_{\mu}}(\beta) =
\int_0^{\beta}a(\tau_1)Q^4(\tau_1)\mathcal{I}^{\kappa_1}(\tau_{1})
\int_{\tau_1}^{\beta}a(\tau_2)Q^4(\tau_2)\mathcal{I}^{\kappa_2}(\tau_{2})\cdots
\int_{\tau_{\mu-1}}^{\beta}a(\tau_{\mu})Q^4(\tau_{\mu})\mathcal{I}^{\kappa_{\mu}}(\tau_{\mu}).
\end {equation}
and the function $\mathcal{D}(\kappa_1,\cdots ,\kappa_{\mu})$ is the extension of the definition (\ref{dfc}) from $\mu=4$ to arbitrary $\mu.$


\begin{thebibliography}{99}

\bibitem{apel-car}T. Appelqui and J. Carrazone, Phys. Rev. D 11 (1975) 2856.
\bibitem{plet}J. Boh\' a\v cik, P. Pre\v snajder, Phys. Lett B 332 (1994) 366.
\bibitem{hs}J. Boh\' a\v cik, P. Pre\v snajder, FIZIKA B 17, (2008) 2, 355-362.
\bibitem{mehler}Mehler F.G., \textit{Ueber die Entwicklung einer Function von beliebing vielen Variabeln nach Laplaceschen Functionen höherer Ordnung}, Journal fur die Reine
    und Angewandte Mathematik (1866), 161 - 176.
\bibitem{hille} Hille E., \textit{A class of reciprocal functions}, Ann. Math. 27 (1926) 427 - 464.
\bibitem{dob} Doob J., \textit{The brownian movement and stochastic equations}, Ann. Math. 43, (1942) 351 - 369, Doob J., \textit{Stochastic processes}, Willey, New York
    1953.
\bibitem{fey} Feynman, R.P., Rev.Mod.Phys. 20 (1948) 367.
\bibitem{feynm} Feynman, R.P., Hibbs, A.P. (1965) \textit{Quantum Machanics and Path Integrals}, New York, McGraw - Hill
\bibitem{kac} Kac, M. \textit{On some connection between probability theory and differential and integral equations}. In: Proceedings of the second Berkeley Symposium on
    Probability and Statistics, J. Neyman ed., Berkeley: University of California Press 1951.
\bibitem{jaffe} Glimm J., Jaffe A. \textit{Quantum Physics}, 1981 by Springer-Verlag New York Inc.
\bibitem{gert}Roepstorff G., \textit{Path Integral  Approach to Quantum Physics}, Springer-Verlag, (1993).
\bibitem{das} A. Das, \textit{Field theory: A path integral
approach},World Scientific Publishing Co. Ptc. Ltd., 2006.
\bibitem{dem} Chaichian M., Demichev A., \textit{Path Integrals in
Physics}, Vol. I, IOP Publishing Ltd. 2001.
\bibitem{bateman} Bateman H., \textit{Higher Transcendental Functions}, Volume
II, Mc Graw-Hill, 1953.
\bibitem{prud}Prudnikov A.P., Britchkov J.A., Marichev O.I., \textit{Integrals and Series},
in Russian Nauka 1981, in English Gordon and Breach, New York 1986.
\bibitem{dattoli} Dattoli G., \textit{Incomplete 2D Hermite polynomials: properties and applications}, J. Math. Anal. Appl. 284 (2003) 447-454.
\bibitem{olver} Olver F.W.J., \textit{Uniform asymptotic expansions for Weber parabolic cylinder functions
of large order}, J. Research NBS, 63B:131-169, 1959.
\bibitem{temme} Temme N.M., \textit{Numerical and Asymptotic Aspect of
Parabolic Cylinder Functions}, J. of Computational and Applied Math., 121 (2000) 221-246.
\bibitem{temme2}Vidunas R., Temme N.M., \textit{Parabolic cylinder functions: Examples of error bounds for asymptotics expansions}, report MAS-R0225 October 31, 2002.
\bibitem{gelY}I. M. Gel`fand and A. M. Yaglom, J. Math. Phys. 1, 48 (1960).
\bibitem{pedrosa} Pedrosa I.A., Phys.Rev. A55, 3219 (1997).
\bibitem{grothaus}Grothaus M., Khandekar D.C., daSilva J.L., Streit L., \textit{The Feynman Integral for Time-dependent Anharmonic Oscillator}, J. Math. Phys. 38, 3278 (1997);
\bibitem{wipf}Wipf A., Statistical Approach to Quantum Field Theory, Lecture Notes in Physics, Springer, Berlin, Heidelberg, ISBN: 978-3-642-33104-6
\bibitem{thesis} Licciardi S.; PhD Thesis: Umbral Calculus, a Different Mathematical Language, 2018, arXiv:1803.03108.
\bibitem{dibucch}Di Bucchianico,A. and Loeb, D.E. (1995, last updated 2000) A selected survey of umbral calculus, Dynamical Survey 3, Electronic J. of Combinatorics.
\bibitem{my}J. Boh\' a\v cik and  P. Pre\v snajder, The functional
integral with unconditional Wiener measure for anharmonic oscillator, J. Math. Phys. 49, 113505 (2008).
\bibitem{appell} Appell P.E., Kamp\'{e} de F\'{e}riet J. \textit{Fonctions hyperg\'{e}om\'{e}triques: polynomes d`Hermite,} Gauthier-Villars, Paris, 1928.








\end{thebibliography}
\end{document}